\documentclass[11pt,a4paper]{article}

\usepackage{epsfig}
\usepackage{amsmath}
\usepackage{amssymb}
\usepackage{verbatim}
\usepackage{bm}
\usepackage{dsfont}
\usepackage[footnotesize]{caption}
\usepackage{subfigure}
\usepackage{cite}
\usepackage{textcomp}
\usepackage{calc}
\usepackage{geometry}
\usepackage{bbm}
\usepackage{xcolor}
\usepackage{multirow}
\usepackage{color}
\usepackage{float}
\usepackage{bbold}
\usepackage{gensymb}

\usepackage{hyperref}
\hypersetup{colorlinks=true,urlcolor=blue,linkcolor=red,citecolor=green!60!black}

\begin{document}

%%%%%%%%%%%%%%%%%%%%%%%%%%%%%%%%%%%%%%%%%%%%%%%%%%%%%%%%%%%%%%%%%%%%%%%%%%%%%%%%%%%%%%%%%%%%%%%%%%%%

\thispagestyle{empty}

\noindent June 2018
\hfill IPMU 18-0118

\vskip 1.5cm

\begin{center}
{\LARGE\bf Axion Isocurvature Perturbations in\\\medskip Low-Scale Models of Hybrid Inflation}

\vskip 2cm

\renewcommand*{\thefootnote}{\fnsymbol{footnote}}

{\large
Kai~Schmitz\,$^{a,\,\hspace{-0.25mm}}$%
\footnote{Corresponding author.
E-mail address: \href{mailto:kai.schmitz@mpi-hd.mpg.de}{kai.schmitz@mpi-hd.mpg.de}}
and Tsutomu T.\ Yanagida\,$^{b,\,\hspace{-0.25mm}}$%
\footnote{Hamamatsu professor.
\,E-mail address: \href{mailto:tsutomu.tyanagida@ipmu.jp}{tsutomu.tyanagida@ipmu.jp}}
}\\[3mm]
{\it{
$^{a}$ Max-Planck-Institut f\"ur Kernphysik (MPIK), 69117 Heidelberg, Germany\\
$^{b}$ Kavli IPMU, WPI, UTIAS, University of Tokyo, Kashiwa, Chiba 277-8583, Japan}}

\end{center}

\vskip 1cm

\renewcommand*{\thefootnote}{\arabic{footnote}}
\setcounter{footnote}{0}

%%%%%%%%%%%%%%%%%%%%%%%%%%%%%%%%%%%%%%%%%%%%%%%%%%%%%%%%%%%%%%%%%%%%%%%%%%%%%%%%%%%%%%%%%%%%%%%%%%%%

\begin{abstract}

%%%%%%%%%%%%%%%%%%%%%%%%%%%%%%%%%%%%%%%%%%%%%%%%%%%%%%%%%%%%%%%%%%%%%%%%%%%%%%%%%%%%%%%%%%%%%%%%%%%%

The QCD axion solves the strong $CP$ problem and represents
an attractive particle candidate for cold dark matter (CDM).
However, quantum fluctuations of the axion field during inflation easily result
in large CDM isocurvature perturbations that are in conflict with
observations of the cosmic microwave background (CMB).
In this paper, we demonstrate how this problem can be solved in low-scale
models of hybrid inflation that may emerge from supersymmetric grand unified theories.
We consider both F-term hybrid inflation (FHI) and D-term
hybrid inflation (DHI) in supergravity, explicitly taking
into account the effect of hidden-sector supersymmetry breaking.
We discuss the production of cosmic strings and show how the
soft terms in the scalar potential readily allow to achieve the
correct scalar spectral index.
In both cases, we are able to identify large regions in parameter space
that are consistent with all constraints.
In particular, we find that evading the CDM isocurvature constraint always
requires a small Yukawa or gauge coupling of $\mathcal{O}\left(10^{-3}\right)$ or smaller.
This translates into upper bounds on the gravitino mass of 
$\mathcal{O}\left(10^5\right)\,\textrm{GeV}$ in FHI and 
$\mathcal{O}\left(10^9\right)\,\textrm{GeV}$ in DHI.
Our results point to interesting scenarios in well-motivated parameter regions
that will be tested in future axion and CMB experiments.

%%%%%%%%%%%%%%%%%%%%%%%%%%%%%%%%%%%%%%%%%%%%%%%%%%%%%%%%%%%%%%%%%%%%%%%%%%%%%%%%%%%%%%%%%%%%%%%%%%%%

\end{abstract}

%%%%%%%%%%%%%%%%%%%%%%%%%%%%%%%%%%%%%%%%%%%%%%%%%%%%%%%%%%%%%%%%%%%%%%%%%%%%%%%%%%%%%%%%%%%%%%%%%%%%

\setcounter{page}{0}

\newpage

\setcounter{page}{1}
\setcounter{tocdepth}{1}

{\hypersetup{linkcolor=black}%\renewcommand{\baselinestretch}{1}
\tableofcontents}

%%%%%%%%%%%%%%%%%%%%%%%%%%%%%%%%%%%%%%%%%%%%%%%%%%%%%%%%%%%%%%%%%%%%%%%%%%%%%%%%%%%%%%%%%%%%%%%%%%%%

\section{Introduction}
\label{sec:introduction}

%%%%%%%%%%%%%%%%%%%%%%%%%%%%%%%%%%%%%%%%%%%%%%%%%%%%%%%%%%%%%%%%%%%%%%%%%%%%%%%%%%%%%%%%%%%%%%%%%%%%

The \textit{Peccei-Quinn} (PQ) mechanism~\cite{Peccei:1977hh,Peccei:1977ur} is a viable and
attractive solution to the strong $CP$ problem in \textit{quantum chromodynamics} (QCD).
It is based on the idea to promote the effective QCD vacuum angle $\bar{\theta}$
to a pseudoscalar field\,---\,known as the axion
$a = f_a\,\bar{\theta}$~\cite{Weinberg:1977ma,Wilczek:1977pj}\,---\,which
dynamically relaxes the QCD vacuum energy until it reaches a ground state that preserves
\textit{charge parity} ($CP$) invariance~\cite{Vafa:1984xg}.
The axion field is almost invisible, \textit{i.e.}, it is a weakly coupled gauge singlet
whose couplings are suppressed by a large decay constant $f_a$.
In concrete realizations of the
PQ mechanism~\cite{Kim:1979if,Shifman:1979if,Dine:1981rt,Zhitnitsky:1980tq},
the axion is identified as the pseudo-Nambu-Goldstone boson of a global $U(1)_{\rm PQ}$
symmetry that exhibits a nonvanishing $SU(3)$ color anomaly
(quantified in terms of an anomaly coefficient $N$) and that is
spontaneously broken at a high energy scale $v_{\rm PQ} = N f_a$.
As has become clear over the years, the QCD axion entails an extremely
rich phenomenology in particle physics, astrophysics, and cosmology
(for reviews, see~\cite{Kuster:2008zz,Kim:2008hd,Marsh:2015xka,Patrignani:2016xqp,Irastorza:2018dyq}),
which makes it a primary target in the hunt for new physics
\textit{beyond the Standard Model} (BSM).
The axion can, in particular, be copiously produced in the early Universe,
which renders it a well-motivated particle candidate for \textit{dark
matter} (DM)~\cite{Preskill:1982cy,Abbott:1982af,Dine:1982ah}.
Together, these observations distinguish the PQ mechanism as a testable and predictive
BSM scenario that not only solves the strong $CP$ problem but that automatically also
accounts for the DM relic density.

%%%%%%%%%%%%%%%%%%%%%%%%%%%%%%%%%%%%%%%%%%%%%%%%%%%%%%%%%%%%%%%%%%%%%%%%%%%%%%%%%%%%%%%%%%%%%%%%%%%%

In the context of inflationary
cosmology~\cite{Starobinsky:1980te,Guth:1980zm,Linde:1981mu,Albrecht:1982wi},
one has to discriminate between two different implementations
of the PQ mechanism, depending on the magnitude of the axion decay constant $f_a$.
First, consider the case in which the Hubble rate during inflation, $H_{\rm inf}$, always
exceeds $f_a$.
In this scenario, spontaneous \textit{PQ symmetry breaking} (PQSB) only occurs after
inflation in the radiation-dominated era.
Similarly, if the maximal temperature in the early Universe, $T_{\rm max}$,
is greater than $f_a$, the PQ symmetry is thermally restored after inflation,
and it only becomes spontaneously broken at lower temperatures
as soon as $T \sim f_a$.
In either case, PQ symmetry breaking occurs at late times, which results
in the production of cosmic strings.
During the QCD phase transitions, these axion strings turn into the boundaries
of domain walls~\cite{Sikivie:1982qv}.
One can show that, for an anomaly coefficient $N$, there are actually $N$ different types
of domain wall solutions, which is why $N$ is also referred to as the domain wall
number.
For $N > 1$, the domain walls are stable, so that they begin to dominate
the total energy density of the Universe soon after their formation.
This is known as the \textit{domain wall problem} of the \textit{postinflationary
PQSB scenario}.
There are several ways out of this problem.
An obvious solution is to simply restrict oneself
to a trivial domain wall number, $N = 1$.
This is, \textit{e.g.}, possible if only one vectorlike exotic quark contributes
to the PQ color anomaly (see~\cite{Ballesteros:2016euj,Ballesteros:2016xej}
for a recent example).
In this case, the domain walls are unstable and the entire string-wall
network decays, which results in a certain fraction of the total
axion DM relic density~\cite{Hiramatsu:2012gg,Kawasaki:2014sqa,Klaer:2017ond,Gorghetto:2018myk}.
Alternatively, one may explicitly break the PQ symmetry by means of higher-dimensional
operators in the effective theory~\cite{Ringwald:2015dsf}, so that
the domain walls become unstable even for a nontrivial domain wall number, $N > 1$.
However, this solution requires some tuning, as the tight upper bound on the effective
QCD theta angle, $\left|\bar\theta\right| \lesssim 10^{-10}$~\cite{Afach:2015sja},
restricts the allowed amount of explicit PQ symmetry breaking.

%%%%%%%%%%%%%%%%%%%%%%%%%%%%%%%%%%%%%%%%%%%%%%%%%%%%%%%%%%%%%%%%%%%%%%%%%%%%%%%%%%%%%%%%%%%%%%%%%%%%

The arguably simplest solution to the domain wall problem is to presume that the
PQ symmetry is already broken during inflation and never becomes restored afterwards.
This \textit{preinflationary PQSB scenario}
corresponds to the second possibility of implementing the PQ mechanism
in the context of inflationary cosmology.
It is realized for large values of the axion decay constant,
\begin{align}
f_a \gtrsim \max\left\{H_{\rm inf}, T_{\rm max}\right\} \,.
\end{align}
In this scenario, all dangerous topological defects that form at early times
are vastly diluted by the exponential expansion during inflation.
We emphasize that this solution neither constrains the value
of the domain wall number $N$ nor requires particular assumptions
about higher-dimensional operators in the effective theory.
Instead, one now has to deal with the implications of a spontaneously broken
global symmetry during inflation and, in particular, with the presence of the massless
axion field.
Just like the inflaton field, the axion field develops
quantum fluctuations during inflation.
These axion fluctuations are nearly scale-invariant and uncorrelated with the adiabatic
curvature perturbations, such that they turn into \textit{cold dark matter} (CDM)
isocurvature perturbations after
inflation~\cite{Axenides:1983hj,Linde:1985yf,Seckel:1985tj,
Lyth:1989pb,Linde:1990yj,Turner:1990uz,Linde:1991km}.
Axion isocurvature perturbations have attracted a great deal of attention in the
last two decades~\cite{Kawasaki:1995ta,Kawasaki:1997ct,
Kasuya:1997td,Kanazawa:1998pa,Kanazawa:1998nk,Fox:2004kb,Beltran:2006sq,Hertzberg:2008wr,
Hamann:2009yf,Kasuya:2009up,Kobayashi:2013nva,Jeong:2013xta,Kawasaki:2013iha,
Higaki:2014ooa,Kawasaki:2014una,Kitajima:2014xla,Kadota:2014hpa,
Nakayama:2015pba,Harigaya:2015hha,Kadota:2015uia,Kawasaki:2015lpf,Nomura:2015xil,
Kearney:2016vqw,Estevez:2016keg,Chung:2016wvv,Kawasaki:2017xwt,
Kawasaki:2017kkr,Chung:2017uzc}.
On the one hand, the prediction of measurable axion isocurvature perturbations
is exciting, as it implies that one may not only be able to probe the
QCD axion in laboratory experiments on Earth but also via
observations of the anisotropies in the \textit{cosmic microwave background} (CMB).
On the other hand, it represents an important restriction of the
preinflationary PQSB scenario, since the amplitude of the
isocurvature power spectrum is tightly constrained by the measurements of
the PLANCK satellite~\cite{Ade:2015xua,Ade:2015lrj}.
This issue is sometimes referred to as the \textit{axion isocurvature perturbations problem}.
The PLANCK constraint on the primordial isocurvature fraction especially implies
an upper bound on the inflationary Hubble rate $H_{\rm inf}$
that is in conflict with typical values of $H_{\rm inf}$ in high-scale models
of inflation.
The preinflationary PQSB scenario therefore calls for low-scale
inflation with a small Hubble rate.

%%%%%%%%%%%%%%%%%%%%%%%%%%%%%%%%%%%%%%%%%%%%%%%%%%%%%%%%%%%%%%%%%%%%%%%%%%%%%%%%%%%%%%%%%%%%%%%%%%%%

In this paper, we will demonstrate that the CDM isocurvature constraint on the inflationary
Hubble scale can be easily satisfied in low-scale models of hybrid
inflation~\cite{Linde:1991km,Linde:1993cn}.
To this end, we will revisit both \textit{F-term hybrid inflation}
(FHI)~\cite{Copeland:1994vg,Dvali:1994ms} and \textit{D-term hybrid inflation}
(DHI)~\cite{Binetruy:1996xj,Halyo:1996pp} in \textit{supergravity} (SUGRA).
These models represent promising inflationary scenarios.
They can be naturally embedded into supersymmetric \textit{grand unified theories}
(GUTs) and, hence, establish a connection between inflation and grand unification.
A particularly attractive feature is that both scenarios end
in a so-called \textit{waterfall transition}, \textit{i.e.}, a rapid
second-order phase transition that can be identified with the
spontaneous breaking of a local GUT symmetry.%
\footnote{The waterfall transition could, \textit{e.g.}, correspond to the spontaneous
breaking of a $U(1)_{B-L}$ gauge symmetry, where $B$ and $L$ denote
baryon and lepton number, respectively.
In this case, hybrid inflation would end in what is known
as the $B\!-\!L$ phase transition~\cite{Buchmuller:2010yy,Buchmuller:2011mw,
Buchmuller:2012wn,Buchmuller:2012bt,Buchmuller:2013lra,Schmitz:2012kaa,
Buchmuller:2013dja,Domcke:2013pma}, a promising framework for a unified picture 
of particle physics and cosmology~\cite{Domcke:2017xvu,Domcke:2017rzu}.
However, for the purposes of this paper, it will not be necessary
to specify the exact nature of the phase transition.}
The key idea behind our analysis is to  explicitly account for the spontaneous
breaking of \textit{supersymmetry} (SUSY) in a hidden sector.
As we will see, hidden-sector SUSY breaking results in a number of soft terms
in the scalar potential that can be used to achieve consistency with the CMB data.
In FHI, the dominant soft term turns out to be a linear tadpole term,
while in DHI, the leading soft term is a quadratic mass term.
In both cases, the size of the soft terms is controlled by
the gravitino mass $m_{3/2}$.
Therefore, by tuning the soft terms against the radiative corrections in
the scalar potential, one is always able to realize
a particularly flat inflaton potential, \textit{i.e.}, a very small
slow-roll parameter $\varepsilon \ll 1$.
At the same time, the energy scale in the tree-level potential, $V_0^{1/4}$,
can always be chosen so as to reproduce the amplitude of the scalar power spectrum,
$A_s \propto V_0/\varepsilon$.
Together, these two relations yield a powerful mechanism to suppress the
inflationary Hubble scale $H_{\rm inf} \propto V_0^{1/2}$.
In addition, the dependence of the slow-roll parameter $\varepsilon$
on $m_{3/2}$ links the gravitino mass to the Hubble rate,
$\varepsilon\left(m_{3/2},\cdots\right) \propto H_{\rm inf}^2$.
For a given $H_{\rm inf}$, we, thus, have to choose a gravitino mass of a certain
magnitude.
Otherwise, the scalar potential will be either too steep or too flat
to obtain the correct value for $A_s$.
For this reason, the CDM isocurvature constraint on $H_{\rm inf}$ can also
be used to derive upper bounds on $m_{3/2}$.

%%%%%%%%%%%%%%%%%%%%%%%%%%%%%%%%%%%%%%%%%%%%%%%%%%%%%%%%%%%%%%%%%%%%%%%%%%%%%%%%%%%%%%%%%%%%%%%%%%%%

To find the viable regions in parameter space, we will study the slow-roll
dynamics of FHI and DHI in a fully analytical fashion.
That is, wherever possible, we will refrain from resorting to the
usual numerical methods that are typically employed in the literature.
On the one hand, this will allow us to determine the implications of the
CDM isocurvature constraint on the model parameters of hybrid inflation
in an analytical and transparent manner.
On the other hand, our analysis will be rather general, so that our results
are actually well suited to be used in further investigations of hybrid inflation,
beyond the question of axion isocurvature perturbations.
The main result of our analysis will be that, in both FHI and DHI,
the inflationary Hubble scale can be pushed down to a
sufficiently small value\,---\,provided that an appropriate
coupling constant is set to a value of $\mathcal{O}\left(10^{-3}\right)$
or smaller.
In FHI, this coupling corresponds to the inflaton Yukawa coupling $\kappa$ in
the superpotential, while in DHI, it typically corresponds to the
gauge coupling $g$ in the waterfall sector.
In both cases, such a small coupling constant is stable against radiative
corrections and, hence, technically natural.
In supersymmetric hybrid inflation, the isocurvature perturbations problem of the QCD axion
can therefore be solved without any unnatural fine-tuning of model parameters.

%%%%%%%%%%%%%%%%%%%%%%%%%%%%%%%%%%%%%%%%%%%%%%%%%%%%%%%%%%%%%%%%%%%%%%%%%%%%%%%%%%%%%%%%%%%%%%%%%%%%

The remainder of this paper is organized as follows.
In the next section, we will review the CDM isocurvature constraint on $H_{\rm inf}$
in the preinflationary PQSB scenario.
In Secs.~\ref{sec:fhi} and \ref{sec:dhi}, we will then discuss in turn the inflationary
dynamics of FHI and DHI. 
In doing so, we will explicitly distinguish between scenarios
with a comparatively large field excursion during inflation
and scenarios with a very small field excursion during inflation.
In Sec.~\ref{sec:benchmark}, we will summarize our main results and discuss
a number of interesting benchmark points in parameter space.
For readers that are primarily interested in our constraints
on parameter space and less interested in the technical details of our slow-roll
analysis, we note that most of the results derived in this paper
are included in one way or another in Fig.~\ref{fig:summary}.
Finally, Sec.~\ref{sec:conclusions} contains our conclusions and a brief outlook.

%%%%%%%%%%%%%%%%%%%%%%%%%%%%%%%%%%%%%%%%%%%%%%%%%%%%%%%%%%%%%%%%%%%%%%%%%%%%%%%%%%%%%%%%%%%%%%%%%%%%

\section{Axion isocurvature perturbations}
\label{sec:axion}

%%%%%%%%%%%%%%%%%%%%%%%%%%%%%%%%%%%%%%%%%%%%%%%%%%%%%%%%%%%%%%%%%%%%%%%%%%%%%%%%%%%%%%%%%%%%%%%%%%%%

We begin by reviewing the CDM isocurvature constraint on $H_{\rm inf}$
in the preinflationary PQSB scenario.
First, we note that most properties of the QCD axion are fixed
by its decay constant $f_a$.
This includes the axion mass $m_a$, which can be obtained via an explicit calculation
in chiral perturbation theory~\cite{diCortona:2015ldu} as well as via
numerical lattice simulations~\cite{Borsanyi:2016ksw}.
The results of both approaches agree within their respective uncertainties and yield the
following expression for $m_a$,
\begin{align}
\label{eq:ma}
m_a \simeq 57.0 \,\textrm{\micro\hspace{0.5pt}eV} \left(\frac{10^{11}\,\textrm{GeV}}{f_a}\right) \,.
\end{align}
Next, let us consider the axion energy density $\Omega_a h^2$.
If the PQ symmetry is already broken before the end of inflation, the only
contribution to the axion abundance in the present epoch follows from the
standard vacuum misalignment mechanism~\cite{Preskill:1982cy,Abbott:1982af,Dine:1982ah}.
In this case, $\Omega_a h^2$ ends up being a function
of the axion decay constant $f_a$ and the initial value of the QCD vacuum angle,
$\bar{\theta}_{\rm ini} = a_{\rm ini}/f_a$, in the observable patch of the Universe.
For a small initial theta angle, $\left|\bar{\theta}_{\rm ini}\right| \ll \pi$, and
assuming that the axion field begins to coherently oscillate 
before the QCD phase transition, one finds~\cite{Ballesteros:2016euj,Ballesteros:2016xej}
\begin{align}
\label{eq:Oah2}
\Omega_a h^2 \simeq 0.65 \left(\frac{\bar{\theta}_{\rm ini}}{10^{-2}}\right)^2
\left(\frac{f_a}{10^{16}\,\textrm{GeV}}\right)^{1.17} \,.
\end{align}
This expression can be further refined by accounting for anharmonic effects
in the vicinity of the local maximum in the axion scalar potential, \textit{i.e.}, for
$\left|\bar{\theta}_{\rm ini}\right| \sim \pi$.
Therefore, following the analyses in~\cite{Visinelli:2009zm,Kobayashi:2013nva},
we shall modify Eq.~\eqref{eq:Oah2} by incorporating a correction factor $C_{\rm anh}$
of the following form,
\begin{align}
\label{eq:Oah2anh}
\Omega_a h^2 \simeq 0.65 \left(\frac{\bar{\theta}_{\rm ini}}{10^{-2}}\right)^2
\left(\frac{C_{\rm anh}\,f_a}{10^{16}\,\textrm{GeV}}\right)^{1.17} \,, \quad
C_{\rm anh} = 1 - \ln\left(1-\frac{\bar{\theta}_{\rm ini}^2}{\pi^2}\right) \,.
\end{align}

%%%%%%%%%%%%%%%%%%%%%%%%%%%%%%%%%%%%%%%%%%%%%%%%%%%%%%%%%%%%%%%%%%%%%%%%%%%%%%%%%%%%%%%%%%%%%%%%%%%%

This prediction needs to be compared with the PLANCK
result for the DM relic density~\cite{Ade:2015xua},
\begin{align}
\label{eq:ODMh2}
\Omega_{\rm DM} h^2 \simeq 0.12 \,.
\end{align}
Suppose that axions make up a fraction $F_{\rm DM}^a \in \left[0,1\right]$
of the total DM abundance.
The PLANCK constraint in Eq.~\eqref{eq:ODMh2} can then be used
to solve Eq.~\eqref{eq:Oah2anh} for the initial theta angle as a function of $f_a$,
\begin{align}
\label{eq:thetaini}
\bar{\theta}_{\rm ini} = \left(F_{\rm DM}^a\right)^{1/2}\,\bar{\theta}_{\rm ini}^{\rm DM} \,, \quad
F_{\rm DM}^a = \frac{\Omega_a}{\Omega_{\rm DM}} \,, \quad
\bar{\theta}_{\rm ini}^{\rm DM} \simeq 4.3 \times 10^{-3}
\left(\frac{10^{16}\,\textrm{GeV}}{f_a}\right)^{0.59} \,,
\end{align}
which is valid and self-consistent in the small-$\bar{\theta}$
regime where $C_{\rm anh} \approx 1$.
Also, note that $\bar{\theta}_{\rm ini}^{\rm DM}$ represents
the initial theta angle that is necessary to achieve pure axion DM.
The main lesson from Eq.~\eqref{eq:thetaini} is that large values 
of the axion decay constant, $f_a \gg 10^{12}\,\textrm{GeV}$, only
lead to viable axion DM if the initial theta angle is somewhat tuned.%
\footnote{An exception to this statement are models with an \textit{extremely}
low Hubble rate, $H_{\rm inf} \lesssim \Lambda_{\rm QCD}$,
where $\Lambda_{\rm QCD}$ denotes the QCD confinement
scale~\cite{Graham:2018jyp,Guth:2018hsa}.
However, in this paper, we will not be interested in this part of parameter space.}
However, it is important to realize that this kind of tuning is very
different from a brute-force tuning of the QCD vacuum angle
in a theory without a dynamical axion field.
First of all, note that, even for an axion decay constant as
large as $f_a \sim M_{\rm Pl}$, the required tuning is only
at the level of $1$ out of roughly $10^4$.
This is certainly less drastic than tuning $\bar{\theta}$ to a value
less than $10^{-10}$ by hand.
But the main \textit{conceptual} difference is that, in the QCD axion
scenario, the initial theta angle becomes susceptible to anthropic reasoning.
In a theory including a dynamical axion field, $\bar{\theta}_{\rm ini}$ controls
the final DM abundance (see Eq.~\eqref{eq:Oah2anh}).
As pointed out by Linde long ago~\cite{Linde:1987bx}, it may, thus, well be that
an apparently tuned theta angle in our observable
Universe is, in fact, the consequence of environmental selection during
inflation~(see also \cite{Tegmark:2005dy,Freivogel:2008qc}).

%%%%%%%%%%%%%%%%%%%%%%%%%%%%%%%%%%%%%%%%%%%%%%%%%%%%%%%%%%%%%%%%%%%%%%%%%%%%%%%%%%%%%%%%%%%%%%%%%%%%

If the axion field is already present during inflation, it will develop
quantum fluctuations that exhibit the typical standard deviation $\sigma_a$ of a massless
scalar field in an expanding de Sitter space,
\begin{align}
\sigma_a = \left<\delta a^2\right>^{1/2} \simeq \frac{H_{\rm inf}}{2\pi} \,,
\end{align}
which translates into the following standard deviation for 
the dynamical theta angle $\bar{\theta} = a/f_a$, 
\begin{align}
\sigma_{\bar{\theta}} = \frac{1}{f_a}\left<\delta a^2\right>^{1/2}
\simeq \frac{H_{\rm inf}}{2\pi f_a} \,.
\end{align}
By virtue of Eq.~\eqref{eq:Oah2anh}, these fluctuations in the initial
theta angle are responsible for the emergence of \textit{CDM density isocurvature} (CDI)
perturbations around the time of the QCD phase transition (\textit{i.e.}, at the onset of the
coherent axion oscillations)~\cite{Axenides:1983hj,Linde:1985yf,
Seckel:1985tj,Lyth:1989pb,Linde:1990yj,Turner:1990uz,Linde:1991km}.
Because the axion fluctuations during inflation are independent
of the quantum fluctuations of the inflaton field, the resulting CDI
perturbations are uncorrelated with the adiabatic curvature perturbations.
Up to corrections of $\mathcal{O}\left(\sigma_{\bar{\theta}}^3\right)$,
the magnitude of the axion isocurvature perturbations at a given length scale, $S_{\rm iso}$,
simply follows from the derivative of the (logarithm of the) axion energy density w.r.t.\ 
the initial theta angle (see, \textit{e.g.},~\cite{Kobayashi:2013nva}),
\begin{align}
S_{\rm iso} = \frac{\delta \Omega_{\rm DM}}{\Omega_{\rm DM}} = 
F_{\rm DM}^a\,\frac{\delta \Omega_a}{\Omega_a} \simeq
F_{\rm DM}^a\,\frac{\partial \ln \Omega_a}{\partial\,\bar{\theta}_{\rm ini}}\,\sigma_{\bar\theta} =
F_{\rm DM}^a\,\frac{2\sigma_{\bar{\theta}}}{\bar{\theta}_{\rm ini}} \,.
\end{align}
The square of this expression yields 
the amplitude of the isocurvature power spectrum $\mathcal{P}_{\rm iso}$,
\begin{align}
\label{eq:Piso}
\mathcal{P}_{\rm iso} = \left|S_{\rm iso}\right|^2 
\simeq \left(F_{\rm DM}^a\,
\frac{2\sigma_{\bar{\theta}}}{\bar{\theta}_{\rm ini}}\right)^2 \simeq 
\left(F_{\rm DM}^a\,\frac{H_{\rm inf}}{\pi f_a \bar{\theta}_{\rm ini}} \right)^2 =
F_{\rm DM}^a \mathcal{P}_{\rm iso}^{\rm DM} \,,
\end{align}
which holds in the small-$\bar{\theta}$ regime and up to corrections
of $\mathcal{O}\left(\sigma_{\bar{\theta}}^4\right)$.
In Eq.~\eqref{eq:Piso}, we again factored out the dependence on the axion DM fraction
$F_{\rm DM}^a$.
In the case of pure axion DM, one has
\begin{align}
\mathcal{P}_{\rm iso}^{\rm DM} \simeq 
\left(\frac{H_{\rm inf}}{\pi f_a \bar{\theta}_{\rm ini}^{\rm DM}} \right)^2 \,.
\end{align}

%%%%%%%%%%%%%%%%%%%%%%%%%%%%%%%%%%%%%%%%%%%%%%%%%%%%%%%%%%%%%%%%%%%%%%%%%%%%%%%%%%%%%%%%%%%%%%%%%%%%

The PLANCK data can be used to place an upper bound
on the primordial isocurvature fraction 
\begin{align}
\beta_{\rm iso}\left(k\right) = \frac{\mathcal{P}_{\rm iso}\left(k\right)}
{\mathcal{P}_{\rm adi}\left(k\right) + \mathcal{P}_{\rm iso}\left(k\right)} \,.
\end{align}
Here, we emphasize that both power spectra $\mathcal{P}_{\rm adi}$ and $\mathcal{P}_{\rm iso}$
are in general scale-dependent and, hence, functions of the wavenumber $k$.
The amplitude of the adiabatic curvature perturbations, $\mathcal{P}_{\rm adi}$,
is fixed by the observed amplitude of the primordial scalar power spectrum,
$\mathcal{P}_{\rm adi} \simeq 2.2 \times 10^{-9}$ at the CMB pivot
scale $ k = 0.05\,\textrm{Mpc}^{-1}$~\cite{Ade:2015lrj}.
For an uncorrelated mixture of adiabatic and CDI modes
and assuming a unit isocurvature spectral index, $n_{\rm iso} = 1$,
the PLANCK 2015 data results in~\cite{Ade:2015lrj},
\begin{align}
\beta_{\rm iso}\left(k\right) < 0.038 \,, \quad k = 0.05\,\textrm{Mpc}^{-1} \quad
\textrm{(95\,\% CL; TT,\,TE,\,EE\,$+$\,lowP)} \,,
\end{align}
which translates into an upper bound on the amplitude of the isocurvature power spectrum of
\begin{align}
\mathcal{P}_{\rm iso} \lesssim 8.7 \times 10^{-11} \,.
\end{align}
Making use of Eq.~\eqref{eq:Piso}, we, thus, obtain the following upper bound
on the inflationary Hubble rate,
\begin{align}
\label{eq:Hinfmax}
H_{\rm inf} \lesssim 1.3 \times 10^9 \,\textrm{GeV}
\left(\frac{1}{F_{\rm DM}^a}\right)^{1/2}
\left(\frac{f_a}{10^{16}\,\textrm{GeV}}\right)^{0.42} \,.
\end{align}
Two comments are in order in view of this bound.
First, we stress that Eq.~\eqref{eq:Hinfmax} is, indeed, a \textit{very tight}
restriction on the allowed set of inflationary models.
To see this more explicitly, recall that, in standard single-field slow-roll inflation,
$H_{\rm inf}$ uniquely determines the tensor-to-scalar ratio $r$,
\begin{align}
\label{eq:rHinf}
r = \frac{A_t}{A_s} = \frac{2}{A_s}\left(\frac{H_{\rm inf}}{\pi\,M_{\rm Pl}}\right)^2 \simeq
1.5 \times 10^{-11} \left(\frac{H_{\rm inf}}{10^9\,\textrm{GeV}}\right)^2 \,,
\end{align}
where $A_s$ and $A_t$ denote the amplitudes of the primordial scalar and tensor
power spectra, respectively.
The small values of $H_{\rm inf}$ that are required by Eq.~\eqref{eq:Hinfmax}
therefore imply that $r$ must be unobservably small.
This can also be formulated by rewriting Eq.~\eqref{eq:Hinfmax}
as an upper bound on $r$,
\begin{align}
\label{eq:rmax}
r \lesssim 2.4 \times 10^{-11} \left(\frac{1}{F_{\rm DM}^a}\right)
\left(\frac{f_a}{10^{16}\,\textrm{GeV}}\right)^{0.83} \,,
\end{align}
which needs to be contrasted with the current upper bound on the tensor-to-scalar ratio,
$r \lesssim 0.1$~\cite{Ade:2015lrj}.
Any detection of nonzero $r$ in the near future would therefore immediately
rule out all low-scale models of inflation that are in accord with
Eq.~\eqref{eq:Hinfmax}.%
\footnote{Of course, this is only true in the context of standard
single-field slow-roll inflation.
In extended scenarios (\textit{e.g.}, in the presence of additional sources of
gravitational waves), it may well be that the relation between $r$ and $H_{\rm inf}$
in Eq.~\eqref{eq:rHinf} no longer holds.
In this case, the tensor-to-scalar ratio may be boosted to large values that are within
reach of upcoming experiments, despite a small inflationary Hubble rate
(for a review of such nonstandard scenarios, see~\cite{Guzzetti:2016mkm}).}
A second comment regarding Eq.~\eqref{eq:Hinfmax} is that the dependence on
the axion DM fraction $F_{\rm DM}^a$ is actually rather mild.
Even if axions only account for, say, 10\,\% of the total DM abundance,
the bound is only relaxed by roughly a factor~$3$.
For this reason, it is impossible to evade the constraint on $H_{\rm inf}$
in high-scale models of inflation (where $H_{\rm inf} \sim 10^{13}\cdots10^{14}\,\textrm{GeV}$)
without completely abandoning the idea of an axion DM fraction.

%%%%%%%%%%%%%%%%%%%%%%%%%%%%%%%%%%%%%%%%%%%%%%%%%%%%%%%%%%%%%%%%%%%%%%%%%%%%%%%%%%%%%%%%%%%%%%%%%%%%

Thus far, we only focused on the small-$\bar{\theta}$ regime, where the anharmonic
correction factor in $C_{\rm anh}$ in Eq.~\eqref{eq:Oah2anh} can be neglected.
However, for completeness, we mention that all of the steps above
can also be repeated including $C_{\rm anh}$.
For $\left|\bar{\theta}_{\rm ini}\right| \sim \pi$, this can be even done analytically.
For small values of the axion decay constant $f_a$ and large values of
$\left|\bar{\theta}_{\rm ini}\right|$, a straightforward calculation yields
\begin{align}
\label{eq:HinfmaxE}
H_{\rm inf} \lesssim 88\,\textrm{GeV} \left(\frac{1}{F_{\rm DM}^a}\right)^{0.15}
e^{-12.64\,\left(E-1\right)} \,, \quad E = \left(\frac{F_{\rm DM}^a}{1}\right)^{0.85}
\left(\frac{10^{10}\,\textrm{GeV}}{f_a}\right) \,.
\end{align}
This is an extremely strong constraint that can only be satisfied
in more or less unconventional scenarios of inflation.
In the following, we will therefore focus our attention 
on the bound in Eq.~\eqref{eq:Hinfmax} and its implications for hybrid inflation.
The bound in Eq.~\eqref{eq:HinfmaxE} will only appear in Fig.~\ref{fig:summary}
where it serves the purpose to mark the boundary of the viable parameter space 
at small values of $f_a$.

%%%%%%%%%%%%%%%%%%%%%%%%%%%%%%%%%%%%%%%%%%%%%%%%%%%%%%%%%%%%%%%%%%%%%%%%%%%%%%%%%%%%%%%%%%%%%%%%%%%%

\section{Low-scale F-term hybrid inflation}
\label{sec:fhi}

%%%%%%%%%%%%%%%%%%%%%%%%%%%%%%%%%%%%%%%%%%%%%%%%%%%%%%%%%%%%%%%%%%%%%%%%%%%%%%%%%%%%%%%%%%%%%%%%%%%%

\subsection{Model setup and scalar potential}

%%%%%%%%%%%%%%%%%%%%%%%%%%%%%%%%%%%%%%%%%%%%%%%%%%%%%%%%%%%%%%%%%%%%%%%%%%%%%%%%%%%%%%%%%%%%%%%%%%%%

We now turn to supersymmetric hybrid inflation and determine the implications
of the CDM isocurvature constraint in Eq.~\eqref{eq:Hinfmax} on its parameter space.
First, we will consider FHI supplemented with a hidden SUSY-breaking
sector~\cite{Buchmuller:2000zm}.
The relevant terms in the superpotential are given as
\begin{align}
\label{eq:WFHI}
W = \kappa\, S\, \Phi \bar{\Phi} - \mu_S^2 S + \mu_X^2 X + w \,,
\end{align}
where $S$ denotes the chiral inflaton field, $\Phi$ and $\bar{\Phi}$ are two
chiral waterfall fields, and $X$ is the Polonyi field.
$\kappa$ is a dimensionless Yukawa coupling, while $\mu_S$
and $\mu_X$ denote the inflaton and Polonyi F-term mass scales, respectively.
$w$ represents a constant contribution to the superpotential that arises
in consequence of $R$ symmetry breaking.
Its value needs to be tuned so as to achieve a vanishingly small
\textit{cosmological constant} (CC) in the true vacuum after inflation.
The first two terms on the \textit{right-hand side} (RHS) of Eq.~\eqref{eq:WFHI}
represent the superpotential of FHI, while the last two terms
coincide with the superpotential of the standard Polonyi model of
spontaneous SUSY breaking~\cite{Polonyi:1977pj}.
For simplicity, we shall assume that all chiral fields possess a canonical
K\"ahler potential to leading order,
\begin{align}
\label{eq:KFHI}
K = S^\dagger S + \Phi^\dagger \Phi + \bar{\Phi}^\dagger \bar{\Phi} + X^\dagger X + 
\frac{\chi}{M_{\rm Pl}^2}\,S^\dagger S\, X^\dagger X + \cdots \,.
\end{align}
Here, we include a higher-dimensional coupling between $S$ and $X$
whose strength is controlled by a dimensionless coefficient $\chi$.
This operator is allowed by all symmetries and expected to be present
in the effective theory at energies below the Planck scale,
$M_{\rm Pl} \simeq 2.44 \times 10^{18}\,\textrm{GeV}$.
As we will see, it contributes to the soft SUSY-breaking parameters
in the scalar inflaton potential.
The ellipsis in Eq.~\eqref{eq:KFHI} stands for further
higher-dimensional operators that are negligible for the present discussion.
We only remark that the K\"ahler potential should also contain a
higher-dimensional self-interaction for the Polonyi field, $K \supset - \left|X\right|^4/M_*^2$
for some high mass scale $M_*$, such that $X$ is always
safely stabilized at the origin in field space.
This can, \textit{e.g.}, be achieved via additional couplings to 
matter fields in the hidden sector (see~\cite{Izawa:1996pk,Intriligator:1996pu,Chacko:1998si}
for an example based on strong gauge dynamics).
Provided that $\left<X\right>  = 0$ for all times during and after inflation,
the parameters $\mu_X$, $w$, and $m_{3/2}$ can be related to each other
based on the requirement that the CC must vanish in the true vacuum after inflation,
\begin{align}
\label{eq:wm32muX}
w = m_{3/2}\, M_{\rm Pl}^2 \,, \quad
m_{3/2} = \frac{\mu_X^2}{\sqrt{3}\,M_{\rm Pl}} \,.
\end{align}

%%%%%%%%%%%%%%%%%%%%%%%%%%%%%%%%%%%%%%%%%%%%%%%%%%%%%%%%%%%%%%%%%%%%%%%%%%%%%%%%%%%%%%%%%%%%%%%%%%%%

The waterfall fields $\Phi$ and $\bar{\Phi}$ transform in
conjugate representations of a gauge group $G$ that may be part
of a larger GUT gauge group, $G \subseteq G_{\rm GUT}$.
The inflaton and the Polonyi field are supposed to transform
as complete singlets under the group $G$.
In the following, we will restrict ourselves to the simplest scenario of an
Abelian gauge group, $G = U(1)$.
In this case, the gauge interactions in the waterfall sector result
in a D-term scalar potential of the form
\begin{align}
V_D = \frac{g^2}{2}
\left[q\left(\left|\phi\right|^2 - \left|\bar{\phi}\right|^2 \right)\right]^2 \,,
\end{align}
where $g$ denotes the $G$ gauge coupling constant and $+q$ and $-q$ are
the $G$ gauge charges of the waterfall fields $\Phi$ and $\bar{\Phi}$.
In the following, we will set $q=1$ without loss of generality.
The situation with arbitrary charge $q$ can always be restored
by redefining the gauge coupling, $g \rightarrow g'= g/q$.
The D-term scalar potential ensures that the \textit{vacuum expectation values}
(VEVs) of the two waterfall fields coincide at all times, 
$\big<\Phi\big> = \big<\bar{\Phi}\big>$.
Apart from this, it is irrelevant for the dynamics of FHI.
During inflation, the two waterfall fields are stabilized at the origin in
field space, $\big<\Phi\big> = \big<\bar{\Phi}\big> = 0$, while after inflation
(\textit{i.e.}, after the waterfall phase transition), both fields acquire
a nonzero VEV,
\begin{align}
\label{eq:PhiVEV}
\big<\Phi\big> = \big<\bar{\Phi}\big> = \frac{v}{\sqrt{2}} \,, \quad
v = \sqrt{2}\,\frac{\mu_S}{\kappa^{1/2}} \,.
\end{align}
The VEV $v$ characterizes the energy scale at which
$G$ becomes spontaneously broken.
It is normalized such that it corresponds to the aligned VEVs of the two
\textit{real} Higgs scalars contained in $\Phi$ and $\bar{\Phi}$.

%%%%%%%%%%%%%%%%%%%%%%%%%%%%%%%%%%%%%%%%%%%%%%%%%%%%%%%%%%%%%%%%%%%%%%%%%%%%%%%%%%%%%%%%%%%%%%%%%%%%

The relevant contribution to the tree-level potential stems from the
F-term scalar potential,
\begin{align}
\label{eq:VF_fhi}
V_F = e^z\left[\left(1-z+z^2\right)\mu_S^4
+ \sqrt{2}\left(2-z\right)\mu_S^2\,m_{3/2}\,s\,\cos\varphi
+ \frac{1-\chi\left(3-z\right)}{2\left(1+\chi\,z\right)}\,m_{3/2}^2\,s^2\right] \,.
\end{align}
This potential is understood to be evaluated along the inflationary trajectory
where $\big<\Phi\big> = \big<\bar{\Phi}\big> = 0$.
The real field variables $s$, $\varphi$, and $z$ are related to
the original complex inflaton field $S$ as follows,
\begin{align}
S = \frac{s}{\sqrt{2}}\,e^{i\varphi} \,, \quad
S^* = \frac{s}{\sqrt{2}}\,e^{-i\varphi} \,, \quad
z = \frac{\left|S\right|^2}{M_{\rm Pl}^2} = \frac{s^2}{2\,M_{\rm Pl}^2} \,.
\end{align}
Remarkably enough, all terms in $V_F$\,---\,except
for the constant contribution to the
vacuum energy density\,---\,correspond to corrections
that only arise in the context of SUGRA.
An investigation of FHI without the proper inclusion of SUGRA effects is therefore
highly incomplete~\cite{Panagiotakopoulos:1997qd,Linde:1997sj}.
At field values below the Planck scale, the F-term scalar potential
in Eq.~\eqref{eq:VF_fhi} can be expanded as follows,
\begin{align}
\label{eq:VF_fhi_Taylor}
V_F = V_F^0 + c_s\left(1+\frac{z}{2}\right)s + 
\frac{1}{2}\,m_s^2\,s^2 + \frac{1}{24}\,\lambda_s\,s^4 + \mathcal{O}\left(s^6\right) \,.
\end{align}
Here, the leading term $V_F^0 = \mu_S^4$ corresponds to the F-term scalar
potential in the global-SUSY limit.
It is constant and sets the inflationary Hubble scale during FHI.
To good approximation, we have
\begin{align}
H_{\rm inf} \simeq \frac{\left(V_F^0\right)^{1/2}}{\sqrt{3}\,M_{\rm Pl}} =
\frac{\mu_S^2}{\sqrt{3}\,M_{\rm Pl}} \,.
\end{align}

%%%%%%%%%%%%%%%%%%%%%%%%%%%%%%%%%%%%%%%%%%%%%%%%%%%%%%%%%%%%%%%%%%%%%%%%%%%%%%%%%%%%%%%%%%%%%%%%%%%%

Eq.~\eqref{eq:VF_fhi_Taylor} also contains a linear tadpole
term whose strength is controlled by the coefficient $c_s$,
\begin{align}
\label{eq:cs}
c_s = 2\sqrt{2}\,\mu_S^2\,m_{3/2}\cos\varphi \,.
\end{align}
This term has important consequences
for the dynamics of FHI~\cite{Buchmuller:2000zm,Rehman:2009nq,
Rehman:2009yj,Nakayama:2010xf,Buchmuller:2014epa}.
In particular, it introduces a dependence on the complex inflaton phase $\varphi$
(through the $\cos\varphi$ factor in $c_s$),
which breaks the rotational invariance in the complex inflaton plane.
FHI consequently turns into a two-field model of inflation whose
full dynamics can only be captured by a comprehensive analysis of
all possible trajectories in the complex plane~\cite{Buchmuller:2014epa}.
However, for the purposes of this paper, we will restrict ourselves
to the case of inflation along the negative real axis where $\varphi = \pi$.
This is the simplest case and motivated by the fact that it
will provide us with the \textit{strongest} bounds on parameter space.
As will become clear later on, our final results are therefore valid and applicable 
for \textit{all} trajectories in the complex plane and do not rely on any assumption
regarding the particular choice of trajectory.
Besides that, we note that also the other coefficients in Eq.~\eqref{eq:VF_fhi}
have an important physical meaning.
$m_s^2$ and $\lambda_s$ denote the inflaton mass and the
inflaton quartic self-coupling, respectively,
\begin{align}
\label{eq:msls}
m_s^2 = \left(1-3\chi\right) m_{3/2}^2 \,, \quad
\lambda_s = 3 \left(\frac{\mu_S}{M_{\rm Pl}}\right)^4 + 
6\left(1-3\chi+3\chi^2\right)\left(\frac{m_{3/2}}{M_{\rm Pl}}\right)^2 \,.
\end{align}
In contrast to $c_s$, these coefficients also depend on the parameter $\chi$
in the K\"ahler potential.
However, since the linear tadpole term in Eq.~\eqref{eq:VF_fhi_Taylor}
will turn out to be most relevant for inflation,
the dependence of $m_s^2$ and $\lambda_s$ on $\chi$
is actually negligible and we can safely set $\chi = 0$ in the remainder
of the section.%
\footnote{The situation will be different in the case of DHI in Sec.~\ref{sec:dhi},
where we will have to set $\chi$ to a value $\chi > 1/3$, so that the inflaton mass
becomes tachyonic, $m_s^2 < 0$.
In the present section, we merely introduced $\chi$ for illustrative purposes.}

%%%%%%%%%%%%%%%%%%%%%%%%%%%%%%%%%%%%%%%%%%%%%%%%%%%%%%%%%%%%%%%%%%%%%%%%%%%%%%%%%%%%%%%%%%%%%%%%%%%%

Next, let us compute the one-loop effective potential $V_{1\ell}$.
In doing so, we shall work in the rigid global-SUSY limit and neglect any
gravitational corrections to the one-loop effective potential.
These corrections are suppressed by combinations of loop factors
and inverse powers of the Planck scale and are, hence, negligible.
$V_{1\ell}$ follows from the standard Coleman-Weinberg formula~\cite{Coleman:1973jx},
which means that we have to determine the mass spectrum
in the waterfall sector in an arbitrary inflaton background.
As for the scalars, we find two complex mass
eigenstates $\phi_\pm$ with masses $m_\pm$,
\begin{align}
\label{eq:mpm_fhi}
m_\pm^2 = m_{\rm eff}^2 \pm m_F^2 \,, \quad 
m_{\rm eff}^2 = \frac{1}{2}\kappa^2 s^2 \,, \quad 
m_F^2 = \kappa\,\mu_S^2 \,.
\end{align}
These masses can also be written as $m_\pm^2 = \kappa^2/2\left(s^2 \pm v^2\right)$,
which illustrates that $\phi_-$ becomes tachyonic
at the critical inflaton field value $s_{\rm crit} = v$.
That is, once the inflaton $s$ reaches its
critical value, the complex scalar $\phi_-$ becomes unstable.
This marks the onset of the waterfall transition.
The mass degeneracy among $\phi_+$ and $\phi_-$ is lifted by $m_F$.
This mass parameter is a direct consequence of
F-term SUSY breaking during inflation, which is evident from
its dependence on the inflaton F-term mass scale $\mu_S$.
The waterfall fermion $\tilde{\phi}$
does not receive any SUSY-breaking mass contributions.
It simply acquires an ordinary Dirac mass, $m_{\tilde{\phi}} = m_{\rm eff}$,
which corresponds to the effective supersymmetric mass
$m_{\rm eff} = \kappa \left<S\right>$ that is induced
by the VEV of the chiral inflaton field $S$ in the
superpotential.
With the mass spectrum at our disposal,
we can immediately write down the one-loop effective potential,
\begin{align}
\label{eq:V1l_fhi}
V_{1\ell} = \frac{1}{2}\,V_{1l}^0\, L\left(x\right) \,, \quad
V_{1\ell}^0 = \frac{m_F^4}{8\pi^2} \,, \quad
x = \left(\frac{s}{s_{\rm crit}}\right)^2 =
\left(\frac{m_{\rm eff}}{m_F}\right)^2 = \bigg(\frac{s}{v}\bigg)^2 \,.
\end{align}
Here, the field variable $x$ measures the distance to the critical
field value $s_{\rm crit}$ in field space.
The constant factor $V_{1\ell}^0$ (which is completely determined
by the SUSY-breaking mass parameter $m_F$) characterizes the overall energy scale,
while the loop function $L$ captures the actual field dependence,
\begin{align}
\label{eq:Lfunc}
L\left(x\right) = \frac{1}{2}\sum_\pm \left(x\pm1\right)^2
\left[\ln\left(x\pm1\right)-\frac{3}{2}\right] - x^2 \left[\ln x - \frac{3}{2}\right] \,.
\end{align}

%%%%%%%%%%%%%%%%%%%%%%%%%%%%%%%%%%%%%%%%%%%%%%%%%%%%%%%%%%%%%%%%%%%%%%%%%%%%%%%%%%%%%%%%%%%%%%%%%%%%

The combination of Eqs.~\eqref{eq:VF_fhi} and \eqref{eq:V1l_fhi} provides us
with the total inflaton potential, $V = V_F + V_{1\ell}$, which sets the stage
for our slow-roll analysis in the following two sections.
However, before turning to the details of inflation, let us
comment on the issue of \textit{cosmic strings} (CSs).
Recall that we assume an Abelian gauge group in the waterfall sector, $G = U(1)$.
For this reason, the spontaneous breaking of $G$ during
the waterfall transition is accompanied
by the production of topological defects in the form of cosmic strings~\cite{Kibble:1976sj,
Vilenkin:1984ib,Hill:1987qx,Hindmarsh:1994re,Jeannerot:2003qv}.
This poses a severe problem for supersymmetric hybrid inflation.
Cosmic strings are expected to leave an imprint in several
cosmological observables, such as the CMB~\cite{Ade:2013xla,Ade:2015xua},
the spectrum of stochastic gravitational
waves~\cite{Sanidas:2012ee,Blanco-Pillado:2017rnf,Ringeval:2017eww},
and the diffuse gamma-ray background~\cite{Mota:2014uka}.
However, no signs of cosmic strings were detected thus far, which allows
to severely constrain the parameter space of supersymmetric hybrid
inflation~\cite{Jeannerot:2005mc,Battye:2006pk,Battye:2010hg}.
The main quantity of interest in the context of cosmic strings
is the cosmic string tension $\mu_{\rm CS}$ (\textit{i.e.}, the
cosmic string energy density per unit length).
A robust and more or less model-independent upper bound on the cosmic string
tension follows from the nonobservation
of cosmic strings in the CMB~\cite{Charnock:2016nzm,Lizarraga:2016onn},
\begin{align}
\label{eq:GmuCSmax}
G\,\mu_{\rm CS}^{\rm max} \sim 1 \times 10^{-7} \,,
\end{align}
where $G = \left(8\pi M_{\rm Pl}^2\right)^{-1}$ denotes Newton's
gravitational constant.

%%%%%%%%%%%%%%%%%%%%%%%%%%%%%%%%%%%%%%%%%%%%%%%%%%%%%%%%%%%%%%%%%%%%%%%%%%%%%%%%%%%%%%%%%%%%%%%%%%%%

The bound in Eq.~\eqref{eq:GmuCSmax} translates into a
strong constraint on the VEV $v$, \textit{i.e.}, on the energy
scale of \textit{spontaneous symmetry breaking} (SSB)
during the waterfall transition.
To see this, it is convenient to rewrite the superpotential
in Eq.~\eqref{eq:WFHI} in terms of the $G$ Higgs multiplet $H$
in unitary gauge,
\begin{align}
\label{eq:WFHIH}
\Phi = \frac{H}{\sqrt{2}}\,e^A \,, \quad
\bar{\Phi} = \frac{H}{\sqrt{2}}\,e^{-A} \qquad\Rightarrow\qquad
W \supset \frac{\kappa}{2}\,S\left(H^2 - v^2\right) \,,
\end{align}
where the multiplet $A$ contains the Goldstone degrees of freedom
of spontaneous $G$ breaking.
Eq.~\eqref{eq:WFHIH} illustrates that the \textit{complex} Higgs boson
contained in $H$ acquires a VEV $\left<H\right> = v$.
This is larger by a factor $\sqrt{2}$ than the complex VEVs
of the fields $\Phi$ and $\bar{\Phi}$ (see Eq.~\eqref{eq:PhiVEV}).
In the broken phase, the physical Higgs boson, thus, obtains a mass 
$m_H^2 = \kappa^2 v^2$, while the vector boson obtains a mass
$m_V^2 = 2\,g^2 v^2$.
These masses allow us to determine the cosmic string tension
(see, \textit{e.g.}~\cite{Hindmarsh:1994re}),
\begin{align}
\label{eq:muCSbeta}
\mu_{\rm CS} = 2\,\pi v^2\,\epsilon_{\rm CS}\left(\beta\right) \,, \quad 
\beta = \left(\frac{m_H}{m_V}\right)^2 = \frac{\kappa^2}{2\,g^2} \,.
\end{align}
Here, the factor $2$ on the RHS stems from the fact that, in FHI, there
are \textit{two} real waterfall fields that participate in the process of
spontaneous symmetry breaking.
This factor is absent in the case of DHI (see Sec.~\ref{subsec:dhi_model}),
and even in the case of FHI, it is sometimes overlooked in the literature.
The factor $\pi v^2$ can be derived analytically and corresponds 
to the cosmic string tension in the so-called Bogomolny limit~\cite{Bogomolny:1975de}
where the Higgs boson is degenerate with the vector boson, such that $\beta = 1$.
For $\beta \neq 1$, the cosmic string tension
needs to be determined numerically.
This is accounted for by the function $\epsilon_{\rm CS}$, which may
be regarded as the cosmic string tension in units of $\pi v^2$
per real Higgs boson. 
In the following, we will approximate $\epsilon_{\rm CS}$ by the numerical
fit function obtained in~\cite{Hill:1987qx},
\begin{align}
\label{eq:epsilonCS}
\epsilon_{\rm CS}\left(\beta\right) \simeq \begin{cases}
1.19 / \left(2/\beta\right)^{0.195} & ;\quad \beta \gtrsim  10^{-2} \\
2.40 / \ln\left(2/\beta\right)      & ;\quad \beta \lesssim 10^{-2} \\
\end{cases} \,,
\end{align}
which is roughly consistent with the Bogomolny limit,
$\epsilon\left(1\right)=1$.
For definiteness, we will also fix the gauge coupling
$g$ at a value that one obtains in typical GUT models,
$g = \left(\pi/6\right)^{1/2} \simeq 0.72$.
This is a rather large value that tends to lead to small $\beta$ values
and, hence, to a more conservative bound on the SSB scale $v$.
In summary, we obtain for the cosmic string
tension in Planck units
\begin{align}
\label{eq:GmuCS}
G\,\mu_{\rm CS} = \frac{1}{4}\left(\frac{v}{M_{\rm Pl}}\right)^2
\epsilon_{\rm CS}\left(\kappa\right)
\,, \quad \epsilon_{\rm CS}\left(\kappa\right) =
\left.\epsilon_{\rm CS}\left(\beta\right)\right|_{\beta = 3/\pi\,\kappa^2} \,.
\end{align}
Making use of Eq.~\eqref{eq:GmuCSmax}, this expression results
in the following upper bound on the SSB scale $v$,
\begin{align}
\label{eq:vCS}
v \lesssim 3.6 \times 10^{15}\,\textrm{GeV}
\left(\frac{0.18}{\epsilon_{\rm CS}}\right)^{1/2}
\left(\frac{G\mu_{\rm CS}^{\rm max}}{10^{-7}}\right)^{1/2} \,.
\end{align}
where we anticipated that $\epsilon_{\rm CS} \simeq 0.18$
for $v \simeq 3.6 \times 10^{15}\,\textrm{GeV}$ (see Eqs.~\eqref{eq:kappaCSFHI}
and \eqref{eq:vCSFHI} in Sec.~\ref{subsec:fhi_close}).

%%%%%%%%%%%%%%%%%%%%%%%%%%%%%%%%%%%%%%%%%%%%%%%%%%%%%%%%%%%%%%%%%%%%%%%%%%%%%%%%%%%%%%%%%%%%%%%%%%%%

Eq.~\eqref{eq:vCS} represents a strong constraint on the parameter space
of FHI.
In the following, we will therefore pursue two different philosophies in parallel.
In one part of our analysis, we will adopt the notion that the bound in Eq.~\eqref{eq:vCS} 
must, indeed, be considered as a serious and physically relevant restriction.
In this case, we will demonstrate how the bound on the cosmic string tension
enables us to constrain the other parameters of our model.
However, in the rest of our analysis, we will simply ignore the bound in
Eq.~\eqref{eq:vCS} and pretend that no cosmic strings are formed during
the waterfall transition.
This is, \textit{e.g.}, possible if, on the one hand, the gauge group $G$ is already
spontaneously broken in a different sector before the end of inflation and if, on
the other hand, this breaking is somehow communicated to the waterfall sector
via marginal couplings in the superpotential or K\"ahler potential
(see \cite{Domcke:2017xvu,Domcke:2017rzu,Evans:2017bjs} for an explicit example
in the context of DHI).
From the perspective of the waterfall fields, the gauge group $G$ is then
explicitly broken to a certain (marginal) degree, such that
no cosmic strings can form in this sector.
Instead, cosmic strings may still form at early times, when $G$ is initially
broken in the hidden sector.
But these cosmic strings will be diluted during the inflationary
expansion, so that they no longer leave any observable signatures in our Universe.
In this case, the bound in Eq.~\eqref{eq:vCS} does not apply any longer,
which permits us to simply ignore it.

%%%%%%%%%%%%%%%%%%%%%%%%%%%%%%%%%%%%%%%%%%%%%%%%%%%%%%%%%%%%%%%%%%%%%%%%%%%%%%%%%%%%%%%%%%%%%%%%%%%%

\subsection{Inflation far away from the waterfall phase transition}
\label{subsec:fhi_far}

%%%%%%%%%%%%%%%%%%%%%%%%%%%%%%%%%%%%%%%%%%%%%%%%%%%%%%%%%%%%%%%%%%%%%%%%%%%%%%%%%%%%%%%%%%%%%%%%%%%%

We are now all set to discuss the inflationary slow-roll dynamics.
Our analysis will be split into two parts.
First, we will consider the case of a relatively large field excursion,
$x \gg 1$, which is realized for larger values of
the inflaton Yukawa coupling, $\kappa \gtrsim \mathcal{O}\left(10^{-3}\right)$.
As shown below, this scenario only complies with
the CDM isocurvature constraint for a very large axion decay
constant, $f_a \sim M_{\rm Pl}$.
In Sec.~\ref{subsec:fhi_close}, we will then turn to the case of a small
field excursion, $ x\simeq 1$, which is
realized for $\kappa \lesssim \mathcal{O}\left(10^{-3}\right)$.
In this regime, we will find viable parameter regions for any
reasonable value of $f_a$.

%%%%%%%%%%%%%%%%%%%%%%%%%%%%%%%%%%%%%%%%%%%%%%%%%%%%%%%%%%%%%%%%%%%%%%%%%%%%%%%%%%%%%%%%%%%%%%%%%%%%

\begin{figure}
\begin{center}

\includegraphics[width=0.485\textwidth]{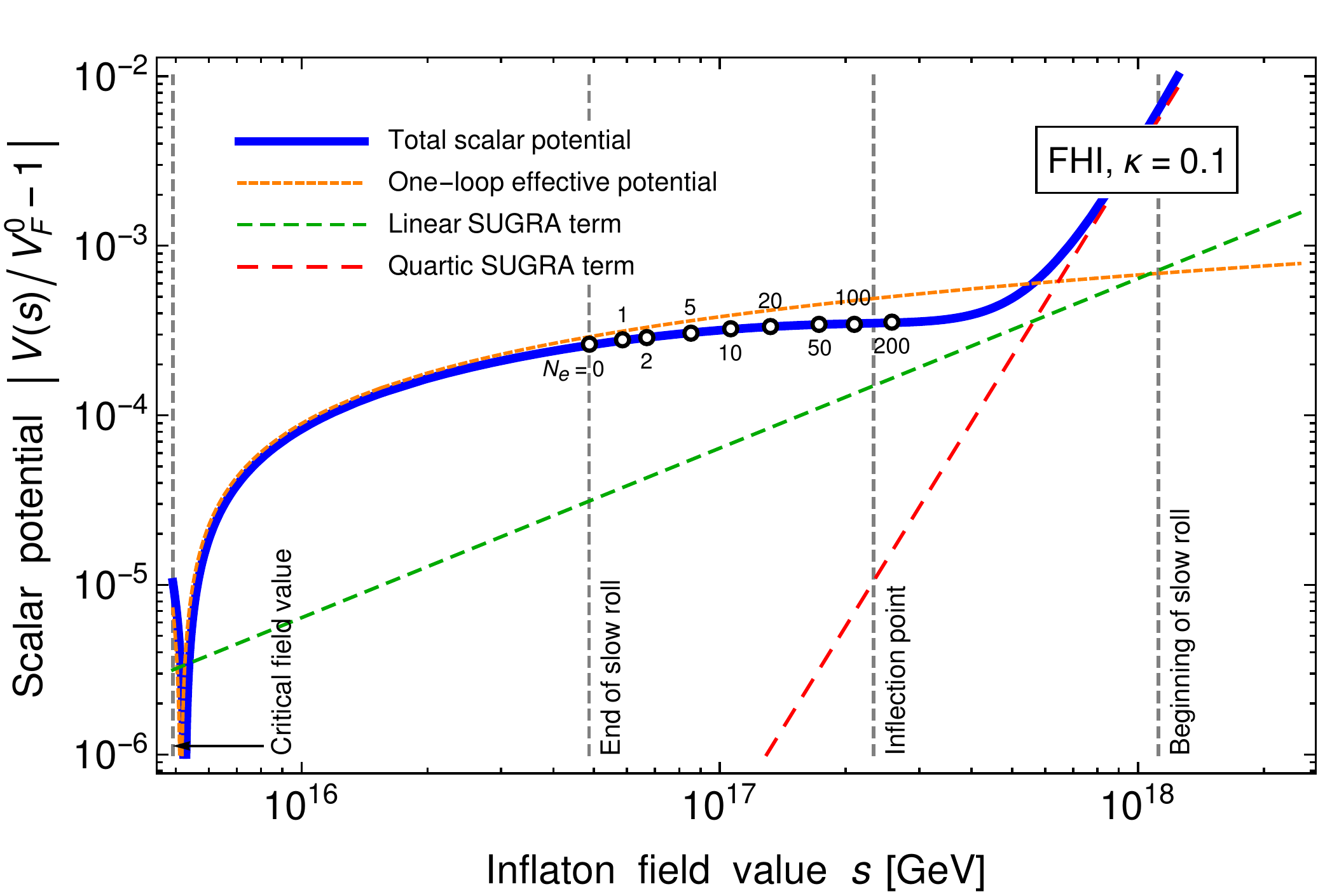}\hfill
\includegraphics[width=0.475\textwidth]{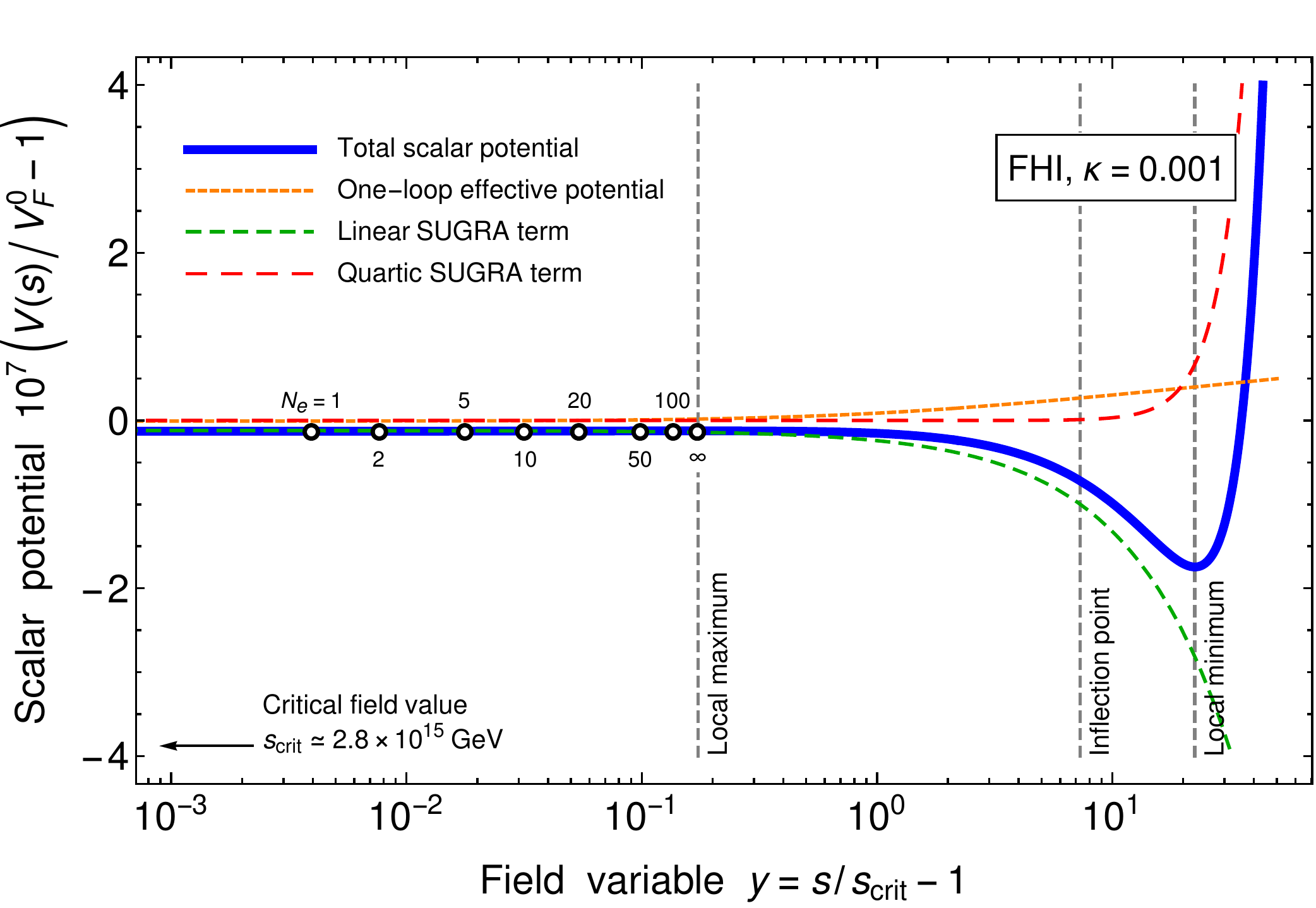}

\caption{Total scalar potential for the real inflaton field $s$ in F-term hybrid inflation
for two representative values of the inflaton Yukawa coupling $\kappa$.
Parameter values: \textbf{(Left panel)} $\kappa = 10^{-1}$,
$\mu_S \simeq 1.1 \times 10^{15}\,\textrm{GeV}$,
$m_{3/2} \simeq 2.7 \times 10^8\,\textrm{GeV}$ and 
\textbf{(Right panel)} $\kappa = 10^{-3}$,
$\mu_S \simeq 6.3 \times 10^{13}\,\textrm{GeV}$,
$m_{3/2} \simeq 6.0 \times 10^3\,\textrm{GeV}$.
Both parameter points are chosen such that they reproduce the measured
CMB observables, $A_s = A_s^{\rm obs}$ and $n_s = n_s^{\rm obs}$.
The left panel represents an example for inflation in the inflection-point regime, while
the right panel represents an example for inflation in the hill-top regime.
In both plots, we also compare the linear, quartic, and radiative contributions
to the total scalar potential.}

\label{fig:potential_fhi}

\end{center}
\end{figure}

In the large-field limit, the loop function $L$ in Eq.~\eqref{eq:Lfunc}
is well approximated by a simple logarithm,
\begin{align}
\label{eq:L_far}
L\left(x\right) = \ln x + \mathcal{O}\left(x^{-2}\right) \,.
\end{align}
The total scalar potential describing inflation in the large-field limit, thus,
takes the following form,%
\footnote{This form of the potential explains the factor $1/2$ in front of
the logarithmic term.
In Eq.~\eqref{eq:V1l_fhi}, we normalized the factor $V_{1\ell}^0$ 
in such a way that the one-loop effective potential reduces to
$V_{1\ell} \simeq V_{1\ell}^0 \ln\left(s/s_{\rm crit}\right)$ in the large-field limit.}
\begin{align}
\label{eq:V_fhi_approx}
V \simeq V_F^0 + c_s\,s +  \frac{1}{24}\,\lambda_s\,s^4 +
\frac{1}{2}\,V_{1l}^0\, \ln\left(x\right) \,.
\end{align}
Here, we omitted the quadratic and cubic terms in Eq.~\eqref{eq:VF_fhi_Taylor}.
The quadratic mass term can be neglected because all viable
inflationary solutions will turn out to require a
small gravitino mass, $m_{3/2}^2 \ll H_{\rm inf}^2$.
Similarly, the cubic term can be neglected compared to the linear tadpole term
because inflation will always take place at sub-Planckian field values, $s\ll M_{\rm Pl}$.
In Fig.~\ref{fig:potential_fhi}, we plot the total scalar potential for two
representative values of the inflaton Yukawa coupling,
$\kappa = 10^{-1}$ and $\kappa = 10^{-3}$, and compare it with the
field-dependent contributions in Eqs.~\eqref{eq:V_fhi_approx}
and \eqref{eq:V_fhi_close} (see below in Sec.~\ref{subsec:fhi_close}).
In both cases, the linear, quartic, and radiative terms are sufficient
to describe the full shape of the scalar potential at field values below
the Planck scale.
Let us now collect a few properties of the scalar
potential $V$ in Eq.~\eqref{eq:V_fhi_approx}.
First of all, we note that the scalar potential always exhibits
an inflection point, $V''\left(s_{\rm flex}\right) = 0$,
whose location is solely determined by the coupling $\kappa$,
\begin{align}
s_{\rm flex} = \left(\frac{2\,V_{1\ell}^0}{\lambda_s}\right)^{1/4} \simeq
\left(\frac{\kappa}{2\sqrt{3}\,\pi}\right)^{1/2} M_{\rm Pl} \simeq
2.3 \times 10^{17}\,\textrm{GeV}\:\bigg(\frac{\kappa}{0.1}\bigg)^{1/2} \,.
\end{align}
The potential gradient at the inflection point, $V'\left(s_{\rm flex}\right)$,
is controlled by the gravitino mass,
\begin{align}
V'\left(s_{\rm flex}\right) = 2\sqrt{2}
\left(m_{3/2}^{\rm crit}-m_{3/2}\right)\mu_S^2 \,, \quad 
m_{3/2}^{\rm crit} = 
\frac{1}{3\,\mu_S^2}\left[2\,\lambda_s\left(V_{1\ell}^0\right)^3\right]^{1/4} \,.
\end{align}
Here, the negative sign in front of $m_{3/2}$ stems from the fact that we are
considering inflation on the negative real axis where $\varphi = \pi$
(see the discussion below Eq.~\eqref{eq:cs}).
$m_{3/2}^{\rm crit}$ denotes the critical value of the gravitino mass
for which the inflection point turns into a saddle point,
$V''\left(s_{\rm flex}\right) = V'\left(s_{\rm flex}\right) = 0$,
\begin{align}
\label{eq:m32_crit_fhi}
m_{3/2}^{\rm crit} \simeq
\left(\frac{\kappa}{\sqrt{3}\,\pi}\right)^{3/2}\frac{\mu_S^2}{4\,M_{\rm Pl}} \simeq
2.6\times 10^8\,\textrm{GeV}\:\bigg(\frac{\kappa}{0.1}\bigg)^{3/2}
\bigg(\frac{\mu_S}{10^{15}\,\textrm{GeV}}\bigg)^2 \,.
\end{align}

%%%%%%%%%%%%%%%%%%%%%%%%%%%%%%%%%%%%%%%%%%%%%%%%%%%%%%%%%%%%%%%%%%%%%%%%%%%%%%%%%%%%%%%%%%%%%%%%%%%%

%%%%%%%%%%%%%%%%%%%%%%%%%%%%%%%%%%%%%%%%%%%%%%%%%%%%%%%%%%%%%%%%%%%%%%%%%%%%%%%%%%%%%%%%%%%%%%%%%%%%

For $m_{3/2} > m_{3/2}^{\rm crit}$, the potential gradient
at the inflection point is negative, $V'\left(s_{\rm flex}\right) < 0$.
This results in the occurrence of a local maximum and a local minimum
in the potential near the inflection point,
$s_{\rm max} < s_{\rm flex} < s_{\rm min}$.
Conversely, for $m_{3/2} < m_{3/2}^{\rm crit}$, the potential gradient
at the inflection point is positive,  $V'\left(s_{\rm flex}\right) > 0$.
In this case, the potential is monotonically increasing
without any local extrema in the vicinity of $s_{\rm flex}$.
To distinguish between these two regimes, \textit{i.e.}, the \textit{hill-top regime}
and the \textit{inflection-point regime}, it is convenient
to introduce the following dimensionless parameter,
\begin{align}
\zeta = \bigg(\frac{m_{3/2}^{\rm crit}}{m_{3/2}}\bigg)^2 = 
\left(\frac{\kappa}{\sqrt{3}\,\pi}\right)^3
\left(\frac{\mu_S^2}{4\,m_{3/2}\,M_{\rm Pl}}\right)^2 \,.
\end{align}
The hill-top and inflection-point regimes then correspond to $\zeta < 1$
and $\zeta \geq 1$, respectively.
Both regimes are suitable for inflation.
In the hill-top regime, inflation can occur near $s_{\rm max}$,
while in the inflection-point regime, it can occur near
$s_{\rm flex}$ if the potential is sufficiently flat.
The parameter $\zeta$ also allows us to write down compact expressions
for $s_{\rm max}$ and $s_{\rm min}$ in the hill-top regime,
\begin{align}
\label{eq:sminmax}
s_{\rm max} = F_-\left(\zeta\right) s_{\rm flex} \,, \quad
s_{\rm min} = F_+\left(\zeta\right) s_{\rm flex} \,,
\end{align}
where $F_+$ and $F_-$ are complicated functions that correspond
to the roots of a quartic polynomial,
\begin{align}
F_\pm\left(\zeta\right) = A^{1/2} \pm
\left[\left(\zeta A\right)^{-1/2}-A\right]^{1/2} \,, \quad
A = \frac{B_+^2 + B_-^2}{2\,B_+\,B_-}  \,, \quad
B_\pm = \left[1\pm\left(1-\zeta^2\right)^{1/2}\right]^{1/6} \,.
\end{align}
If inflation occurs at $s \lesssim s_{\rm max}$, the quartic term in
Eq.~\eqref{eq:V_fhi_approx} is typically subdominant.
This allows us to expand $s_{\rm max}$ in
Eq.~\eqref{eq:sminmax} for small values of $\zeta$.
Up to corrections of $\mathcal{O}\left(\zeta^{5/2}\right)$, this results in
\begin{align}
s_{\rm max} \simeq \frac{3}{4}\,\zeta^{1/2}\,s_{\rm flex} = 
- \frac{V_{1\ell}^0}{c_s} = \frac{\kappa^2}{16\sqrt{2}\,\pi^2}\frac{\mu_S^2}{m_{3/2}} \,,
\end{align}
which coincides with the result that one obtains if one sets $\lambda_s \rightarrow 0$
in Eq.~\eqref{eq:V_fhi_approx} from the outset.

%%%%%%%%%%%%%%%%%%%%%%%%%%%%%%%%%%%%%%%%%%%%%%%%%%%%%%%%%%%%%%%%%%%%%%%%%%%%%%%%%%%%%%%%%%%%%%%%%%%%

Next, after these remarks on the potential,
let us compute the slow-roll parameters $\varepsilon$ and $\eta$,
\begin{align}
\varepsilon = \frac{M_{\rm Pl}^2}{2}\left(\frac{V'}{V}\right)^2 \,, \quad
\eta = M_{\rm Pl}^2\,\frac{V''}{V} \,, \quad
V' = \frac{\partial V}{\partial s} \,, \quad
V'' = \frac{\partial^2 V}{\partial s^2} \,.
\end{align}
For simplicity, we shall work in the $\lambda_s \rightarrow 0$ limit from now on,
which will yield acceptable results as long as $\zeta \lesssim \mathcal{O}\left(1\right)$.
In fact, we will justify the small-$\lambda_s$ approximation \textit{a posteriori}
by an explicit numerical analysis that demonstrates the validity of our analytical results.
For small $\lambda_s$, we obtain
\begin{align}
\label{eq:ee_fhi}
\varepsilon = \frac{1}{2}\left(\frac{c_s s + V_{1\ell}^0}{V_F^0}\right)^2
\left(\frac{M_{\rm Pl}}{s}\right)^2 \,, \quad
\eta = - \frac{V_{1\ell}^0}{V_F^0}\left(\frac{M_{\rm Pl}}{s}\right)^2 \,.
\end{align}
Note that $\varepsilon$ is suppressed by a
factor $V_{1\ell}^0/V_F^0$ compared to $\eta$.
As usual in supersymmetric hybrid inflation,
the duration of inflation is therefore controlled by $\eta$\,---\,slow-roll
inflation only occurs as long as $\eta$ is small.
To make this statement more precise, let us impose the following
condition on $\eta$,
\begin{align}
\label{eq:etamax}
\left|\eta\right| \lesssim \eta_{\rm max} = 10^{-0.5} \,.
\end{align}
The transition between slow-roll inflation and the subsequent fast-roll stage
is therefore reached at
\begin{align}
\label{eq:sfastFHI}
s_{\rm fast} = \left(\frac{V_{1\ell}^0}{m_{\rm max}^2}\right)^{1/2} =
\frac{\kappa\,M_{\rm Pl}}{2\sqrt{2}\,\pi\,\eta_{\rm max}^{1/2}} \,, \quad
m_{\rm max}^2 = \eta_{\rm max}\,\frac{V_F^0}{M_{\rm Pl}^2} \,.
\end{align}
At this field value, $\left|\eta\right|$ saturates the upper bound in Eq.~\eqref{eq:etamax}.
The mass parameter $m_{\rm max}^2$ in Eq.~\eqref{eq:sfastFHI}
denotes the maximal curvature of the potential, $V''$, that is
allowed by the upper bound on $\eta$.
Given the expression for $s_{\rm fast}$ in Eq.~\eqref{eq:sfastFHI}, we are now
able to determine the end point of inflation.
Slow-roll inflation either ceases once the inflaton field
enters the fast-roll regime (\textit{i.e.}, at
$s = s_{\rm fast}$) or once it reaches the critical point in field space
that triggers the waterfall transition (\textit{i.e.}, at
$s = s_{\rm crit}$),
\begin{align}
\label{eq:send}
s_{\rm end} = \max\left\{s_{\rm fast},s_{\rm crit}\right\} \,.
\end{align}
The slow-roll parameters in Eq.~\eqref{eq:ee_fhi} also allow us to compute
the inflationary CMB observables,
\begin{align}
\label{eq:Asns}
A_s = \frac{1}{24\,\pi^2}\frac{V}{\varepsilon\,M_{\rm Pl}^4} \,, \quad
n_s = 1 + 2\,\eta - 6\,\varepsilon \,,
\end{align} 
where $A_s$ and $n_s$ denote the amplitude and the spectral index of
the scalar power spectrum, respectively.
An important step in our analysis will be to identify the parameter regions
that manage to reproduce the measured values of these observables.
According to the PLANCK 2015 data~\cite{Ade:2015lrj},
\begin{align}
A_s^{\rm obs} \simeq 2.2 \times 10^{-9} \,, \quad
n_s^{\rm obs} \simeq 0.9645 \quad
\textrm{(TT,\,TE,\,EE\,$+$\,lowP)} \,.
\end{align}
We will not be interested in the tensor-to-scalar ratio $r$.
This observable is predicted to be unobservably small in the entire parameter
space of interest (see the discussion related to Eq.~\eqref{eq:rmax}).
The expressions in Eq.~\eqref{eq:Asns} can be used to compute 
theoretical predictions for $A_s$ and $n_s$.
To this end, the slow-roll parameters $\varepsilon$ and $\eta$
need to be evaluated at $s = s_*$, \textit{i.e.}, at the inflaton field value
that corresponds to the horizon exit of the CMB pivot scale $N_*$ $e$-folds before the end
of inflation,
\begin{align}
N_* \simeq 47.4 + \frac{1}{3}\ln\left(\frac{H_{\rm inf}}{10^9\,\textrm{GeV}}\right)
+ \frac{1}{3}\ln\left(\frac{T_{\rm rh}}{10^9\,\textrm{GeV}}\right) \,,
\end{align}
where $T_{\rm rh}$ denotes the reheating temperature after inflation.
If not specified otherwise, we will use $T_{\rm rh} \simeq 10^9\,\textrm{GeV}$ as a
benchmark in the following, which is motivated by thermal leptogenesis~\cite{Fukugita:1986hr}.

%%%%%%%%%%%%%%%%%%%%%%%%%%%%%%%%%%%%%%%%%%%%%%%%%%%%%%%%%%%%%%%%%%%%%%%%%%%%%%%%%%%%%%%%%%%%%%%%%%%%

The dynamics of the inflaton
field are governed by the following slow-roll equation of motion,
\begin{align}
\label{eq:eom_fhi_far}
s' = \Delta \left(\frac{s_0}{s}-1\right) s_0 \,, \quad
s' = \frac{ds}{N_e} \,, \quad
s_0 = - \frac{V_{1\ell}^0}{c_s} \,.
\end{align}
Here, $s'$ stands for the derivative of the inflaton field $s$
w.r.t.\ the number of $e$-folds $N_e$ until the end of inflation.
The reference field value $s_0$ corresponds to the (would-be)
position of the local maximum in the scalar potential.
That is, $s_0$ is defined through the relation $s_0 = -V_{1\ell}^0/c_s$, which coincides
with $s_{\rm max}$ in the hill-top regime (\textit{i.e.}, for $\zeta < 1$).
The parameter $\Delta$ in Eq.~\eqref{eq:eom_fhi_far} measures
the strength of the linear SUGRA term in the scalar potential
in relation to the radiative corrections,
\begin{align}
\label{eq:Delta_fhi}
\Delta = \frac{c_s^2 M_{\rm Pl}^2}{V_{1\ell}^0 V_F^0} =
\frac{1}{3}\left(\frac{8\pi}{\kappa}\right)^2\left(\frac{m_{3/2}}{H_{\rm inf}}\right)^2 \,.
\end{align}
Given the boundary condition that the field $s$ must reach $s_{\rm end}$
for $N_e = 0$, Eq.~\eqref{eq:eom_fhi_far} has a unique solution in terms of the 
(principal branch of the) Lambert W function or product logarithm $W_0$,
\begin{align}
\label{eq:s_fhi_far}
s\left(N_e\right) = s_0 \left(1 + W\right) \,, \quad
W = W_0\left[\left(\frac{s_{\rm end}}{s_0}-1\right)
\exp\left(\frac{s_{\rm end}}{s_0}-1\right)e^{-\Delta\,N_e}\right] \,.
\end{align}
$W_0$ is the inverse function of the product function $X e^X$ and, thus, features
the following properties,
\begin{align}
\label{eq:Lambert}
X = W_0\left(X\right) e^{W_0\left(X\right)} \,, \quad 
W_0\left(Xe^X\right) = X \,, \quad 
W_0\left(0\right) = 0 \,, \quad
W_0\left(X\right) \geq -1 \,.
\end{align}
The solution in Eq.~\eqref{eq:s_fhi_far} can also be written as a function
of the three parameters $\eta_{\rm max}$, $N_e$, and $\Delta$,
\begin{align}
\label{eq:_s_fhi_far_W}
s\left(N_e\right) = s_0 \left(1 + W\right) \,, \quad
W = W_0\left(Xe^{X-\Delta\,N_e}\right) \,, \quad
X = \left(\frac{\Delta}{\eta_{\rm max}}\right)^{1/2}-1 \,.
\end{align}

%%%%%%%%%%%%%%%%%%%%%%%%%%%%%%%%%%%%%%%%%%%%%%%%%%%%%%%%%%%%%%%%%%%%%%%%%%%%%%%%%%%%%%%%%%%%%%%%%%%%

With the aid of Eq.~\eqref{eq:_s_fhi_far_W}, the slow-roll parameters
$\varepsilon$ and $\eta$ in Eq.~\eqref{eq:ee_fhi} can be written as follows,
\begin{align}
\label{eq:ee_fhi_W}
\varepsilon = \bigg(\frac{\kappa}{4\pi}\bigg)^2\left(\frac{W}{1+W}\right)^2\Delta \,, \quad 
\eta = - \left(\frac{1}{1+W}\right)^2\Delta \,.
\end{align}
These explicit expressions illustrate once more that $\varepsilon$ is suppressed
by a loop factor compared to $\eta$.
In the computation of the scalar spectral index $n_s$, we can therefore neglect $\varepsilon$
and simply use
\begin{align}
n_s \approx 1 + 2\,\eta = 1 - \frac{2\Delta}{\left(1+W\right)^2} \,.
\end{align}
This relation allows us to compute $n_s$ as a function of
$\eta_{\rm max}$, $N_e$, and $\Delta$.
Or in other words, for given values of $\eta_{\rm max}$ and $N_e$, the measured value
$n_s^{\rm obs}$ directly translates into a specific value for $\Delta$,
\begin{align}
\label{eq:Deltaobs}
\eta_{\rm max} = 10^{-0.5} \,, \quad N_* = 50 \,, \quad n_s = n_s^{\rm obs}
\qquad\Rightarrow\qquad \Delta \simeq 7.1 \times 10^{-3} \,.
\end{align}
This is an important result that eliminates one free parameter from our 
analysis.
First of all, we note that the numerical values in Eq.~\eqref{eq:Deltaobs}
fix the field value $s_*$ at the time of CMB horizon exit,
\begin{align}
\label{eq:sstar_fhi}
\eta_{\rm max} = 10^{-0.5} \,, \quad N_* = 50 \,, \quad \Delta \simeq 7.1 \times 10^{-3} 
\qquad\Rightarrow\qquad W \simeq -0.37 \,, \quad s_* \simeq 0.63\,s_{\rm max} \,.
\end{align}
But more importantly, the measured value of $\Delta$ also fixes
the relation between $m_{3/2}$ and $H_{\rm inf}$,
\begin{align}
\label{eq:m32Hinf}
m_{3/2} = \frac{\kappa}{8\pi}\sqrt{3}\,\Delta^{1/2} H_{\rm inf} \simeq
5.8 \times 10^{-4}\,H_{\rm inf} \,\bigg(\frac{\kappa}{0.1}\bigg) \,.
\end{align}
Evidently, the gravitino mass needs to be several orders of magnitude
smaller than the inflationary Hubble rate in order to explain the observed
scalar spectral index, $m_{3/2} \ll H_{\rm inf}$.
This conclusion justifies our decision to neglect the
quadratic mass term in Eq.~\eqref{eq:V_fhi_approx}.
Moreover, the relation in Eq.~\eqref{eq:m32Hinf} also results
in a numerical expression for the parameter $\zeta$ as a function of the coupling $\kappa$,
\begin{align}
\label{eq:zeta}
\zeta = \frac{4\,\kappa}{3\sqrt{3}\,\pi\,\Delta} \simeq 3.5 \,\bigg(\frac{\kappa}{0.1}\bigg) \,.
\end{align}
For $\kappa \geq  3/4\sqrt{3}\,\pi\,\Delta \simeq 2.9 \times 10^{-2}$, inflation therefore occurs
in the inflection-point regime, while for smaller $\kappa$ values, it occurs in the 
hill-top regime.
According to Eq.~\eqref{eq:zeta}, we also expect that our analysis in the small-$\lambda_s$
approximation should be reliable as long as $\kappa \lesssim 0.1$,
so that $\zeta \lesssim \mathcal{O}\left(1\right)$.

%%%%%%%%%%%%%%%%%%%%%%%%%%%%%%%%%%%%%%%%%%%%%%%%%%%%%%%%%%%%%%%%%%%%%%%%%%%%%%%%%%%%%%%%%%%%%%%%%%%%

In addition to $n_s^{\rm obs}$, we can also use the observed value
of the scalar spectral amplitude, $A_s^{\rm obs}$,
to eliminate yet another parameter from the analysis.
Making use of Eqs.~\eqref{eq:Asns} and \eqref{eq:ee_fhi_W}, we can write
\begin{align}
A_s = \frac{2}{3\,\kappa^2\,\Delta}\left(\frac{1+W}{W}\right)^2
\left(\frac{\mu_S}{M_{\rm Pl}}\right)^4 \,.
\end{align}
The condition $A_s = A_s^{\rm obs}$ can then be solved
for the inflaton F-term mass scale as a function of $\kappa$,
\begin{align}
\label{eq:muS_fhi_far}
\mu_S = \left(\frac{3}{2}\,A_s^{\rm obs}\Delta\right)^{1/4}
\left(\frac{\kappa\left|W\right|}{1+W}\right)^{1/2} M_{\rm Pl}
\simeq 1.3 \times 10^{15}\,\textrm{GeV} \:\bigg(\frac{\kappa}{0.1}\bigg)^{1/2} \,.
\end{align}
This result immediately fixes the SSB scale of the waterfall transition
at the end of inflation,
\begin{align}
\label{eq:v_fhi}
v = \left(6\,A_s^{\rm obs}\Delta\right)^{1/4}
\left(\frac{\left|W\right|}{1+W}\right)^{1/2} M_{\rm Pl}
\simeq 5.8 \times 10^{15}\,\textrm{GeV} \,,
\end{align}
which is remarkably close to the GUT scale in typical SUSY
GUT scenarios, $\Lambda_{\rm GUT} \sim 10^{16}\,\textrm{GeV}$.
The numerical result in Eq.~\eqref{eq:v_fhi} therefore serves as another
indication that FHI is, indeed, well suited to be embedded into a bigger GUT framework.
Eq.~\eqref{eq:v_fhi} also fixes the cosmic string tension,
\begin{align}
\label{eq:Gmu_fhi_far}
G\,\mu_{\rm CS} = \left(\frac{3}{8}\,A_s^{\rm obs}\Delta\right)^{1/2}
\frac{\epsilon_{\rm CS}\left|W\right|}{1+W} \simeq
6.4 \times 10^{-7} \:\bigg(\frac{\epsilon_{\rm CS}}{0.45}\bigg) \,,
\end{align}
where we used that $\epsilon_{\rm CS}\simeq 0.45$ for $\kappa = 10^{-1}$
(see Eqs.~\eqref{eq:epsilonCS} and \eqref{eq:GmuCS}).
In view of Eq.~\eqref{eq:Gmu_fhi_far}, we conclude that FHI in the large-$\kappa$
regime produces cosmic strings with a large tension that is conflict with the observational
bound in Eq.~\eqref{eq:GmuCSmax}.
Therefore, if we take the bound in Eq.~\eqref{eq:GmuCSmax} seriously,
FHI in the large-$\kappa$ regime is ruled out.
Alternatively, we can simply presume that the gauge symmetry $G$ already becomes
broken in a different sector before the end of inflation.
In this case, we do not need to worry about the large cosmic string tension in 
Eq.~\eqref{eq:Gmu_fhi_far} (see the discussion below Eq.~\eqref{eq:vCS}).

%%%%%%%%%%%%%%%%%%%%%%%%%%%%%%%%%%%%%%%%%%%%%%%%%%%%%%%%%%%%%%%%%%%%%%%%%%%%%%%%%%%%%%%%%%%%%%%%%%%%

In consequence of the two conditions $n_s = n_s^{\rm obs}$ and $A_s = A_s^{\rm obs}$,
the viable parameter space of FHI shrinks to a one-dimensional hypersurface that can be 
parametrized in terms of the Yukawa coupling $\kappa$.
The Hubble rate $H_{\rm inf}$, \textit{e.g.}, follows immediately
from the expression for $\mu_S$ in Eq.~\eqref{eq:muS_fhi_far},
\begin{align}
\label{eq:Hinf_fhi_far}
H_{\rm inf} = \left(\frac{1}{2}\,A_s^{\rm obs}\Delta\right)^{1/2}
\frac{\kappa\left|W\right|}{1+W}\,M_{\rm Pl}
\simeq 4.0 \times 10^{11}\,\textrm{GeV} \:\bigg(\frac{\kappa}{0.1}\bigg) \,.
\end{align}
Thanks to the relation in Eq.~\eqref{eq:m32Hinf}, this result for $H_{\rm inf}$
determines in turn the gravitino mass $m_{3/2}$,
\begin{align}
\label{eq:m32_fhi_far}
m_{3/2} = \left(\frac{3}{2}\,A_s^{\rm obs}\right)^{1/2}\frac{\kappa^2}{8\pi}
\frac{\Delta\left|W\right|}{1+W}\,M_{\rm Pl}
\simeq 2.3 \times 10^8\,\textrm{GeV} \:\bigg(\frac{\kappa}{0.1}\bigg)^2 \,.
\end{align}
At this point, we emphasize that Eq.~\eqref{eq:m32_fhi_far}
corresponds to the solution for $m_{3/2}$ on the negative real axis.
As shown in~\cite{Buchmuller:2014epa}, more complicated trajectories
in the complex inflaton plane also lead to successful inflation\,---\,however,
keeping the value of $H_{\rm inf}$ fixed, these alternative solutions are all
associated with a larger value of $m_{3/2}$.
In this sense, the expression in Eq.~\eqref{eq:m32_fhi_far} should
be regarded as a \textit{lower} bound on the gravitino mass in FHI
(see the discussion below Eq.~\eqref{eq:cs}).
Furthermore, given the $\kappa$ dependence of $\mu_S$ and $m_{3/2}$
in Eqs.~\eqref{eq:muS_fhi_far} and \eqref{eq:m32_fhi_far}, we are now able
to compute the critical $\kappa$ value that separates the large-$\kappa$ 
regime (where $x_* \gg 1$) from the small-$\kappa$ regime (where $x_* \simeq 1$),
\begin{align}
\label{eq:kappacrit_fhi}
s_{\rm scrit} = s_{\rm max} \qquad\Rightarrow\qquad
\kappa_0 = 4 \left[\pi^2\,\frac{m_{3/2}\left(\kappa_0\right)}
{\mu_S\left(\kappa_0\right)}\right]^{2/5} \simeq 1.8 \times 10^{-3} \,.
\end{align}
As anticipated at the beginning of this section, the critical $\kappa$
value is, indeed, of $\mathcal{O}\left(10^{-3}\right)$.

%%%%%%%%%%%%%%%%%%%%%%%%%%%%%%%%%%%%%%%%%%%%%%%%%%%%%%%%%%%%%%%%%%%%%%%%%%%%%%%%%%%%%%%%%%%%%%%%%%%%

\begin{figure}
\begin{center}

\includegraphics[width=0.475\textwidth]{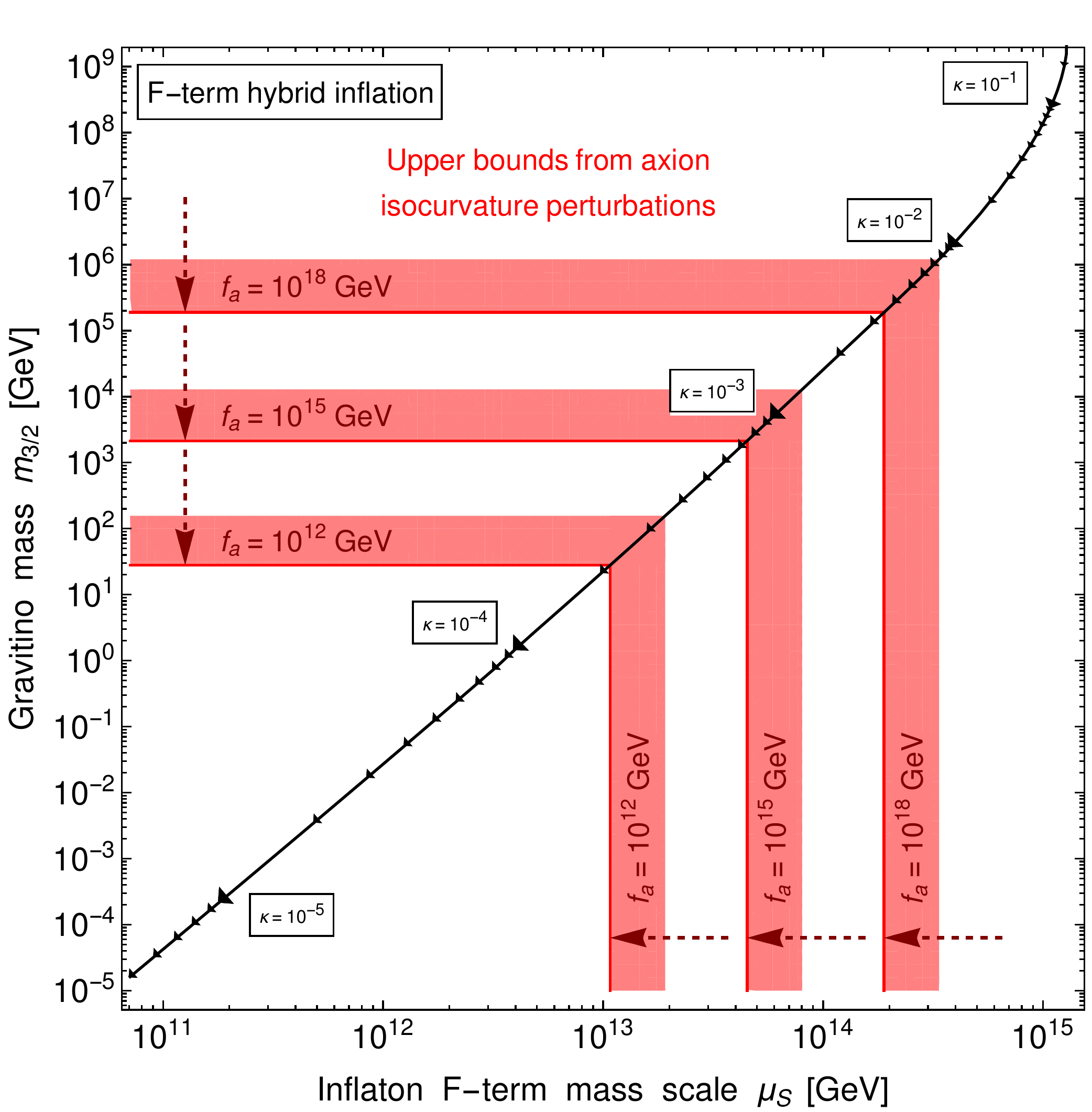}\hfill
\includegraphics[width=0.475\textwidth]{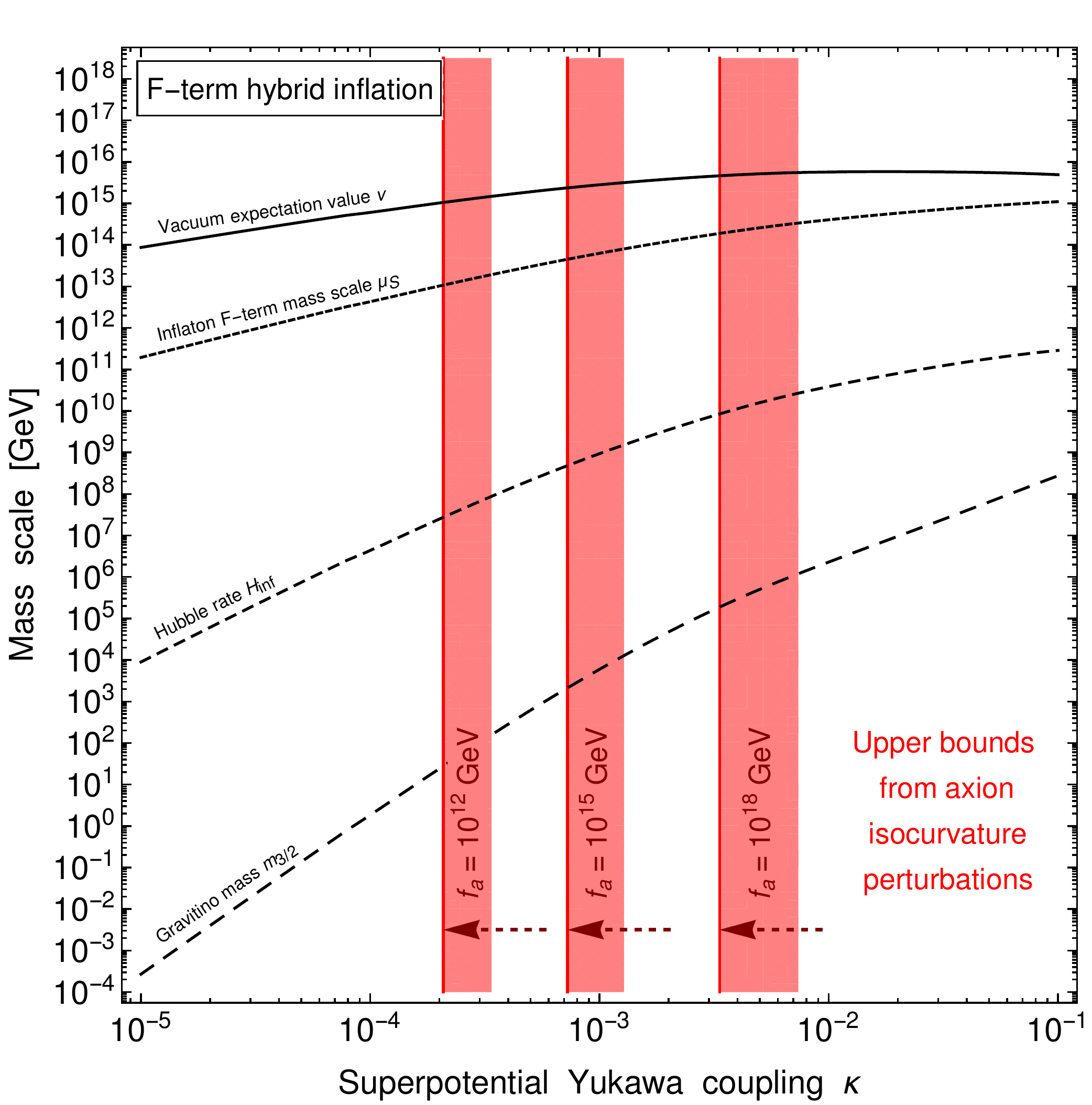}

\caption{Parameter values for F-term hybrid inflation that
reproduce the CMB data, $A_s = A_s^{\rm obs}$ and $n_s = n_s^{\rm obs}$,
in combination with the CDM isocurvature constraint for several values of the
axion decay constant and different assumptions regarding the axion DM fraction.
The stronger [weaker] bounds correspond to $F_{\rm DM}^a = 1$ [$F_{\rm DM}^a = 0.1$].
\textbf{(Left panel)} One-dimensional hypersurface in the $\mu_S$--$m_{3/2}$ plane
that manages to reproduce the observed CMB data.
\textbf{(Right panel)} Various mass scales that are relevant in the description of FHI
as functions of the Yukawa coupling $\kappa$.
Both plots are based on a numerical analysis that accounts for
the complete scalar potential in Eqs.~\eqref{eq:VF_fhi} and \eqref{eq:V1l_fhi}.}

\label{fig:bounds_fhi}

\end{center}
\end{figure}

The expressions for $H_{\rm inf}$ and $m_{3/2}$ in Eqs.~\eqref{eq:Hinf_fhi_far}
and \eqref{eq:m32_fhi_far} mark the main technical results in this section.
Based on these results, we can now determine the implications of
the CDM isocurvature constraint on the parameters of FHI in the large-$\kappa$ regime.
Confronting our explicit expression for $H_{\rm inf}$ in
Eq.~\eqref{eq:Hinf_fhi_far} with the upper bound in
Eq.~\eqref{eq:Hinfmax}, we arrive at the following upper bound on $\kappa$,
\begin{align}
\label{eq:kappa_max_fhi_far}
\boxed{
\kappa \lesssim 3.1 \times 10^{-3}
\left(\frac{1}{F_{\rm DM}^a}\right)^{1/2} \left(\frac{f_a}{M_{\rm Pl}}\right)^{0.42}}
\end{align}
This is a tight constraint on the inflaton Yukawa coupling $\kappa$.
In fact, only for a very large axion decay constant, $f_a \sim M_{\rm Pl}$,
the bound in Eq.~\eqref{eq:kappa_max_fhi_far} manages to exceed the critical $\kappa$
value in Eq.~\eqref{eq:kappacrit_fhi}.
In view of this result, it is important to remember that a Planck-scale axion decay
constant is questionable for both theoretical and phenomenological reasons.
On the one hand, string theory suggests that it is impossible to realize an
axion decay constant larger than the Planck scale.
Values as large as $f \sim M_{\rm Pl}$ are therefore only marginally feasible.
Instead, string theory rather points to axion decay constants of the order of
$f_a \sim 10^{16}\cdots10^{17}\,\textrm{GeV}$~\cite{Svrcek:2006yi,Arvanitaki:2009fg,Cicoli:2012sz}.
On the other hand, spin measurements of stellar black holes
allow to constrain $f_a$ based on the phenomenon of black
hole superradiance.
At present, these measurements exclude $f_a$ values in the range
$3 \times 10^{17}\,\textrm{GeV} \lesssim f_a \lesssim
10^{19}\,\textrm{GeV}$~\cite{Arvanitaki:2009fg,Arvanitaki:2010sy,Arvanitaki:2014wva}.
We, thus, conclude that FHI in the large-$\kappa$ regime is highly constrained
by the CDM isocurvature bound.
A viable region in parameter space survives only if $f_a \sim M_{\rm Pl}$ for one
reason or another.
Making use of Eq.~\eqref{eq:m32_fhi_far}, the bound
in Eq.~\eqref{eq:kappa_max_fhi_far} can also be formulated as an upper bound on $m_{3/2}$,
\begin{align}
\label{eq:m32_max_fhi_far}
\boxed{
m_{3/2} \lesssim 2.2 \times 10^5\,\textrm{GeV}
\left(\frac{1}{F_{\rm DM}^a}\right) \left(\frac{f_a}{M_{\rm Pl}}\right)^{0.83}}
\end{align}
Again, we stress that this is an \textit{inclusive} upper bound on $m_{3/2}$ that guarantees
that the CDM isocurvature constraint in Eq.~\eqref{eq:Hinfmax} is satisfied,
no matter which inflationary trajectory is chosen in the complex inflaton plane.
A more extensive analysis assessing the dependence on the chosen trajectory
is much more involved and beyond the scope of this work.
In Fig.~\ref{fig:bounds_fhi}, we show the upper bounds on $\mu_S$,
$m_{3/2}$, etc.\ for a few representative values of $f_a$.
The plots in Fig.~\ref{fig:bounds_fhi} are based on a fully numerical
analysis of slow-roll inflation in the \textit{complete} scalar potential of FHI
(see Eqs.~\eqref{eq:VF_fhi} and \eqref{eq:V1l_fhi}).
The comparison between these plots and the analytical results derived in this section
demonstrates that our analytical calculations reproduce the exact results very well.
This observation serves as a cross-check and validates the various
approximations in the above discussion.

%%%%%%%%%%%%%%%%%%%%%%%%%%%%%%%%%%%%%%%%%%%%%%%%%%%%%%%%%%%%%%%%%%%%%%%%%%%%%%%%%%%%%%%%%%%%%%%%%%%%

Finally, we use the expression for $H_{\rm inf}$ in Eq.~\eqref{eq:Hinf_fhi_far}
to determine the parameter region in which the PQ symmetry actually remains
intact during inflation. 
In this case, the requirement $H_{\rm inf} > f_a$ results in lower bounds on 
the Yukawa coupling $\kappa$ and the gravitino mass $m_{3/2}$,
\begin{align}
\label{eq:kappa_min_fhi_far}
\kappa \gtrsim 2.5 \times 10^{-3}
\left(\frac{f_a}{10^{10}\,\textrm{GeV}}\right) \,, \quad 
m_{3/2} \gtrsim 1.5 \times 10^5\,\textrm{GeV}
\left(\frac{f_a}{10^{10}\,\textrm{GeV}}\right)^2 \,.
\end{align}
Thus, for low values of the axion decay constant, $f_a \sim 10^{10}\,\textrm{GeV}$,
the PQ symmetry is only spontaneously broken after inflation,
which may result in the production of dangerous domain walls.

%%%%%%%%%%%%%%%%%%%%%%%%%%%%%%%%%%%%%%%%%%%%%%%%%%%%%%%%%%%%%%%%%%%%%%%%%%%%%%%%%%%%%%%%%%%%%%%%%%%%

%%%%%%%%%%%%%%%%%%%%%%%%%%%%%%%%%%%%%%%%%%%%%%%%%%%%%%%%%%%%%%%%%%%%%%%%%%%%%%%%%%%%%%%%%%%%%%%%%%%%

\subsection{Inflation close to the waterfall phase transition}
\label{subsec:fhi_close}

%%%%%%%%%%%%%%%%%%%%%%%%%%%%%%%%%%%%%%%%%%%%%%%%%%%%%%%%%%%%%%%%%%%%%%%%%%%%%%%%%%%%%%%%%%%%%%%%%%%%

In the previous section, we saw that FHI in the large-$\kappa$ regime
is highly constrained by the nonobservation of axion isocurvature perturbations
and the upper bound on the cosmic string tension.
This situation changes in the small-$\kappa$ regime, which opens up 
the possibility to lower the inflationary Hubble scale to smaller values,
$H_{\rm inf} \lesssim \mathcal{O}\left(10^9\right)\,\textrm{GeV}$.
This scenario is therefore compatible with values of the axion 
decay constant significantly below the Planck scale, $f_a \ll M_{\rm Pl}$.
However, it is clear from the outset that this improvement over
the large-$\kappa$ regime is not for free.
The price one has to pay is an additional tuning in the initial conditions
of inflation. 
In the small-$\kappa$ regime, the local maximum in the scalar potential
is located in the direct vicinity of the critical field value that triggers
the waterfall transition.
The initial field value $s_{\rm ini}$, thus, has to be tuned to lie
in the small interval in between $s_{\rm crit}$ and $s_{\rm max}$ to ensure
that inflation proceeds in the correct direction in field space.
Otherwise, \textit{i.e.}, for $s_{\rm ini} > s_{\rm max}$, the inflaton will
roll towards the false vacuum at $s = s_{\rm min}$, so that inflation never ends
(see the right panel of Fig.~\ref{fig:potential_fhi}).
On the other hand, we stress that a fine-tuning of the initial conditions
is of a different conceptual quality than a fine-tuning of model parameters.
One may, \textit{e.g.}, speculate that the evolution of the inflaton field
prior to inflation is responsible for a \textit{dynamical} selection of initial
field values close to $s_{\rm max}$ for one reason or another. 
In addition, as shown in~\cite{Buchmuller:2014epa}, the issue
of initial conditions in FHI becomes relaxed if one accounts for all possible trajectories
in the complex plane. 
In this case, it is possible that the inflaton trajectory starts out at large field values
and then bends in just the right way to avoid the local minimum at $s = s_{\rm min}$.
Finally, we point out that the small-$\kappa$ regime does not require
any unnatural fine-tuning of model parameters.
In the $\kappa \rightarrow 0$ limit, the waterfall fields
cease to participate in Yukawa interactions.
This restores a global $U(1) \times U(1)$ symmetry in the waterfall
sector that contains the local gauge symmetry $G$ as a subgroup.
Small $\kappa$ values are therefore natural in the sense of 't Hooft~\cite{tHooft:1979rat}.

%%%%%%%%%%%%%%%%%%%%%%%%%%%%%%%%%%%%%%%%%%%%%%%%%%%%%%%%%%%%%%%%%%%%%%%%%%%%%%%%%%%%%%%%%%%%%%%%%%%%

In the small-$\kappa$ regime, we can no longer use the large-field expansion
of the loop function $L$ in Eq.~\eqref{eq:L_far}.
Instead, we now have to evaluate $L$ in the vicinity of the critical
field value $s_{\rm crit}$,
\begin{align}
\label{eq:LL2}
L\left(x\right) = L_2\left(y\right) + \mathcal{O}\left(y^3\right) \,, \quad
x = \left(\frac{s}{s_{\rm crit}}\right)^2 \,, \quad y = \frac{s}{s_{\rm crit}} -1 \,.
\end{align}
Here, $L_2$ encompasses the leading contributions to $L$ up to second order
in the new field variable $y$,
\begin{align}
\label{eq:L_close}
L_2\left(y\right) = c_0  + c_1\,y + \frac{1}{2}\left(c_2 + \bar{c}_2 \ln y\right) y^2 \,.
\end{align}
The coefficients $c_0$, $c_1$, $c_2$, and $\bar{c}_2$ can be determined analytically,
\begin{align}
c_0  = 2\ln2 - \frac{3}{2} \,, \quad
c_1  = 4\ln2 \,, \quad 
c_2  = 6\left(2\ln2-1\right) \,, \quad 
\bar{c}_2 = 4 \,.
\end{align}
Given this expansion of the radiative one-loop corrections, the inflaton
potential now reads
\begin{align}
\label{eq:V_fhi_close}
V \simeq V_F^0 + c_s\,s + \frac{1}{2}\,V_{1l}^0\, L_2\left(y\right) \,,
\end{align}
where we again neglected the quartic SUGRA term.
Correspondingly, $\varepsilon$ and $\eta$ in Eq.~\eqref{eq:ee_fhi} turn into
\begin{align}
\label{eq:ee_fhi_close}
\varepsilon = \frac{1}{2} \left(\frac{c_s s_{\rm crit} + c_1 V_{1\ell}^0/2}{V_F^0}\right)^2
\left(\frac{M_{\rm Pl}}{s_{\rm crit}}\right)^2 \,, \quad
\eta = \frac{V_{1\ell}^0}{2\,V_F^0}\left(c_2 + \frac{3}{2}\,\bar{c}_2 + \bar{c}_2
\ln y\right)\left(\frac{M_{\rm Pl}}{s_{\rm crit}}\right)^2 \,.
\end{align}
Also the slow-roll equation of motion in Eq.~\eqref{eq:eom_fhi_far}
obtains a new form.
To leading order, we can write
\begin{align}
s'= \sqrt{2}\,\varepsilon^{1/2} M_{\rm Pl} \,.
\end{align}
This equation can be readily integrated, resulting in the following expression for
the inflaton field $s$,
\begin{align}
s\left(N_e\right) = s_{\rm crit} + \sqrt{2}\,\varepsilon^{1/2} M_{\rm Pl}\, N_e \,.
\end{align}

%%%%%%%%%%%%%%%%%%%%%%%%%%%%%%%%%%%%%%%%%%%%%%%%%%%%%%%%%%%%%%%%%%%%%%%%%%%%%%%%%%%%%%%%%%%%%%%%%%%%

As before, we shall now eliminate two free parameters 
by making use of the conditions $A_s = A_s^{\rm obs}$ and $n_s = n_s^{\rm obs}$.
To this end, we first solve $A_s = A_s^{\rm obs}$ for the slow-roll parameter
$\varepsilon$ (see Eq.~\eqref{eq:Asns}),
\begin{align}
\varepsilon = \frac{1}{24\pi^2}\frac{V_F^0}{A_s^{\rm obs} M_{\rm Pl}^4} \,.
\end{align}
Then, we equate this result with the expression for
$\varepsilon$ in Eq.~\eqref{eq:ee_fhi_close} and solve for the gravitino mass,
\begin{align}
\label{eq:m32_fhi_e}
m_{3/2} = \left[\ln2\,\kappa^{5/2} - \frac{2\sqrt{2}\,\pi}{\sqrt{3}\left(A_s^{\rm obs}\right)^{1/2}}
\left(\frac{\mu_S}{M_{\rm Pl}}\right)^3\right]\frac{\mu_S}{16\pi^2} \,.
\end{align}
Next, we make use of the condition $n_s = n_s^{\rm obs}$.
Again, we approximate $n_s \approx 1 + 2\,\eta$, such that
\begin{align}
n_s \approx 1 + 2\,\ln\left(8y_*\right)\delta \,.
\end{align}
Here, the dimensionless parameter $\delta$ characterizes the curvature
of $V_{1\ell}$ close to the critical field value,
\begin{align}
\label{eq:delta_def}
\delta = \frac{2V_{1\ell}^0}{V_F^0}\left(\frac{M_{\rm Pl}}{s_{\rm crit}}\right)^2
= \frac{\kappa^3}{8\pi^2}\left(\frac{M_{\rm Pl}}{\mu_S}\right)^2 \,.
\end{align}
Meanwhile, $y_*$ stands for the field variable $y$ evaluated at the time of CMB horizon exit,
\begin{align}
y_* = \frac{s_*}{s_{\rm crit}} - 1 =
\sqrt{2}\,\varepsilon^{1/2} N_*\,\frac{M_{\rm Pl}}{s_{\rm crit}} =
\frac{\kappa^2}{8\sqrt{3}\,\pi^2} \frac{N_*}{\left(A_s^{\rm obs}\delta\right)^{1/2}} \,.
\end{align}
Putting everything together, we find that the scalar spectral index can be written
as follows,%
\footnote{A similar formula appears in~\cite{Buchmuller:2014epa}.
Here, we extend the analysis in~\cite{Buchmuller:2014epa} by explicitly solving
$n_s = n_s^{\rm obs}$ for $\delta$.}
\begin{align}
n_s = 1 - \ln\left(\frac{3\pi^4}{\kappa^4}\frac{A_s^{\rm obs}}{N_*^2}\,\delta\right)\delta \,.
\end{align}
In the next step, we explicitly solve the condition $n_s = n_s^{\rm obs}$
for the curvature parameter $\delta$,
\begin{align}
\label{eq:delta}
\delta = \frac{1-n_s^{\rm obs}}{W_0\left(Y\right)} \,, \quad
Y = \left(1-n_s^{\rm obs}\right) \frac{3\pi^4}{\kappa^4}\frac{A_s^{\rm obs}}{N_*^2} \,,
\end{align}
where $W_0$ again denotes the Lambert W function (see Eq.~\eqref{eq:Lambert}).
The result in Eq.~\eqref{eq:delta} enables us to compute $\delta$ as a function of $N_*$ and $\kappa$.
For $N_* = 47.5$ and $\kappa = 10^{-5}$, we find, \textit{e.g.}, $\delta \simeq 0.002$.
The dependence of $\delta$ on the Yukawa coupling $\kappa$ is in general rather weak.
For $\kappa$ values in between $10^{-7}$ and $10^{-3}$, the parameter $\delta$
varies only by roughly an order of magnitude, $0.001 \lesssim \delta \lesssim 0.02$.

%%%%%%%%%%%%%%%%%%%%%%%%%%%%%%%%%%%%%%%%%%%%%%%%%%%%%%%%%%%%%%%%%%%%%%%%%%%%%%%%%%%%%%%%%%%%%%%%%%%%

The definition of $\delta$ in Eq.~\eqref{eq:delta_def} can be
solved for the inflaton F-term mass scale.
We, thus, obtain
\begin{align}
\label{eq:muS_fhi_close}
\mu_S = \left(\frac{\kappa^3}{8\pi^2\delta}\right)^{1/2} M_{\rm Pl}
\simeq 1.9 \times 10^{11}\,\textrm{GeV} \left(\frac{0.002}{\delta}\right)^{1/2}
\bigg(\frac{\kappa}{10^{-5}}\bigg)^{3/2} \,.
\end{align}
Again, this result immediately translates into an expression for the SSB scale $v$,
\begin{align}
\label{eq:v_fhi_close}
v = \frac{\kappa\,M_{\rm Pl}}{2\pi\,\delta^{1/2}} 
\simeq 8.7 \times 10^{13}\,\textrm{GeV} \left(\frac{0.002}{\delta}\right)^{1/2}
\bigg(\frac{\kappa}{10^{-5}}\bigg) \,,
\end{align}
which now turns out to be parametrically suppressed compared to the GUT scale,
$\Lambda_{\rm GUT} \sim 10^{16}\,\textrm{GeV}$.
Unlike Eq.~\eqref{eq:v_fhi}, Eq.~\eqref{eq:v_fhi_close} results in a parameter-dependent expression for
the cosmic string tension,
\begin{align}
\label{eq:Gmu_fhi_close}
G\,\mu_{\rm CS} = \bigg(\frac{\kappa}{4\pi}\bigg)^2 \:\frac{\epsilon_{\rm CS}}{\delta}
\simeq 3.2 \times 10^{-11} \left(\frac{0.002}{\delta}\right)\bigg(\frac{\kappa}{10^{-5}}\bigg)^2
\bigg(\frac{\epsilon_{\rm CS}}{0.10}\bigg) \,,
\end{align}
where we used that $\epsilon_{\rm CS}\simeq 0.10$ for $\kappa = 10^{-5}$
(see Eqs.~\eqref{eq:epsilonCS} and \eqref{eq:GmuCS}).
Therefore, for a sufficiently small value of the Yukawa coupling $\kappa$,
there is no problem to satisfy the bound on the cosmic
string tension in Eq.~\eqref{eq:GmuCSmax}.
As mentioned above, the only price to pay is an increased tuning in the initial
conditions for inflation.
Eq.~\eqref{eq:muS_fhi_close} also results in an expression for the inflationary Hubble rate,
\begin{align}
\label{eq:Hinf_fhi_close}
H_{\rm inf} = \frac{\kappa^3\,M_{\rm Pl}}{8\sqrt{3}\,\pi^2\delta} \simeq 
9.0 \times 10^3\,\textrm{GeV} \left(\frac{0.002}{\delta}\right)
\bigg(\frac{\kappa}{10^{-5}}\bigg)^3 \,,
\end{align}
which now scales more strongly with $\kappa$ than in the large-$\kappa$ regime
(see Eq.~\eqref{eq:Hinf_fhi_far}).
Similarly, we can use the results in
Eqs.~\eqref{eq:m32_fhi_e} and \eqref{eq:muS_fhi_close} to obtain an expression for $m_{3/2}$
as a function of $\kappa$,
\begin{align}
m_{3/2} = \left[\ln2-\frac{\kappa^2}{8\sqrt{3}\,\pi^2\left(A_s^{\rm obs}\right)^{1/2}\delta^{3/2}}\right]
\frac{\kappa^4 M_{\rm Pl}}{32\sqrt{2}\,\pi^3\,\delta^{1/2}} \,.
\end{align}
For small $\kappa$, the $\ln 2$ term dominates
the square brackets on the RHS of this expression, such that
\begin{align}
\label{eq:m32_fhi_close_approx}
m_{3/2} \approx
\frac{\ln2\,\kappa^4 M_{\rm Pl}}{32\sqrt{2}\,\pi^3\,\delta^{1/2}}
\simeq 2.7\times10^{-4}\,\textrm{GeV} \left(\frac{0.002}{\delta}\right)^{1/2}
\bigg(\frac{\kappa}{10^{-5}}\bigg)^4 \,.
\end{align}

%%%%%%%%%%%%%%%%%%%%%%%%%%%%%%%%%%%%%%%%%%%%%%%%%%%%%%%%%%%%%%%%%%%%%%%%%%%%%%%%%%%%%%%%%%%%%%%%%%%%

With the above results at hand, we can again use the CDM isocurvature bound
in Eq.~\eqref{eq:Hinfmax} to constrain the parameter space of FHI.
However, this time, we need to determine all bounds numerically because
of the complicated $\kappa$ dependence of the parameter $\delta$ (see Eq.~\eqref{eq:delta}).
First, we compare our result for $H_{\rm inf}$ in Eq.~\eqref{eq:Hinf_fhi_close}
with Eq.~\eqref{eq:Hinfmax} to determine an upper bound on $\kappa$,
\begin{align}
\label{eq:kappa_max_fhi_close}
\boxed{
\kappa \lesssim 1.0 \times 10^{-3}
\left(\frac{1}{F_{\rm DM}^a}\right)^{0.21}\left(\frac{f_a}{10^{16}\,\textrm{GeV}}\right)^{0.17}}
\end{align}
This constraint is consistent with the critical $\kappa$ value in Eq.~\eqref{eq:kappacrit_fhi}
that separates the small-$\kappa$ regime from the large-$\kappa$ regime.
In particular, as we are working with small values of $\kappa$ in this section,
the axion decay constant $f_a$ can now be chosen to be
significantly smaller than the Planck scale.
Combining our results in Eqs.~\eqref{eq:m32_fhi_close_approx}
and \eqref{eq:kappa_max_fhi_close}, we are also able to deduce an upper bound on $m_{3/2}$,
\begin{align}
\label{eq:m32_max_fhi_close}
\boxed{
m_{3/2} \lesssim  9.4 \times 10^3\,\textrm{GeV}
\left(\frac{1}{F_{\rm DM}^a}\right)^{0.76}
\left(\frac{f_a}{10^{16}\,\textrm{GeV}}\right)^{0.63}} 
\end{align}
Just like the bound in Eq.~\eqref{eq:m32_max_fhi_far}, this bound is again an
absolute upper bound that guarantees that the CDM isocurvature constraint is satisfied
for all possible trajectories in the complex plane.
The (quasi-) analytical result in Eq.~\eqref{eq:m32_max_fhi_close} needs to be compared
to the fully numerical result in Fig.~\ref{fig:bounds_fhi}. 
Again, we find excellent agreement, which confirms the validity
of the above analytical discussion.
Eq.~\eqref{eq:m32_fhi_close_approx}
can also be used to translate the upper bound on the cosmic string 
tension in Eq.~\eqref{eq:GmuCSmax} into an upper bound on 
the gravitino mass.
The combination of Eqs.~\eqref{eq:GmuCSmax}, \eqref{eq:Gmu_fhi_close},
and \eqref{eq:m32_fhi_close_approx} results in
\begin{align}
\label{eq:kappaCSFHI}
G\,\mu_{\rm CS} < G\,\mu_{\rm CS}^{\rm max} \qquad\Rightarrow\qquad
\kappa  \lesssim 1.8 \times 10^{-3} \,, \quad 
m_{3/2} \lesssim 3.2 \times 10^4\,\textrm{GeV} \,.
\end{align}
where we used that $\epsilon_{\rm CS}\simeq 0.18$ for $\kappa = 1.8 \times 10^{-3}$
(see Eqs.~\eqref{eq:epsilonCS} and \eqref{eq:GmuCS}).
Note that the upper bound on $\kappa$ accidentally coincides with
the critical $\kappa$ value in Eq.~\eqref{eq:kappacrit_fhi}.
By coincidence, the region in parameter space where
$G\,\mu_{\rm CS} < G\,\mu_{\rm CS}^{\rm max}$ therefore happens
to be identical with the small-$\kappa$ regime.
Thanks to  Eqs.~\eqref{eq:muS_fhi_close}, \eqref{eq:v_fhi_close},
and \eqref{eq:Hinf_fhi_close}, the bounds in Eq.~\eqref{eq:kappaCSFHI}
also result in the following constraints,
\begin{align}
\label{eq:vCSFHI}
v           \lesssim 3.6 \times 10^{15}\,\textrm{GeV} \,, \quad
\mu_S       \lesssim 1.1 \times 10^{14}\,\textrm{GeV} \,, \quad
H_{\rm inf} \lesssim 2.8 \times 10^9   \,\textrm{GeV}  \,.
\end{align}
This result is consistent with the bound on the SSB scale $v$ in Eq.~\eqref{eq:vCS}.

%%%%%%%%%%%%%%%%%%%%%%%%%%%%%%%%%%%%%%%%%%%%%%%%%%%%%%%%%%%%%%%%%%%%%%%%%%%%%%%%%%%%%%%%%%%%%%%%%%%%

Finally, similarly to the large-$\kappa$ case, we conclude by determining
the region in parameter space where the PQ symmetry remains intact during
inflation.
Combining Eqs.~\eqref{eq:Hinf_fhi_close} and \eqref{eq:m32_fhi_close_approx}
with the requirement that $H_{\rm inf}$ must exceed $f_a$, we obtain the following
lower bounds on $\kappa$ and $m_{3/2}$,
\begin{align}
\label{eq:kappa_min_fhi_close}
\kappa \gtrsim 3.4 \times 10^{-4}
\left(\frac{f_a}{10^8\,\textrm{GeV}}\right)^{0.39} \,, \quad
m_{3/2} \gtrsim  1.8 \times 10^2\,\textrm{GeV}
\left(\frac{f_a}{10^8\,\textrm{GeV}}\right)^{1.47} \,.
\end{align}
Thus, for small values of the axion decay constant $f_a$, there are also
viable parameter combinations in the small-$\kappa$ regime that are
compatible with the postinflationary PQSB scenario.

%%%%%%%%%%%%%%%%%%%%%%%%%%%%%%%%%%%%%%%%%%%%%%%%%%%%%%%%%%%%%%%%%%%%%%%%%%%%%%%%%%%%%%%%%%%%%%%%%%%%

\section{Low-scale D-term hybrid inflation}
\label{sec:dhi}

%%%%%%%%%%%%%%%%%%%%%%%%%%%%%%%%%%%%%%%%%%%%%%%%%%%%%%%%%%%%%%%%%%%%%%%%%%%%%%%%%%%%%%%%%%%%%%%%%%%%

\subsection{Model setup and scalar potential}
\label{subsec:dhi_model}

%%%%%%%%%%%%%%%%%%%%%%%%%%%%%%%%%%%%%%%%%%%%%%%%%%%%%%%%%%%%%%%%%%%%%%%%%%%%%%%%%%%%%%%%%%%%%%%%%%%%

In Sec.~\ref{sec:fhi}, we discussed the slow-roll dynamics of FHI and the compatibility
with the CDM isocurvature constraint in Eq.~\eqref{eq:Hinfmax}.
We found an absolute upper bound on the Yukawa coupling $\kappa$ of
$\mathcal{O}\left(10^{-3}\right)$ (see Eq.~\eqref{eq:kappa_max_fhi_far})
and a corresponding bound on the gravitino mass $m_{3/2}$ of 
$\mathcal{O}\left(10^5\right)\,\textrm{GeV}$ (see Eq.~\eqref{eq:m32_max_fhi_far}).
Moreover, we concluded that the large-$\kappa$ regime of FHI is strongly
constrained by the nonobservation of axion isocurvature perturbations
and the upper bound on the cosmic string tension.
Likewise, we concluded that the small-$\kappa$ regime of FHI manages to 
avoid these constraints\,---\,however, at the price of a moderate fine-tuning
of the initial conditions of inflation.
In addition, we recall that both regimes of FHI actually need to be described
as a two-field model of inflation.
As shown in~\cite{Buchmuller:2014epa}, this includes the possibility
of inflaton trajectories in the complex plane
that fail to reach the critical field value $s_{\rm crit}$.
FHI therefore requires an additional selection mechanism among all
possible trajectories ensuring that
inflaton ends in a successful waterfall transition.

%%%%%%%%%%%%%%%%%%%%%%%%%%%%%%%%%%%%%%%%%%%%%%%%%%%%%%%%%%%%%%%%%%%%%%%%%%%%%%%%%%%%%%%%%%%%%%%%%%%%

In this section, we will now show that most of the above problems
related to FHI are absent in the case of DHI.
The reason for this is twofold: 
First of all, DHI is a standard single-field model of inflation. 
The inflaton field does not possess an F term and, hence, the rotational
invariance in the complex plane remains unbroken.
Thus, there are no problems related to the proper choice of trajectory in field space.
Second, in contrast to FHI, the dynamics of DHI are controlled by the magnitude
of the gauge coupling constant $g$.
This provides a larger parametric freedom that can be used to achieve a low Hubble
rate even in the large-$\kappa$ regime.
In DHI, it is therefore possible to satisfy the CDM isocurvature constraint
without any fine-tuning of the initial conditions.
Only the issue of cosmic string formation during the waterfall transitions
remains more or less unaffected.
Also in DHI, the cosmic string tension can only be successfully suppressed
if the inflaton Yukawa coupling $\kappa$ is set to a small value,
$\kappa \lesssim \mathcal{O}\left(10^{-4}\right)$.
However, we reiterate that this constraint becomes null if cosmic strings already
form before the end of inflation (see the discussion below Eq.~\eqref{eq:vCS}).

%%%%%%%%%%%%%%%%%%%%%%%%%%%%%%%%%%%%%%%%%%%%%%%%%%%%%%%%%%%%%%%%%%%%%%%%%%%%%%%%%%%%%%%%%%%%%%%%%%%%

We begin by describing the setup of our model and collecting a few important
properties of the scalar potential.
Again, we will incorporate the effect of spontaneous SUSY breaking in the form
of a hidden Polonyi sector that couples to the inflaton sector only via
gravitational interactions.
The superpotential of our model, thus, follows from Eq.~\eqref{eq:WFHI}
after setting the inflaton F-term mass scale to zero, $\mu_S \rightarrow 0$.
The K\"ahler potential remains unchanged and is the same
as in FHI (see Eq.~\eqref{eq:KFHI}),
\begin{align}
W = \kappa\, S\, \Phi \bar{\Phi} + \mu_X^2 X + w  \,, \quad
K = S^\dagger S + \Phi^\dagger \Phi + \bar{\Phi}^\dagger \bar{\Phi} + X^\dagger X + 
\frac{\chi}{M_{\rm Pl}^2}\,S^\dagger S\, X^\dagger X + \cdots \,.
\end{align}
We continue to assume that $X$ is safely stabilized 
at the origin in field space, $\left<X\right> = 0$, such that the relations
in Eq.~\eqref{eq:wm32muX} remain valid also in the case of DHI.
The crucial difference between FHI and DHI is that, instead of an
inflaton F term in the superpotential, DHI features a nonvanishing
\textit{Fayet-Iliopoulos} (FI) D term~\cite{Fayet:1974jb}.
This results in an FI parameter $\xi$ in the D-term scalar potential,
\begin{align}
\label{eq:VD_dhi}
V_D = \frac{g^2}{2} \left[q_0\,\xi
- q\left(\left|\phi\right|^2 - \left|\bar{\phi}\right|^2 \right)\right]^2 \,.
\end{align}
For definiteness, we will assume $\xi > 0$.
The gauge charge $q_0$ in front of $\xi$ serves as a rescaling factor that
can take different values depending on the dynamical origin of the FI parameter.
Without loss of generality, we will simply set $q_0 = q = 1$ in the following.
This is possible since the case of general gauge charges $q_0$ and $q$
can always be restored by the following reparametrization of $g$ and $\xi$,
\begin{align}
g \rightarrow g' = \frac{g}{q} \,, \quad
\xi  \rightarrow \xi' = \frac{q}{q_0}\,\xi \,.
\end{align}
The origin of the FI parameter $\xi$ in Eq.~\eqref{eq:VD_dhi} has been the
subject of a long debate in the literature.
In particular, it has been pointed out that it is not possible to consistently
embed a genuine (\textit{i.e.}, constant) FI parameter $\xi$ into
SUGRA~\cite{Komargodski:2009pc,Dienes:2009td}.
Therefore, $\xi$ needs to be an effective FI parameter that depends
on the VEVs of scalar moduli.
This can, \textit{e.g.}, be achieved in string
theory~\cite{Dine:1987xk,Atick:1987gy} via the Green-Schwarz mechanism
of anomaly cancellation~\cite{Green:1984sg} or in strongly coupled
gauge theories via the effect of dimensional
transmutation~\cite{Domcke:2014zqa} (see~\cite{Domcke:2017xvu,Domcke:2017rzu}
for an explicit DHI model).
Besides that, there have recently been various proposals for nonstandard 
FI terms that can be consistently embedded into SUGRA
after all~\cite{Cribiori:2017laj,Kuzenko:2018jlz} (see~\cite{Antoniadis:2018cpq}
for an explicit DHI model).
However, in this paper, we will not delve into the details of this issue.
Instead, we will simply assume that an appropriate ultraviolet completion\,---\,presumably
related to one of the mechanisms listed above\,---\,results
in an effective FI term that can be treated as a constant for the purposes of inflation.
Any further speculations regarding the origin of the FI parameter
$\xi$ are beyond the scope of this work.

%%%%%%%%%%%%%%%%%%%%%%%%%%%%%%%%%%%%%%%%%%%%%%%%%%%%%%%%%%%%%%%%%%%%%%%%%%%%%%%%%%%%%%%%%%%%%%%%%%%%

The waterfall fields are again stabilized at zero during inflation,
$\big<\Phi\big> = \big<\bar{\Phi}\big> = 0$.
However, in DHI, only one field obtains a VEV during the waterfall transition.
Given our sign conventions,
\begin{align}
\label{eq:v_dhi}
\big<\Phi\big> = \frac{v}{\sqrt{2}} \,, \quad \big<\bar{\Phi}\big> = 0 \,, \quad
v = \sqrt{2\,\xi} \,,
\end{align}
where $v$ is again normalized such that it corresponds
to the VEV of the \textit{real} Higgs scalar contained in $\Phi$.
The F-term scalar potential of DHI simply follows from setting $\mu_S \rightarrow 0$
in Eq.~\eqref{eq:VF_fhi},
\begin{align}
\label{eq:VF_dhi}
V_F = e^z\,\frac{1-\chi\left(3-z\right)}{2\left(1+\chi\,z\right)}\,m_{3/2}^2\,s^2 \,, \quad
z = \frac{s^2}{2\,M_{\rm Pl}^2} \,.
\end{align}
The disappearance of the inflaton F term also eliminates the 
dependence on the complex inflaton phase $\varphi$.
DHI is therefore, indeed, a single-field model
that preserves the rotational invariance in the complex inflaton plane.
Moreover, the F-term scalar potential in Eq.~\eqref{eq:VF_dhi}
no longer contains odd powers of the real inflaton field $s$. 
Most notably, the linear tadpole term that is crucial for the dynamics of FHI
(see Eq.~\eqref{eq:cs}) is now absent.
The only terms that survive at small field values are the quadratic
mass term and the quartic self-interaction.
Analogously to Eq.~\eqref{eq:VF_fhi_Taylor}, we can write
\begin{align}
\label{eq:VF_dhi_Taylor}
V_F = \frac{1}{2}\,m_s^2\,s^2 + \frac{1}{24}\,\lambda_s\,s^4 + \mathcal{O}\left(s^6\right) \,,
\end{align}
where the coefficients $m_s^2$ and $\lambda_s$
are identical to the expressions in Eq.~\eqref{eq:msls} in the limit $\mu_S \rightarrow 0$,
\begin{align}
\label{eq:msls_dhi}
m_s^2  = \left(1-3\chi\right) m_{3/2}^2 \,, \quad 
\lambda_s = 6\left(1-3\chi+3\chi^2\right)\left(\frac{m_{3/2}}{M_{\rm Pl}}\right)^2 \,.
\end{align}
Evidently, the mass squared $m_s^2$ remains unchanged, while the
quartic self-coupling constant $\lambda_s$ no longer receives a
contribution from the superpotential in the inflation sector.
DHI only manages to reproduce the correct scalar 
spectral index, $n_s = n_s^{\rm obs}$, if the F-term scalar potential
yields a negative contribution to the slow-roll parameter $\eta$.
For this reason, we must require that $\chi > 1/3$.
In fact, we will simply set $\chi = 1$ in the
remainder of our analysis for definiteness.
The exact value of the quartic coupling $\lambda_s$ will be
irrelevant in the viable region of parameter space.
In this sense, we can set $\chi = 1$ even without loss of generality,
since any alternative value of $\chi$ (larger than $1/3$) would simply
correspond to a rescaling of the gravitino mass,
$m_{3/2}\rightarrow m_{3/2}' = \left[2/\left(3\chi-1\right)\right]^{1/2}m_{3/2}$.
The F-term scalar potential in Eq.~\eqref{eq:VF_dhi} also no longer contains
a constant SUSY-breaking contribution $V_F^0$.
Instead, the vacuum energy density driving inflation is now provided by
the constant contribution to the D-term scalar potential
along the inflationary trajectory (where $\big<\Phi\big> = \big<\bar{\Phi}\big> = 0$),
\begin{align}
\label{eq:VD0_dhi}
V_D^0 = \frac{1}{2}\,g^2\xi^2 \,.
\end{align}
To good approximation, the inflationary Hubble rate $H_{\rm inf}$ during DHI is
therefore given by
\begin{align}
\label{eq:Hinf_dhi}
H_{\rm inf} \simeq  \frac{\left(V_D^0\right)^{1/2}}{\sqrt{3}\,M_{\rm Pl}} =
\frac{g\,\xi}{\sqrt{6}\,M_{\rm Pl}} \,.
\end{align}

%%%%%%%%%%%%%%%%%%%%%%%%%%%%%%%%%%%%%%%%%%%%%%%%%%%%%%%%%%%%%%%%%%%%%%%%%%%%%%%%%%%%%%%%%%%%%%%%%%%%

Next, let us determine the mass spectrum of the waterfall sector in the global-SUSY
limit and compute the one-loop effective potential.
For the scalars, we find masses similar to those in Eq.~\eqref{eq:mpm_fhi},
\begin{align}
\label{eq:mpm_dhi}
m_\pm^2 = m_{\rm eff}^2 \pm m_D^2 \,, \quad 
m_{\rm eff}^2 = \frac{1}{2}\kappa^2 s^2 \,, \quad 
m_D^2 = g^2\xi \,,
\end{align}
which can also be written as
$m_\pm^2 = \kappa^2/2\left(s^2 \pm g^2/\kappa^2\,v^2\right)$.
From this expression, we read off the 
critical inflaton field value, $s_{\rm crit} = g/\kappa\,v$,
which now exhibits a slightly more complicated parameter dependence than in the case of FHI
(where one simply has $s_{\rm crit} = v$).
In the following, we shall restrict ourselves to parameter
values that lead to sub-Planckian values of $s_{\rm crit}$.
This is motivated by the fact that, at larger $s_{\rm crit}$,
the dynamics of inflation become sensitive to Planck-suppressed
operators in the K\"ahler potential over which we only have limited control.
The requirement of a sub-Planckian critical field value,
$s_{\rm crit} \lesssim 10^{-0.5}\,M_{\rm Pl}$, can be used to
constrain the gauge coupling $g$ from above,
\begin{align}
\label{eq:g_max_scrit}
g \lesssim \frac{\kappa s_{\rm crit}^{\rm max}}{v} \simeq 7.7 \times 10^{-2}
\:\bigg(\frac{\kappa}{10^{-5}}\bigg)\left(\frac{10^{14}\,\textrm{GeV}}{v}\right)
\left(\frac{s_{\rm crit}^{\rm max}}{10^{-0.5}M_{\rm Pl}}\right) \,,
\end{align}
which restricts part of the parameter space in the small-$\kappa$ regime.
Of course, this bound can be avoided as soon as
one is willing to make additional assumptions regarding the structure of
the K\"ahler potential at super-Planckian field values.
Large values of $s_{\rm crit}$ can, \textit{e.g.}, be achieved in combination
with a shift symmetry along the inflaton direction in the K\"ahler
potential~\cite{Buchmuller:2014rfa,Buchmuller:2014dda}.
In this case, a significant amount of inflation can even
occur at subcritical field values, $s < s_{\rm crit}$, while the combined
inflaton-waterfall-field system slowly rolls towards the true vacuum
(see also~\cite{Bryant:2016tzg,Bryant:2016sjj,Ishiwata:2018dxg}).
However, in this paper, we will neglect this possibility and simply
focus on the standard scenario of inflation \textit{prior} to the waterfall transition.
In addition to the scalar mass eigenvalues in Eq.~\eqref{eq:mpm_dhi},
we also need to know the mass of the waterfall fermion $\tilde{\phi}$.
Again, $\tilde{\phi}$ acquires a Dirac mass that coincides with
the effective supersymmetric mass induced by the inflaton VEV in
the superpotential, $m_{\tilde{\phi}}^2 = m_{\rm eff}^2$.
The one-loop effective potential $V_{1\ell}$ can, thus, be brought
into (almost) the same form as in FHI,
\begin{align}
\label{eq:V1l_dhi}
V_{1\ell} = \frac{1}{2}\,V_{1l}^0\, L\left(x\right) \,, \quad
V_{1\ell}^0 = \frac{m_D^4}{8\pi^2} \,, \quad
x = \left(\frac{s}{s_{\rm crit}}\right)^2 = 
\left(\frac{m_{\rm eff}}{m_D}\right)^2 =
\left(\frac{\kappa}{g}\right)^2\bigg(\frac{s}{v}\bigg)^2 \,.
\end{align}
This result differs from the expression in Eq.~\eqref{eq:V1l_fhi} only in terms 
of two minor details.
First of all, the overall energy scale (characterized by the
constant factor $V_{1\ell}^0$) is now determined by the
D-term-induced mass parameter $m_D$ instead of the
F-term-induced mass parameter $m_F$.
Second, the parameter dependence of the field variable $x$ is slightly
different because of the more complicated expression for $s_{\rm crit}$.
However, the loop function $L$ remains unchanged
and is still given as in Eq.~\eqref{eq:Lfunc}.

%%%%%%%%%%%%%%%%%%%%%%%%%%%%%%%%%%%%%%%%%%%%%%%%%%%%%%%%%%%%%%%%%%%%%%%%%%%%%%%%%%%%%%%%%%%%%%%%%%%%

Finally, we comment on the production of cosmic strings
in the waterfall transition at the end of inflation.
In the case of DHI (and for our sign conventions), the chiral waterfall
field $\Phi$ plays the role of both the symmetry-breaking
Higgs multiplet $H$ and the Goldstone multiplet $A$
(see the discussion around Eq.~\eqref{eq:WFHIH}).
For this reason, the mass of the physical Higgs boson, $m_H$, and
the mass of the vector boson, $m_V$, automatically coincide
with each other after the waterfall transition, $m_H^2 = m_V^2 = 2\,g^2\xi$.
As a consequence, DHI always saturates the Bogomolny limit, such that $\beta = 1$
and $\epsilon_{\rm CS} = 1$ (see Eq.~\eqref{eq:muCSbeta}).
Furthermore, there is only one real Higgs scalar that participates 
in the process of spontaneous symmetry breaking.
In DHI, the cosmic string tension is therefore simply given by the analytical
Bogomolny expression, $\mu_{\rm CS} = \pi v^2$.
In Planck units, this can be written as
\begin{align}
\label{eq:Gmu_dhi}
G\,\mu_{\rm CS} = \frac{1}{8}\left(\frac{v}{M_{\rm Pl}}\right)^2 = 
\frac{1}{4}\left(\frac{\sqrt{\xi}}{M_{\rm Pl}}\right)^2 \,.
\end{align}
Together with Eq.~\eqref{eq:GmuCSmax}, this expression results
in the following upper bound on the FI parameter $\xi$,
\begin{align}
\label{eq:ximaxCS}
\sqrt{\xi} \lesssim 1.5 \times 10^{15}\,\textrm{GeV}
\left(\frac{G\mu_{\rm CS}^{\rm max}}{10^{-7}}\right)^{1/2} \,.
\end{align}
In the following, we will again discuss two different interpretations of this bound.
On the one hand, we will explicitly illustrate its consequence for the other
parameters of DHI.
On the other hand, we will simply ignore it and explore \textit{all} of parameter
space, including the regions that violate Eq.~\eqref{eq:ximaxCS}.

%%%%%%%%%%%%%%%%%%%%%%%%%%%%%%%%%%%%%%%%%%%%%%%%%%%%%%%%%%%%%%%%%%%%%%%%%%%%%%%%%%%%%%%%%%%%%%%%%%%%

\subsection{Inflation far away from the waterfall phase transition}
\label{subsec:dhi_far}

%%%%%%%%%%%%%%%%%%%%%%%%%%%%%%%%%%%%%%%%%%%%%%%%%%%%%%%%%%%%%%%%%%%%%%%%%%%%%%%%%%%%%%%%%%%%%%%%%%%%

Let us now turn to the slow-roll dynamics of DHI.
Similarly as in Sec.~\ref{sec:fhi}, we will split our analysis into two parts
and discuss the regimes of large and small $\kappa$ values separately.
However, this time, the distinction between large and small
$\kappa$ values will be less crucial than for FHI.
The dynamics of DHI are controlled by the interplay between
the Yukawa coupling $\kappa$ and the gauge coupling $g$.
This provides us with a larger parametric freedom
that we can use to satisfy the CDM isocurvature constraint
for a broad range of axion decay constants for \textit{both} large and
small $\kappa$ values.
In this section, we will first consider the large-$\kappa$ regime.
The small-$\kappa$ regime will be discussed in Sec.~\ref{subsec:dhi_close}.

%%%%%%%%%%%%%%%%%%%%%%%%%%%%%%%%%%%%%%%%%%%%%%%%%%%%%%%%%%%%%%%%%%%%%%%%%%%%%%%%%%%%%%%%%%%%%%%%%%%%

\begin{figure}
\begin{center}

\includegraphics[width=0.490\textwidth]{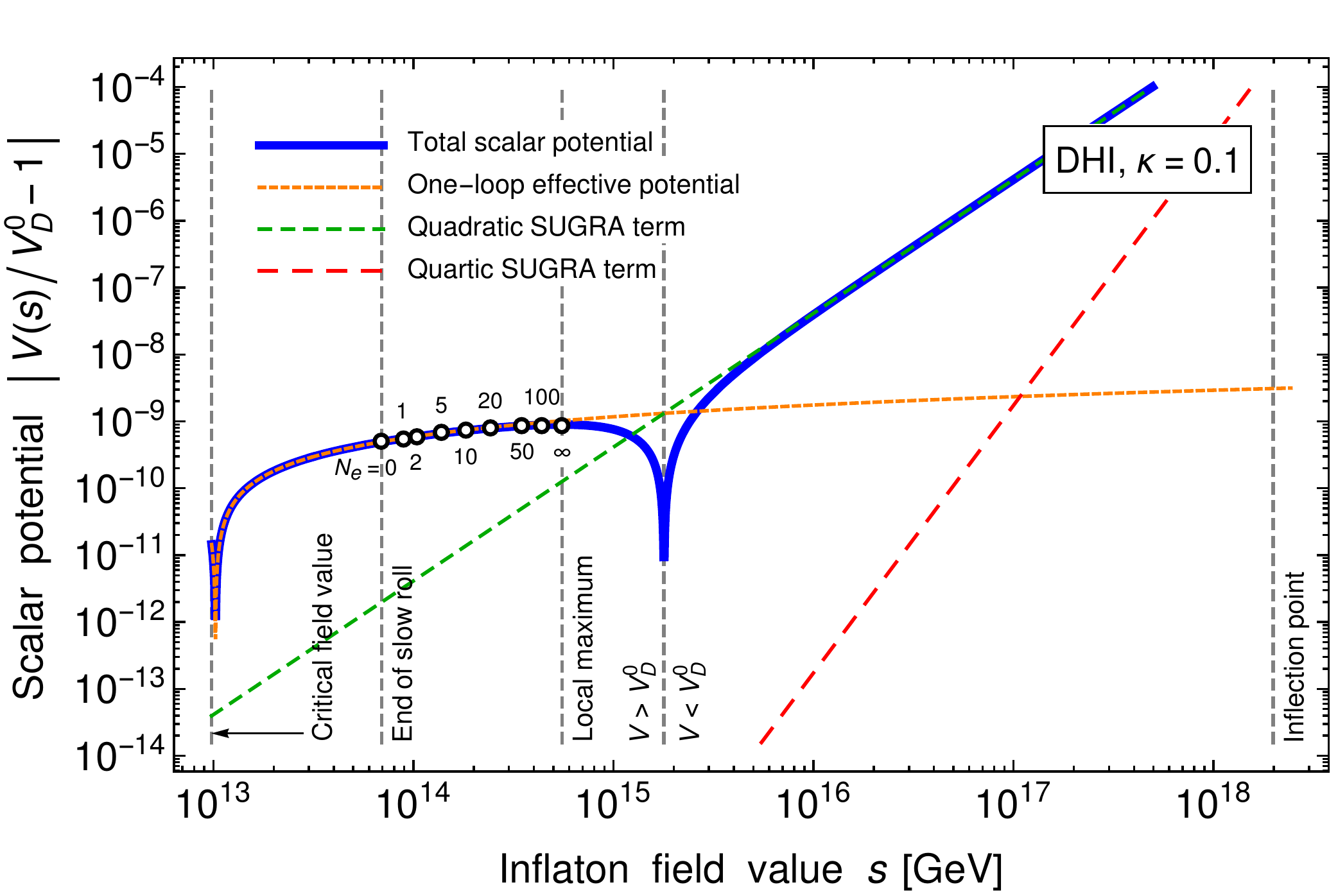}\hfill
\includegraphics[width=0.475\textwidth]{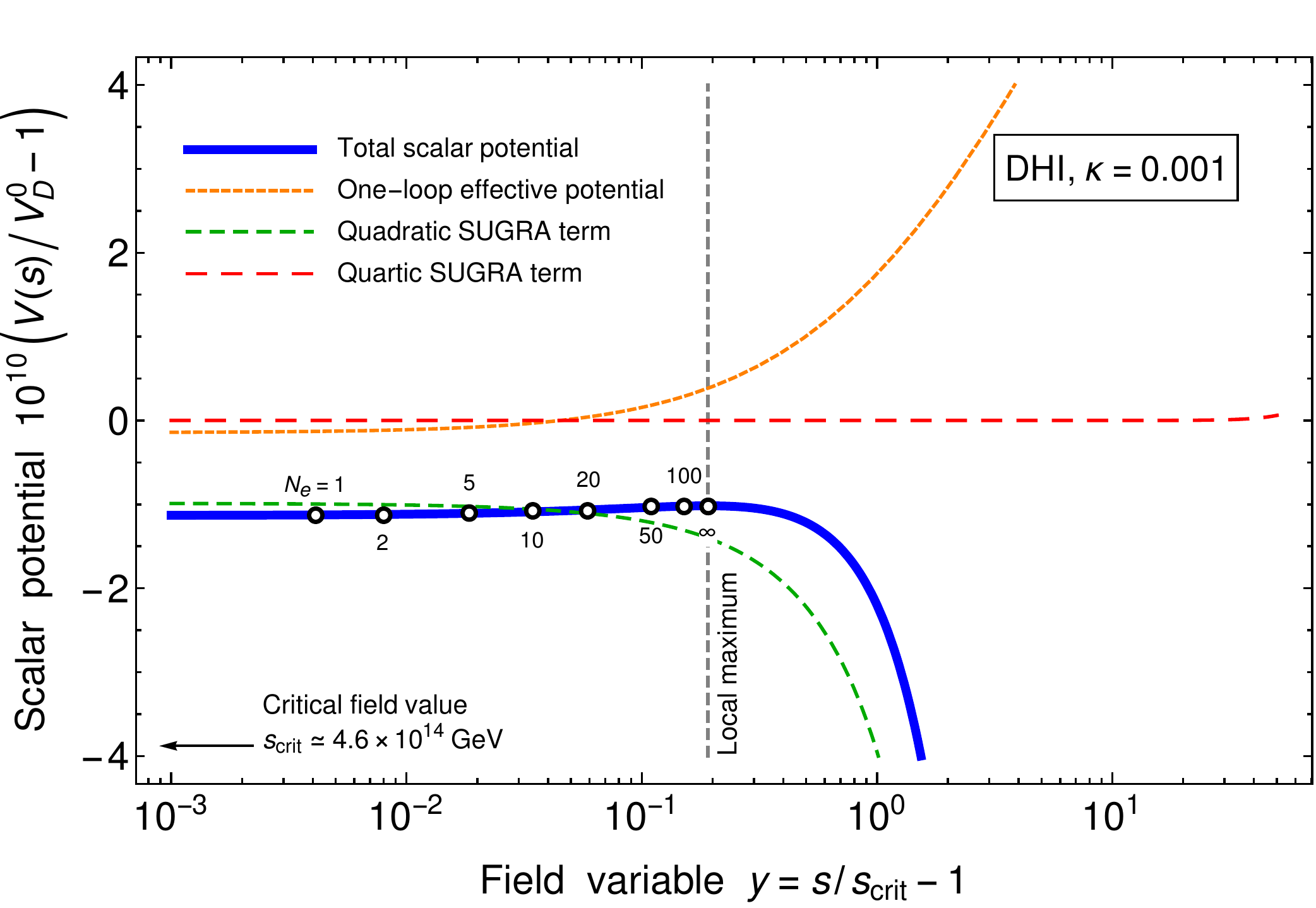}

\caption{Total scalar potential for the real inflaton field $s$ in D-term hybrid inflation
for two representative values of the inflaton Yukawa coupling $\kappa$.
Parameter values: \textbf{(Left panel)} $\kappa = 10^{-1}$,
$\sqrt{\xi} \simeq 6.9 \times 10^{15}\,\textrm{GeV}$,
$m_{3/2} \simeq 6.9 \times 10^7\,\textrm{GeV}$ and 
\textbf{(Right panel)} $\kappa = 10^{-3}$,
$\sqrt{\xi} \simeq 3.3 \times 10^{15}\,\textrm{GeV}$,
$m_{3/2} \simeq 1.6 \times 10^7\,\textrm{GeV}$.
In both panels, we set $g = 10^{-4}$, which results in a hill top
in the potential.
Both parameter points are chosen such that they reproduce the measured
CMB observables, $A_s = A_s^{\rm obs}$ and $n_s = n_s^{\rm obs}$.
The potential is always bounded from below and
positive at large field values.
In both plots, we also compare the quadratic, quartic, and radiative contributions
to the total scalar potential.}

\label{fig:potential_dhi}

\end{center}
\end{figure}

Large $\kappa$ values result again in a large inflaton
field excursion from the critical field value.
We can therefore use Eq.~\eqref{eq:L_far} again to approximate the loop function
$L$ in the one-loop effective potential by a simple logarithm.
The combination of Eqs.~\eqref{eq:VF_dhi_Taylor}, \eqref{eq:VD0_dhi},
and \eqref{eq:V1l_dhi} then yields the following approximate expression for the total
scalar potential far away from the critical field value,
\begin{align}
\label{eq:V_dhi_approx}
V \simeq V_D^0 + \frac{1}{2}\,m_s^2\,s^2 +
\frac{1}{24}\,\lambda_s\,s^4 + \frac{1}{2}\,V_{1l}^0\, \ln\left(x\right) \,.
\end{align}
In Fig.~\ref{fig:potential_dhi}, we plot the full
scalar potential for two representative $\kappa$ values, $\kappa = 10^{-1}$ and $\kappa = 10^{-3}$,
and compare it with the field-dependent contributions in Eqs.~\eqref{eq:V_dhi_approx}
and \eqref{eq:V_dhi_close} (see below in Sec.~\ref{subsec:dhi_close}).
In both cases, we conclude that the quadratic, quartic and radiative terms are
adequate to describe the full shape of the scalar potential at
field values below the Planck scale.
We also find that the scalar potential always features an
inflection point.
To see this, recall that our choice for the $\chi$ parameter, $\chi = 1$,
results in a tachyonic inflaton mass, $m_s^2 = -2\,m_{3/2}^2 < 0$ (see the discussion
below Eq.~\eqref{eq:msls_dhi}). 
Thus, there is always a point in field space, $s_{\rm flex}$, where the
positive curvature due to the quartic self-interaction term is balanced
by the negative curvature due to the logarithmic one-loop term and the quadratic mass term,
such that $V''\left(s_{\rm flex}\right) = 0$.
For a certain critical gravitino mass,
this inflection point turns again into a saddle point.
Analogously to Eq.~\eqref{eq:m32_crit_fhi}, we now have
\begin{align}
m_{3/2}^{\rm crit} & = \left(\frac{\lambda_s V_{1\ell}^0}{-3\,m_s^2}\right)^{1/2} = 
\frac{\left(V_{1\ell}^0\right)^{1/2}}{M_{\rm Pl}} 
\\ \nonumber
& =
\frac{g^2}{2\sqrt{2}\,\pi}\frac{\xi}{M_{\rm Pl}} \simeq
4.6\times10^4\,\textrm{GeV}\,\bigg(\frac{g}{10^{-4}}\bigg)^2
\left(\frac{\sqrt{\xi}}{10^{16}\,\textrm{GeV}}\right)^2 \,.
\end{align}
As in Sec.~\ref{sec:fhi}, $m_{3/2}^{\rm crit}$ allows us to
distinguish between a hill-top and an inflection-point regime.
Again, we introduce a parameter $\zeta$ that is less [greater]
than unity in the hill-top [inflection-point] regime,
\begin{align}
\label{eq:zeta_dhi}
\zeta = \bigg(\frac{m_{3/2}^{\rm crit}}{m_{3/2}}\bigg)^2 = 
\frac{g^4}{8\pi^2}\left(\frac{\xi}{m_{3/2}\,M_{\rm Pl}}\right)^2 \,.
\end{align}
Making use of this definition, we derive compact expressions
for the location of the inflection point,
\begin{align}
s_{\rm flex} = \left[1+\left(1+3\,\zeta\right)^{1/2}\right]^{1/2}
\frac{M_{\rm Pl}}{\sqrt{3}} \,,
\end{align}
as well as for the positions of the local extrema, $s_{\rm max}$ and $s_{\rm min}$, 
in the hill-top regime (\textit{i.e.}, for $\zeta < 1$), 
\begin{align}
s_{\rm  max} = \left[1-\left(1-\zeta\right)^{1/2}\right]^{1/2} M_{\rm Pl} \,, \quad
s_{\rm  min} = \left[1+\left(1-\zeta\right)^{1/2}\right]^{1/2} M_{\rm Pl} \,.
\end{align}
Note that all three field values converge to the Planck scale in the saddle-point
limit, $\zeta \rightarrow 1$.
In the following, we will, however, mostly be interested in the small-$\zeta$ regime,
which is automatically realized for small values of the gauge coupling $g$
(see Eq.~\eqref{eq:zeta_dhi}).
In this regime, we can simplify the expression for $s_{\rm max}$ by expanding
in small values of $\zeta$.
Up to corrections of $\mathcal{O}\left(\zeta^{3/2}\right)$, we obtain
\begin{align}
\label{eq:smax_dhi}
s_{\rm max} \simeq \frac{\zeta^{1/2}}{\sqrt{2}}\,M_{\rm Pl} =
\left(\frac{V_{1\ell}^0}{-m_s^2}\right)^{1/2} = \frac{g^2}{4\pi} \frac{\xi}{m_{3/2}} \,.
\end{align}
This expression coincides with the result that one obtains
if one neglects the quartic self interaction in
Eq.~\eqref{eq:V_dhi_approx} from the beginning, $\lambda_s \rightarrow 0$.
In fact, in the following, we will exclusively consider the hill-top regime for small values of $g$,
such that inflation always occurs in between the critical field value and the local maximum
in the scalar potential, $s \in \left[s_{\rm crit},s_{\rm max}\right]$.
In this part of field\,/\,parameter space, the quartic term can be safely neglected,
which is why we will set $\lambda_s \rightarrow 0$ from now on.

%%%%%%%%%%%%%%%%%%%%%%%%%%%%%%%%%%%%%%%%%%%%%%%%%%%%%%%%%%%%%%%%%%%%%%%%%%%%%%%%%%%%%%%%%%%%%%%%%%%%

%%%%%%%%%%%%%%%%%%%%%%%%%%%%%%%%%%%%%%%%%%%%%%%%%%%%%%%%%%%%%%%%%%%%%%%%%%%%%%%%%%%%%%%%%%%%%%%%%%%%

In the next step, we compute the slow-roll parameters $\varepsilon$ and $\eta$.
In parallel to Eq.~\eqref{eq:ee_fhi}, we obtain
\begin{align}
\label{eq:ee_dhi}
\varepsilon = \frac{1}{2} \left(\frac{m_s^2s^2 + V_{1\ell}^0}{V_D^0}\right)^2
\left(\frac{M_{\rm Pl}}{s}\right)^2 \,, \quad 
\eta = \Delta -\frac{V_{1\ell}^0}{V_D^0} \left(\frac{M_{\rm Pl}}{s}\right)^2 \,.
\end{align}
As usual in supersymmetric hybrid inflation, the slow-roll parameter $\varepsilon$ is
suppressed by an additional factor $V_{1\ell}^0/V_D^0$ compared to the slow-roll parameter $\eta$.
The parameter $\Delta$ in Eq.~\eqref{eq:ee_dhi} accounts for the SUGRA correction
to $\eta$ in consequence of the tachyonic mass term in Eq.~\eqref{eq:V_dhi_approx}
(see also Eq.~\eqref{eq:Delta_fhi}),
\begin{align}
\label{eq:Delta_dhi}
\Delta = M_{\rm Pl}^2\,\frac{m_s^2}{V_D^0} =
- \frac{2}{3} \left(\frac{m_{3/2}}{H_{\rm inf}}\right)^2 \,.
\end{align}
Slow-roll inflation ends and transitions into a fast-roll stage as soon
as $\eta$ reaches $\eta_{\rm max}$ (see Eq.~\eqref{eq:etamax}),
\begin{align}
s_{\rm fast} = \left(\frac{V_{1\ell}^0}{m_s^2 + m_{\rm max}^2}\right)^{1/2} = 
\frac{g\,M_{\rm Pl}}{\left(4\pi^2\,\eta_{\rm max} - 2\,g^2/\zeta\right)^{1/2}} \,, \quad
m_{\rm max}^2 = \eta_{\rm max}\,\frac{V_D^0}{M_{\rm Pl}^2} \,.
\end{align}
Here, $m_{\rm max}^2$ denotes again the maximal curvature of the scalar potential, $V''$,
that is allowed by the slow-roll bound on the parameter $\eta$.
Similarly as in Sec.~\ref{sec:fhi}, inflation ends as soon as the inflaton
field ceases to slowly roll in the scalar potential (\textit{i.e.}, at
$s = s_{\rm fast}$) or once it reaches the critical point in field space
that triggers the waterfall transition (\textit{i.e.}, at
$s = s_{\rm crit}$), $s_{\rm end} = \max\left\{s_{\rm fast},s_{\rm crit}\right\}$.

%%%%%%%%%%%%%%%%%%%%%%%%%%%%%%%%%%%%%%%%%%%%%%%%%%%%%%%%%%%%%%%%%%%%%%%%%%%%%%%%%%%%%%%%%%%%%%%%%%%%

In the hill-top regime, the slow-roll equation of motion takes the
following form (see also Eq.~\eqref{eq:eom_fhi_far}),
\begin{align}
\left(s^2\right)' = 2\,\Delta \left(s^2 - s_{\rm max}^2\right) \,, \quad
\left(s^2\right)' = 2ss' \,, \quad s_{\rm max}^2 = - \frac{V_{1\ell}^0}{m_s^2} \,.
\end{align}
In combination with the boundary condition $s = s_{\rm end}$ at $N_e = 0$,
this first-order ordinary differential equation has a unique
solution that varies exponentially with the number of $e$-folds $N_e$,
\begin{align}
\label{eq:s_dhi_far}
s^2\left(N_e\right) = s_{\rm max}^2 \left(1 + \overline{W}\right) \,, \quad 
\overline{W} = 
\bigg[\left(\frac{s_{\rm end}}{s_{\rm max}}\right)^2 - 1\bigg]\: e^{2\,\Delta\,N_e} \,.
\end{align}
Here, the function $\overline{W}$ plays a role similar to the Lambert W function
in Eq.~\eqref{eq:s_fhi_far}.
The solution in Eq.~\eqref{eq:s_dhi_far} can also be written as a function
of the three parameters $\eta_{\rm max}$, $N_e$, and $\Delta$,
\begin{align}
\label{eq:s_dhi_far_W}
s^2\left(N_e\right) = s_{\rm max}^2 \left(1 + \overline{W}\right) \,, \quad 
\overline{W} = -\left(1+\frac{\Delta}{\eta_{\rm max} + \Delta}\right)
e^{2\,\Delta\,N_e} \,.
\end{align}
Together with Eq.~\eqref{eq:ee_dhi}, this function results in
the following compact expressions for $\varepsilon$ and $\eta$,
\begin{align}
\label{eq:ee_dhi_W}
\varepsilon = \left(\frac{s_{\rm max}}{M_{\rm Pl}}\right)^2 
\frac{\left(\overline{W}\Delta\right)^2}{2\left(1+\overline{W}\right)} \,, \quad
\eta = \frac{\left(2+\overline{W}\right)\Delta}{1+\overline{W}} \,,
\end{align}
from which it is evident that $\varepsilon$ is suppressed w.r.t.\ $\eta$
by a factor $\Delta \left(s_{\rm max}/M_{\rm Pl}\right)^2$.
Therefore, to compute the scalar spectral index $n_s$, we only need
to take into account the slow-roll parameter $\eta$,
\begin{align}
\label{eq:ns_dhi}
n_s \approx 1 + 2\,\eta = 1 + \frac{2\left(2+\overline{W}\right)\Delta}{1+\overline{W}} \,.
\end{align}

%%%%%%%%%%%%%%%%%%%%%%%%%%%%%%%%%%%%%%%%%%%%%%%%%%%%%%%%%%%%%%%%%%%%%%%%%%%%%%%%%%%%%%%%%%%%%%%%%%%%

For given values of $\eta_{\rm max}$ and $N_e$ and requiring that DHI in the
large-$\kappa$ regime must result in the correct scalar
spectral index, $n_s = n_s^{\rm obs}$, Eq.~\eqref{eq:ns_dhi} can
be used to determine the parameter $\Delta$,
\begin{align}
\label{eq:Delta_value_dhi}
\eta_{\rm max} = 10^{-0.5} \,, \quad N_* = 47.5 \,, \quad n_s = n_s^{\rm obs}
\qquad\Rightarrow\qquad 
\Delta \simeq - 4.9 \times 10^{-3} \,.
\end{align}
In contrast to FHI, we now obtain a negative value for $\Delta$.
This is a direct consequence of the definition in Eq.~\eqref{eq:Delta_dhi}
and the negative sign of the inflaton mass squared in Eq.~\eqref{eq:V_dhi_approx}.
Thanks to Eq.~\eqref{eq:s_dhi_far_W}, the numerical result in Eq.~\eqref{eq:Delta_value_dhi}
fixes the inflaton field value $s_*$ at the time of CMB horizon exit,
\begin{align}
\eta_{\rm max} = 10^{-0.5} \,, \quad N_* = 47.5 \,, \quad \Delta \simeq - 4.9 \times 10^{-3} 
\qquad\Rightarrow\qquad 
\overline{W} \simeq -0.62 \,, \quad s_* \simeq 0.62\,s_{\rm max} \,.
\end{align}
Accidentally, the ratio $s_*/s_{\rm max}$ obtains almost the same value as
in the case of FHI (see Eq.~\eqref{eq:sstar_fhi}).
Furthermore, we can use the numerical value for $\Delta$ to fix the relation
between $m_{3/2}$ and $H_{\rm inf}$,
\begin{align}
\label{eq:m32Hinf_dhi}
m_{3/2} = \left(\frac{3}{2} \left|\Delta\right|\right)^{1/2} H_{\rm inf}
\simeq 8.6 \times 10^{-2}\,H_{\rm inf} \,.
\end{align}
This relation is analogous to Eq.~\eqref{eq:m32Hinf}.
Now, however, we find that the gravitino mass must only be mildly suppressed
compared to the Hubble rate.
This underlines the importance of the quadratic SUGRA term in the
scalar potential\,---\,in DHI, the soft inflaton mass term is
supposed to result in a relative variation of the
slow-roll parameter $\eta$ of $\mathcal{O}\left(1\right)$ in order to 
achieve the correct value for $n_s$.
Finally, the numerical $\Delta$ value also provides us with a numerical
expression for the parameter $\zeta$,
\begin{align}
\zeta = \frac{g^2}{2\pi^2\left|\Delta\right|} \simeq 1.0 \times 10^{-7}
\:\bigg(\frac{g}{10^{-4}}\bigg)^2 \,.
\end{align}
Therefore, for sufficiently small values of $g$, we are always deep inside the hill-top
regime.
Only for $g \geq \sqrt{2}\,\pi\left|\Delta\right|^{1/2} \simeq 0.31$,
we enter the inflection-point regime.
However, such large values of $g$ will be less interesting for us,
as they turn out to be incompatible with the CDM isocurvature constraint.

%%%%%%%%%%%%%%%%%%%%%%%%%%%%%%%%%%%%%%%%%%%%%%%%%%%%%%%%%%%%%%%%%%%%%%%%%%%%%%%%%%%%%%%%%%%%%%%%%%%%

Eq.~\eqref{eq:m32Hinf_dhi} eliminates the gravitino mass as a free parameter
from our analysis.
Similarly, we can use the observed value of the scalar
spectral amplitude, $A_s^{\rm obs}$, to eliminate the FI parameter $\xi$.
Combining Eqs.~\eqref{eq:Asns}, \eqref{eq:VD0_dhi}, \eqref{eq:smax_dhi},
\eqref{eq:Delta_dhi}, and \eqref{eq:ee_dhi_W}, we find the following compact expression,
\begin{align}
A_s = \frac{1+\overline{W}}{6\left|\Delta\right|\overline{W}^2}
\left(\frac{\sqrt{\xi}}{M_{\rm Pl}}\right)^4 \,.
\end{align}
The requirement $A_s = A_s^{\rm obs}$, thus, fixes $\sqrt{\xi}$ to
a unique value in direct proximity to the GUT scale,
\begin{align}
\label{eq:xi_num_dhi}
\sqrt{\xi} = \left(6\,A_s^{\rm obs}\left|\Delta\right|\frac{\overline{W}^2}
{1+\overline{W}}\right)^{1/4} M_{\rm Pl}
\simeq 6.9 \times 10^{15}\,\textrm{GeV} \,.
\end{align}
Remarkably enough, this result is independent of the coupling constants $\kappa$ and $g$.
This differs from the situation in FHI, where the
F-term mass scale $\mu_S$ scales like $\mu_S \propto \kappa^{1/2}$
in the large-$\kappa$ regime (see Eq.~\eqref{eq:muS_fhi_far}).
Meanwhile, the SSB scale $v$ again obtains a constant value
just like in FHI (see Eq.~\eqref{eq:v_fhi}),
\begin{align}
\label{eq:v_num_dhi}
v = \left(24\,A_s^{\rm obs}\left|\Delta\right| \frac{\overline{W}^2}
{1+\overline{W}}\right)^{1/4} M_{\rm Pl}
\simeq 9.8 \times 10^{15}\,\textrm{GeV} \,.
\end{align}
DHI saturates the Bogomolny limit (see Eq.~\eqref{eq:Gmu_dhi}).
Eq.~\eqref{eq:v_num_dhi}, thus, fixes the cosmic string tension,
\begin{align}
\label{eq:Gmu_dhi_far}
G\,\mu_{\rm CS} = \left(\frac{3}{8}\,A_s^{\rm obs}\left|\Delta\right|\frac{\overline{W}^2}
{1+\overline{W}}\right)^{1/2} \simeq 2.0 \times 10^{-6} \,.
\end{align}
This value violates the upper bound in Eq.~\eqref{eq:GmuCSmax} by an order
of magnitude.
For this reason, we are again facing two options.
We can either presume that the gauge symmetry $G$ already becomes broken 
before the end of inflation or have to resort to a different part of 
parameter space where the cosmic string tension is sufficiently suppressed
(see the discussion below Eq.~\eqref{eq:vCS}).

%%%%%%%%%%%%%%%%%%%%%%%%%%%%%%%%%%%%%%%%%%%%%%%%%%%%%%%%%%%%%%%%%%%%%%%%%%%%%%%%%%%%%%%%%%%%%%%%%%%%

An important result of our analysis is that the phenomenology of DHI is obviously
insensitive to the precise value of $\kappa$ in the large-$\kappa$ regime.
The two conditions $n_s = n_s^{\rm obs}$ and $A_s = A_s^{\rm obs}$ therefore
reduce the viable parameter space again to a one-dimensional hypersurface.
However, this time, this hypersurface is parametrized in terms
of the gauge coupling $g$ rather than the Yukawa coupling $\kappa$.
Thanks to the numerical result in Eq.~\eqref{eq:xi_num_dhi}, we obtain, \textit{e.g.},
for the inflationary Hubble rate
\begin{align}
\label{eq:Hinf_dhi_far}
H_{\rm inf} = \left|\overline{W}\right|
\left(\frac{A_s^{\rm obs}\left|\Delta\right|}{1+\overline{W}}\right)^{1/2}
g\,M_{\rm Pl} \simeq 8.0 \times 10^8\,\textrm{GeV} \:\bigg(\frac{g}{10^{-4}}\bigg) \,.
\end{align}
This expression scales linearly with $g$,
which is a completely free parameter for the time being.
As a consequence, it is straightforward to reduce the Hubble scale
of DHI by lowering $g$.
At this point, recall that the beta function of the gauge
coupling $g$ is proportional to $g$ itself (at one loop, $\beta_g^{1\ell} \propto g^3$).
Thus, small $g$ values are stable under renormalization group running and, hence,
technically natural.
The combination of Eqs.~\eqref{eq:m32Hinf_dhi} and \eqref{eq:Hinf_dhi_far}
results in the following expression for the gravitino mass,
\begin{align}
\label{eq:m32_dhi_far}
m_{3/2} = \Delta\,\overline{W}
\left[\frac{3\,A_s^{\rm obs}}{2\left(1+\overline{W}\right)}\right]^{1/2}
g\,M_{\rm Pl}
\simeq 6.9 \times 10^7\,\textrm{GeV} \:\bigg(\frac{g}{10^{-4}}\bigg) \,.
\end{align}
This explicit expression allows us to determine the critical $\kappa$
value $\kappa_0$ that separates the large-$\kappa$ regime from the small-$\kappa$ regime.
As in the case of FHI, we demand that, for $\kappa \lesssim \kappa_0,$ the local maximum
in the scalar potential is located in the direct vicinity
of the critical field value $s_{\rm crit}$,
\begin{align}
\label{eq:kappacrit_dhi}
s_{\rm scrit} = s_{\rm max} \qquad\Rightarrow\qquad
\kappa_0 = \frac{4\sqrt{2}\,\pi\,m_{3/2}\left(g\right)}{g\sqrt{\xi}} \simeq 1.8 \times 10^{-3} \,.
\end{align}
By accident, this value coincides with the critical $\kappa$
value in FHI (see Eq.~\eqref{eq:kappacrit_fhi}).

%%%%%%%%%%%%%%%%%%%%%%%%%%%%%%%%%%%%%%%%%%%%%%%%%%%%%%%%%%%%%%%%%%%%%%%%%%%%%%%%%%%%%%%%%%%%%%%%%%%%

Eqs.~\eqref{eq:Hinf_dhi_far} and \eqref{eq:m32_dhi_far} mark the main technical
results in this section.
Confronting our result for $H_{\rm inf}$ with the CDM isocurvature constraint
in Eq.~\eqref{eq:Hinfmax}, we obtain the following upper bound on $g$,
\begin{align}
\label{eq:g_max_dhi_far}
\boxed{
g \lesssim 1.6 \times 10^{-4}\left(\frac{1}{F_{\rm DM}^a}\right)^{1/2}
\left(\frac{f_a}{10^{16}\,\textrm{GeV}}\right)^{0.42}} 
\end{align}
This bound is independent of the Yukawa coupling $\kappa$ and can, hence, be
satisfied for any sensible value of $f_a$ without leaving the large-$\kappa$ regime.
This is a characteristic advantage of DHI over FHI.
Moreover, we find that Planck-scale values of $f_a$
result in an upper bound on $g$ of $\mathcal{O}\left(10^{-3}\right)$,
which is of the same order of magnitude as the upper bound
on $\kappa$ in Eq.~\eqref{eq:kappa_max_fhi_far}.
This statement remains unaffected if one also
accounts for the upper bound on $f_a$ from black hole superradiance.
Together with Eq.~\eqref{eq:m32_dhi_far}, the upper bound in 
Eq.~\eqref{eq:g_max_dhi_far} can be used to obtain an upper bound
on $m_{3/2}$,
\begin{align}
\label{eq:m32_max_dhi_far}
\boxed{
m_{3/2} \lesssim 1.1 \times 10^8\,\textrm{GeV}
\left(\frac{1}{F_{\rm DM}^a}\right)^{1/2}
\left(\frac{f_a}{10^{16}\,\textrm{GeV}}\right)^{0.42}}
\end{align}
which is weaker than the corresponding bound in
Eq.~\eqref{eq:m32_max_fhi_far} by several orders of magnitude.
This is easily explained by the fact that, unlike FHI, DHI requires a
large $m_{3/2}$--$H_{\rm inf}$ ratio in order to reproduce the observed
value of the scalar spectral index
(see the discussion below Eq.~\eqref{eq:m32Hinf_dhi}).
In Fig.~\ref{fig:bounds_dhi}, we illustrate the implications of
the CDM isocurvature constraint for the parameter space of DHI.
The plots in this figure are based on a fully numerical
analysis of slow-roll inflation in the \textit{complete} scalar potential of DHI
(see Eqs.~\eqref{eq:VF_dhi}, \eqref{eq:VD0_dhi}, and \eqref{eq:V1l_dhi}).
Again, we find excellent agreement between the numerical data and the
analytical results derived in this section.

%%%%%%%%%%%%%%%%%%%%%%%%%%%%%%%%%%%%%%%%%%%%%%%%%%%%%%%%%%%%%%%%%%%%%%%%%%%%%%%%%%%%%%%%%%%%%%%%%%%%

Finally, we use the expressions in Eqs.~\eqref{eq:Hinf_dhi_far} and \eqref{eq:m32_dhi_far}
to identify the region in parameter space where the PQ symmetry remains
unbroken during inflation.
In analogy to Eq.~\eqref{eq:kappa_min_fhi_far}, we find
\begin{align}
\label{eq:g_min_dhi_far}
g \gtrsim 1.3 \times 10^{-3}
\left(\frac{f_a}{10^{10}\,\textrm{GeV}}\right) \,, \quad
m_{3/2} \gtrsim 8.6 \times 10^8\,\textrm{GeV}
\left(\frac{f_a}{10^{10}\,\textrm{GeV}}\right) \,.
\end{align}
For small $f_a$ and large values of $g$ and $m_{3/2}$, one might therefore encounter a
domain wall problem.

%%%%%%%%%%%%%%%%%%%%%%%%%%%%%%%%%%%%%%%%%%%%%%%%%%%%%%%%%%%%%%%%%%%%%%%%%%%%%%%%%%%%%%%%%%%%%%%%%%%%

\subsection{Inflation close to the waterfall phase transition}
\label{subsec:dhi_close}

%%%%%%%%%%%%%%%%%%%%%%%%%%%%%%%%%%%%%%%%%%%%%%%%%%%%%%%%%%%%%%%%%%%%%%%%%%%%%%%%%%%%%%%%%%%%%%%%%%%%

\begin{figure}
\begin{center}

\includegraphics[width=0.475\textwidth]{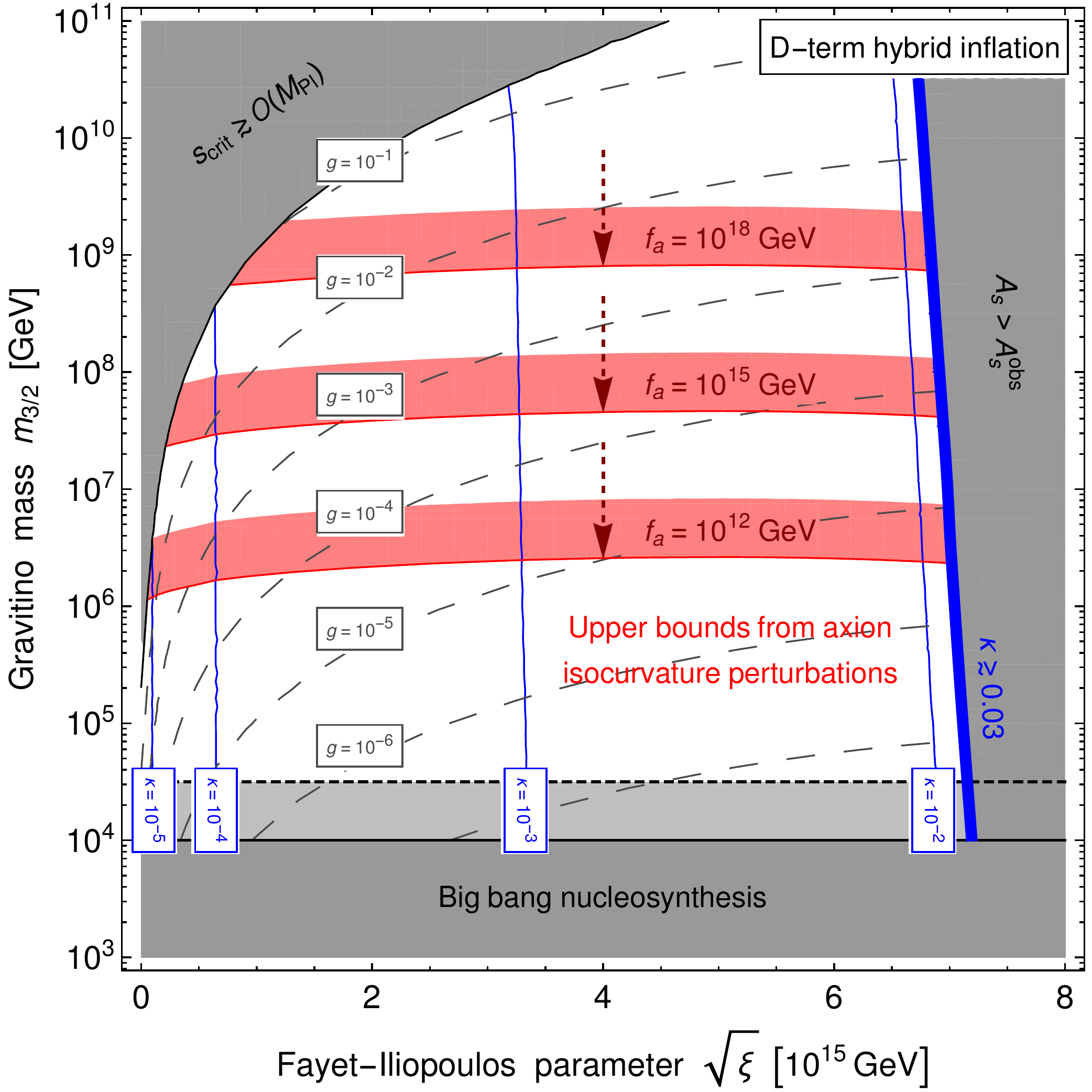}\hfill
\includegraphics[width=0.475\textwidth]{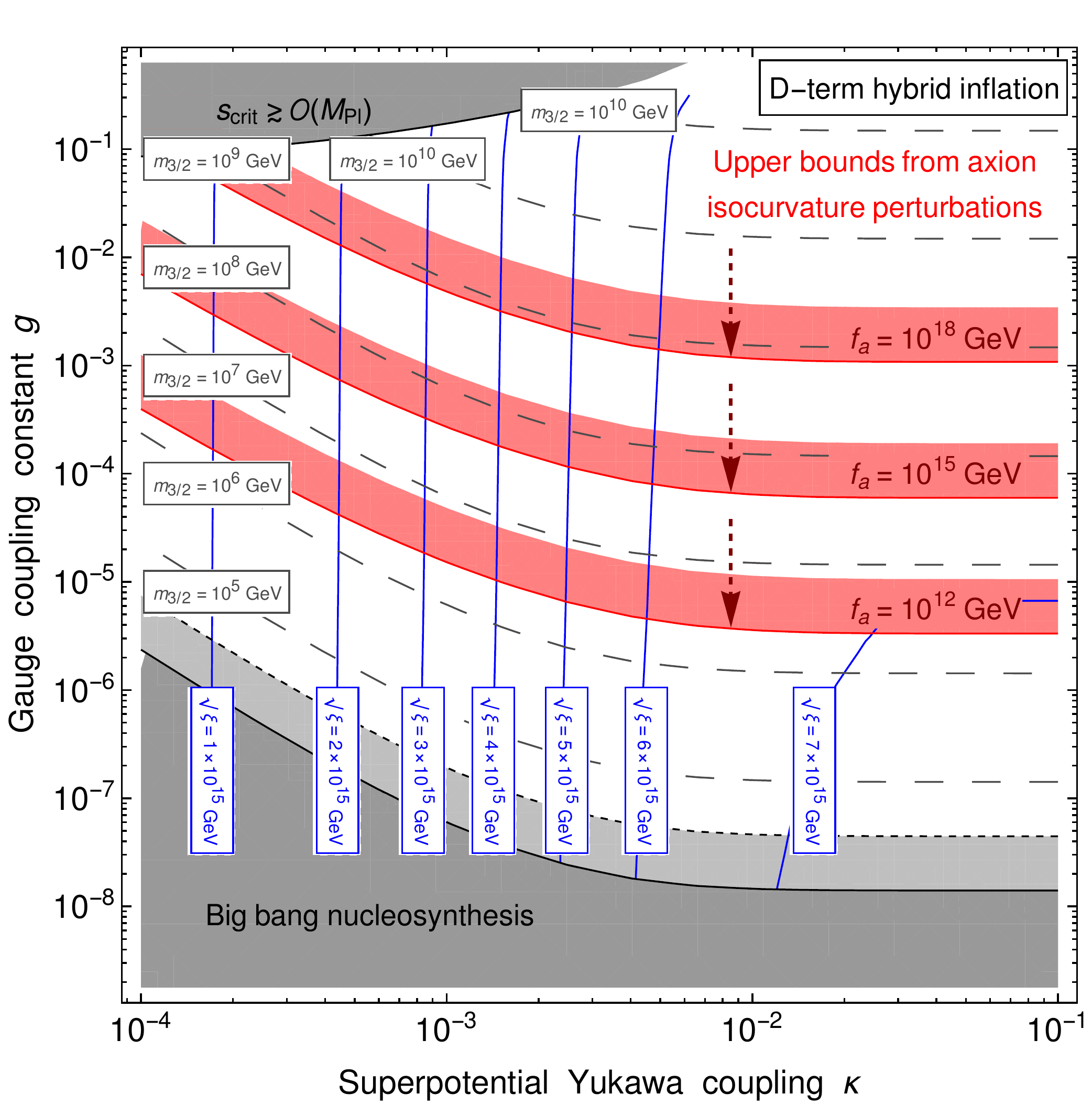}

\caption{Parameter values for D-term hybrid inflation that
reproduce the CMB data, $A_s = A_s^{\rm obs}$ and $n_s = n_s^{\rm obs}$,
in combination with the CDM isocurvature constraint for several values of the
axion decay constant and different assumptions regarding the axion DM fraction.
The stronger [weaker] bounds correspond to $F_{\rm DM}^a = 1$ [$F_{\rm DM}^a = 0.1$].
\textbf{(Left panel)} Two-dimensional region in the $\sqrt{\xi}$--$m_{3/2}$ plane
that manages to reproduce the observed CMB data.
\textbf{(Right panel)} FI scale $\sqrt{\xi}$ and gravitino mass $m_{3/2}$
as functions of the Yukawa coupling $\kappa$ and the gauge coupling $g$.
Both plots are based on a numerical analysis that accounts for
the complete scalar potential in Eqs.~\eqref{eq:VF_dhi} and \eqref{eq:V1l_dhi}.}

\label{fig:bounds_dhi}

\end{center}
\end{figure}

%%%%%%%%%%%%%%%%%%%%%%%%%%%%%%%%%%%%%%%%%%%%%%%%%%%%%%%%%%%%%%%%%%%%%%%%%%%%%%%%%%%%%%%%%%%%%%%%%%%%

In the case of FHI, the CDM isocurvature constraint forces one 
to venture into the small-$\kappa$ regime for all but the largest $f_a$ values.
As we saw in the previous section, this is no longer necessary in DHI, where small $g$
values allow one to avoid large axion isocurvature perturbations
even in the large-$\kappa$ regime.
Nonetheless, we shall also study the dynamics of DHI for small $\kappa$ values.
On the one hand, this will serve the purpose of completing our systematic
study of supersymmetric hybrid inflation for large and small Yukawa couplings.
On the other hand, small $\kappa$ values will again turn out to be the means of choice to
suppress the cosmic string tension.
At the same time, the small-$\kappa$ regime of DHI faces the same challenges
w.r.t.\ the initial conditions of inflation as the small-$\kappa$ regime of FHI
(see the first paragraph of Sec.~\ref{subsec:fhi_close}
and the right panel of Fig.~\ref{fig:potential_dhi}).
This means that a suppressed cosmic string tension can again only be achieved at the cost
of a somewhat tuned initial field value.

%%%%%%%%%%%%%%%%%%%%%%%%%%%%%%%%%%%%%%%%%%%%%%%%%%%%%%%%%%%%%%%%%%%%%%%%%%%%%%%%%%%%%%%%%%%%%%%%%%%%

To obtain the scalar potential in the small-$\kappa$ regime, we are able to proceed 
in the same way as in Sec.~\ref{subsec:fhi_close}.
That is, we have to replace the logarithm $\ln\left(x\right)$
in Eq.~\eqref{eq:V_dhi_approx} by the function $L_2\left(y\right)$,
\begin{align}
\label{eq:V_dhi_close}
V \simeq V_D^0 + \frac{1}{2}\,m_s^2\,s^2 + \frac{1}{2}\,V_{1\ell}^0\,L_2\left(y\right) \,,
\end{align}
where we again neglected the quartic SUGRA term.
Correspondingly, $\varepsilon$ and $\eta$ in Eq.~\eqref{eq:ee_dhi} turn into
\begin{align}
\label{eq:ee_dhi_close}
\varepsilon = \frac{1}{2}\left(\frac{m_s^2 s_{\rm crit} s + c_1 V_{1\ell}^0/2}{V_D^0}\right)^2
\left(\frac{M_{\rm Pl}}{s_{\rm crit}}\right)^2 \,, \quad 
\eta = \Delta + \frac{V_{1\ell}^0}{2\,V_D^0}\left(c_2 + \frac{3}{2}\,\bar{c}_2 + \bar{c}_2
\ln y\right)\left(\frac{M_{\rm Pl}}{s_{\rm crit}}\right)^2 \,.
\end{align}
In contrast to FHI in the small-$\kappa$ regime, the
parameter $\varepsilon$ now receives a field-dependent
contribution from the quadratic mass term in Eq.~\eqref{eq:V_dhi_close}.
This contribution comes with a negative sign (recall that $m_s^2 < 0$),
which is responsible for the presence of the local maximum at $s_{\rm max}$.
In the small-$\kappa$ regime, the field value $s_{\rm max}$ follows from
the requirement that $\varepsilon$ in Eq.~\eqref{eq:ee_dhi_close} must vanish
at $s = s_{\rm max}$,
\begin{align}
\label{eq:smax_dhi_close}
s_{\rm max} = - \frac{c_1V_{1\ell}^0}{2\,s_{\rm crit}\,m_s^2} =
\frac{\ln2\,g^3\kappa}{8\sqrt{2}\,\pi^2} \left(\frac{\sqrt{\xi}}{m_{3/2}}\right)^2\sqrt{\xi} \,.
\end{align}
This expression comes in handy when writing down the slow-roll equation of motion
for the inflaton,
\begin{align}
\label{eq:eom_dhi_close}
s' = \Delta\left(s-s_{\rm max}\right) \,,
\end{align}
where $\Delta$ is still defined as in Eq.~\eqref{eq:Delta_dhi}.
Together with the boundary condition $s = s_{\rm crit}$
at $N_e = 0$, the differential equation in Eq.~\eqref{eq:eom_dhi_close} has
a unique solution in terms of a simple exponential function,
\begin{align}
s\left(N_e\right) = s_{\rm max} + \left(s_{\rm crit} - s_{\rm max}\right) e^{\Delta\,N_e} \,.
\end{align}
This result allows us to write down explicit expressions
for $\varepsilon$ and $\eta$ as functions of $y_{\rm max}$, $N_e$, and $\Delta$,
\begin{align}
\label{eq:ee_dhi_y}
\varepsilon & = \bigg(\frac{g}{4\pi}\bigg)^2
\left(\frac{y_{\rm max}}{1+y_{\rm max}}\right)^2 c_1\,e^{2\,\Delta\,N_e} \left|\Delta\right| \,,
\\ \nonumber
\eta & = \Delta - \Delta \left(y_{\rm max}+1\right)\left[\frac{c_2}{c_1}+ \frac{3\,\bar{c}_2}{2\,c_1}
+ \frac{\bar{c}_2}{c_1} \ln\left(y_{\rm max}\left(1-e^{\Delta\,N_e}\right)\right)\right] \,,
\end{align}
where $y_{\rm max} = s_{\rm max}/s_{\rm crit}-1$ (see also Eq.~\eqref{eq:LL2}).
Again, we notice that $\varepsilon$ is suppressed w.r.t.\ $\eta$.

%%%%%%%%%%%%%%%%%%%%%%%%%%%%%%%%%%%%%%%%%%%%%%%%%%%%%%%%%%%%%%%%%%%%%%%%%%%%%%%%%%%%%%%%%%%%%%%%%%%%

Next, we use the two conditions
$A_s = A_s^{\rm obs}$ and $n_s = n_s^{\rm ons}$
to determine the two parameters $y_{\rm max}$ and $\Delta$.
First, let us consider the amplitude of the scalar power spectrum.
Combining our results in Eqs.~\eqref{eq:Asns}, \eqref{eq:VD0_dhi},
\eqref{eq:Delta_dhi}, \eqref{eq:smax_dhi_close}, and \eqref{eq:ee_dhi_y},
a straightforward calculation provides us with
\begin{align}
\label{eq:As_dhi_close}
A_s = \frac{c_1}{3}\left(\frac{\kappa}{4\pi}\right)^4 \frac{e^{-2\,\Delta\,N_e}}
{\left(y_{\rm max} + 1\right)y_{\rm max}^2\left|\Delta\right|^3} \approx 
\frac{c_1}{3}\left(\frac{\kappa}{4\pi}\right)^4 \frac{e^{-2\,\Delta\,N_e}}
{y_{\rm max}^2\left|\Delta\right|^3} \,.
\end{align}
Here, we made use of the fact that $y_{\rm max}$ is much smaller than unity
in the small-$\kappa$ regime, $y_{\rm max} \ll 1$.
Imposing the condition that $A_s$ must reproduce $A_s^{\rm obs}$,
we are able to solve Eq.~\eqref{eq:As_dhi_close} for $y_{\rm max}$,
\begin{align}
\label{eq:ymax}
y_{\rm max} = \left(\frac{c_1}{3A_s^{\rm obs}}\right)^{1/2}\bigg(\frac{\kappa}{4\pi}\bigg)^{2}
\:\frac{e^{-\Delta\,N_e}} {\left|\Delta\right|^{3/2}} \,,
\end{align}
which is suppressed by two powers of the small factor $\kappa/\left(4\pi\right)$.
Together with Eq.~\eqref{eq:ee_dhi_y}, this result allows us to write down
the scalar spectral index $n_s$ as a function of $N_e$, $\Delta$, and $\kappa$.
As before, we will neglect the slow-roll parameter $\varepsilon$ and simply approximate
$n_s$ by $n_s \approx 1 + 2\,\eta$.
We, thus, obtain
\begin{align}
n_s \approx 1 + 2\,\Delta - 2\,\Delta\left[\frac{c_2}{c_1}
+ \frac{3\,\bar{c}_2}{2\,c_1}
+ \frac{\bar{c}_2}{c_1}\ln\bigg(\hspace{-3pt}\left(\frac{c_1}{3A_s^{\rm obs}}\right)^{1/2}
\bigg(\frac{\kappa}{4\pi}\bigg)^{2}
\:\frac{e^{-\Delta\,N_e}-1}{\left|\Delta\right|^{3/2}}\bigg)\right] \,.
\end{align}
For a fixed value of $N_e$, the condition $n_s = n_s^{\rm obs}$ can be numerically solved 
for $\Delta$ as a function of $\kappa$,
\begin{align}
\label{eq:Delta_value_dhi_close}
\quad N_e = 47.5 \,, \quad n_s = n_s^{\rm obs}
\qquad\Rightarrow\qquad 
\Delta \simeq - 1.2 \times 10^{-3} \:\bigg(\frac{\kappa}{10^{-5}}\bigg)^p \,, \quad
p = 0.14 \,.
\end{align}
Evidently, $\Delta$ depends only very weakly on the Yukawa coupling $\kappa$.
The slight numerical uncertainty in the exponent $p$ is therefore irrelevant for all
practical purposes.
In fact, we checked that the power law in Eq.~\eqref{eq:Delta_value_dhi_close}
represents an adequate fit to the exact numerical result for all relevant $\kappa$ values
deep inside the small-$\kappa$ regime, $\kappa \lesssim \mathcal{O}\left(10^{-4}\right)$.
Together with Eq.~\eqref{eq:ymax}, we now obtain for $y_{\rm max}$,
\begin{align}
\label{eq:ymax_num}
y_{\rm max} \simeq 3.1 \times 10^{-4} \:\bigg(\frac{\kappa}{10^{-5}}\bigg)^{1.79} \,,
\end{align}
which is, indeed, much smaller than unity as long as $\kappa$ is sufficiently small.

%%%%%%%%%%%%%%%%%%%%%%%%%%%%%%%%%%%%%%%%%%%%%%%%%%%%%%%%%%%%%%%%%%%%%%%%%%%%%%%%%%%%%%%%%%%%%%%%%%%%

With the numerical expressions in Eqs.~\eqref{eq:Delta_value_dhi_close}
and \eqref{eq:ymax_num} at our disposal, we are now ready to compute
the mass scales that are relevant in the description of
DHI for small $\kappa$ values.
As in Sec.~\ref{subsec:dhi_far}, we first consider the FI mass scale $\sqrt{\xi}$.
Making use of Eqs.~\eqref{eq:Delta_dhi} and \eqref{eq:smax_dhi_close},
we find 
\begin{align}
\label{eq:xi_dhi_close}
\sqrt{\xi} = \left(\frac{c_1}{\left(1+y_{\rm max}\right)\left|\Delta\right|}\right)^{1/2}
\frac{\kappa}{4\pi}\,M_{\rm Pl}
\simeq 9.2\times 10^{13}\,\textrm{GeV} \:\bigg(\frac{\kappa}{10^{-5}}\bigg)^{0.93} \,. 
\end{align}
In contrast to the value in Eq.~\eqref{eq:xi_num_dhi}, this result
is suppressed by the small Yukawa coupling $\kappa$ and, hence,
parametrically smaller than $\Lambda_{\rm GUT}$.
The same applies to the value of the SSB scale $v$,
\begin{align}
\label{eq:v_dhi_close}
v = \left(\frac{2\,c_1}{\left(1+y_{\rm max}\right)\left|\Delta\right|}\right)^{1/2}
\frac{\kappa}{4\pi}\,M_{\rm Pl}
\simeq 1.3\times 10^{14}\,\textrm{GeV} \:\bigg(\frac{\kappa}{10^{-5}}\bigg)^{0.93} \,.
\end{align}
As a consequence, the cosmic string tension turns out to be suppressed by almost
two powers of $\kappa$,
\begin{align}
\label{eq:Gmu_dhi_close}
G\,\mu_{\rm CS} = \frac{c_1}{4\left(1+y_{\rm max}\right)\left|\Delta\right|}
\:\bigg(\frac{\kappa}{4\pi}\bigg)^2 \simeq 3.5 \times 10^{-10}
\:\bigg(\frac{\kappa}{10^{-5}}\bigg)^{1.86} \,.
\end{align}
Similarly as in FHI, it is therefore always possible to avoid the cosmic string bound
in Eq.~\eqref{eq:GmuCSmax} by choosing $\kappa$ small enough.
This resolves the cosmic string problem of DHI at the price of fine-tuned initial 
conditions.
We also note that the mass scales $\sqrt{\xi}$ and $v$ are solely controlled by $\kappa$.
This is no longer the case for the Hubble rate $H_{\rm inf}$, which depends on
both $\kappa$ and $g$ (see Eq.~\eqref{eq:Hinf_dhi}),
\begin{align}
\label{eq:Hinf_dhi_close}
H_{\rm inf} = \frac{c_1\,g}{\sqrt{6}\left(1+y_{\rm max}\right)\left|\Delta\right|}
\:\bigg(\frac{\kappa}{4\pi}\bigg)^2 M_{\rm Pl}
\simeq 1.4 \times 10^7\,\textrm{GeV}
\:\bigg(\frac{g}{10^{-2}}\bigg)\bigg(\frac{\kappa}{10^{-5}}\bigg)^{1.86} \,.
\end{align}
This result illustrates that, at small $\kappa$ values, the Hubble rate
approximately scales like $H_{\rm inf} \propto g \kappa^2$.
It is, thus, possible to suppress $H_{\rm inf}$ by a small $\kappa$ value
while keeping $g$ moderately large.
This is an important difference from the large-$\kappa$ regime
where $H_{\rm inf}$ can only be suppressed by small values
of $g$ (see Eq.~\eqref{eq:Hinf_dhi_far}).
The situation is similar for the gravitino mass for which we obtain
(see Eq.~\eqref{eq:m32Hinf_dhi})
\begin{align}
\label{eq:m32_dhi_close}
m_{3/2} = \frac{c_1\,g}{2\left(1+y_{\rm max}\right)\left|\Delta\right|^{1/2}}
\:\bigg(\frac{\kappa}{4\pi}\bigg)^2 M_{\rm Pl}
\simeq 6.1 \times 10^5\,\textrm{GeV}
\:\bigg(\frac{g}{10^{-2}}\bigg)\bigg(\frac{\kappa}{10^{-5}}\bigg)^{1.93} \,.
\end{align}

%%%%%%%%%%%%%%%%%%%%%%%%%%%%%%%%%%%%%%%%%%%%%%%%%%%%%%%%%%%%%%%%%%%%%%%%%%%%%%%%%%%%%%%%%%%%%%%%%%%%

For the fourth and last time, we are now able to use our results
and constrain the viable parameter space by means of
the CDM isocurvature constraint in Eq.~\eqref{eq:Hinfmax}.
Together, Eqs.~\eqref{eq:Hinfmax} and \eqref{eq:Hinf_dhi_close} yield
\begin{align}
\label{eq:g_max_dhi_close}
\boxed{
g \lesssim 2.0 \times 10^{-2}
\left(\frac{10^{-5}}{\kappa}\right)^{1.86}
\left(\frac{1}{F_{\rm DM}^a}\right)^{1/2}
\left(\frac{f_a}{10^{12}\,\textrm{GeV}}\right)^{0.42}}
\end{align}
This bound is weaker than the one in Eq.~\eqref{eq:g_max_dhi_far}, which
reflects the fact that, now, the Hubble rate $H_{\rm inf}$ is suppressed
by the small values of both $\kappa$ and $g$.
In particular, we note that $g$ can now even be larger than
$\mathcal{O}\left(10^{-3}\right)$.
However, it is important to remember that this is only possible as long as we
are in the small-$\kappa$ regime, \textit{i.e.}, as long as $\kappa \lesssim
\mathcal{O}\left(10^{-3}\right)$ (see Eq.~\eqref{eq:kappacrit_dhi}).
In summary, we therefore conclude that, also in the small-$\kappa$ regime,
at least one coupling constant must not be larger than $\mathcal{O}\left(10^{-3}\right)$.
This completes our analysis of the upper bounds on $\kappa$ and $g$ in consequence
of the CDM isocurvature constraint.
Our main result according to Eqs.~\eqref{eq:kappa_max_fhi_far},
\eqref{eq:kappa_max_fhi_close}, \eqref{eq:g_max_dhi_far}, and \eqref{eq:g_max_dhi_close}
is that
\begin{quote}
\textit{Supersymmetric hybrid inflation complies with the CDM isocurvature
constraint if an appropriate (Yukawa or gauge) coupling constant
is set to a value of $\mathit{\mathcal{O}\left(10^{-3}\right)}$ or smaller.}
\end{quote}
The upper bound in Eq.~\eqref{eq:g_max_dhi_close} can also be used 
to derive an upper bound on $m_{3/2}$ (see Eq.~\eqref{eq:m32_dhi_close}),
\begin{align}
\label{eq:m32_max_dhi_close}
\boxed{
m_{3/2} \lesssim 1.2 \times 10^6\,\textrm{GeV}
\:\bigg(\frac{\kappa}{10^{-5}}\bigg)^{0.07}
\left(\frac{1}{F_{\rm DM}^a}\right)^{1/2}
\left(\frac{f_a}{10^{12}\,\textrm{GeV}}\right)^{0.42}}
\end{align}
This bound is stronger than the one in Eq.~\eqref{eq:m32_max_dhi_far}, which
is consistent with the fact that, in the small-$\kappa$ regime,
all mass scales are subject to an additional suppression by the small value of $\kappa$.
Eq.~\eqref{eq:m32_max_dhi_close} completes our analysis of the upper bounds on $m_{3/2}$.
Similarly as for $\kappa$ and $g$,
we are now able to compare and summarize the bounds in Eq.~\eqref{eq:m32_max_fhi_far},
\eqref{eq:m32_max_fhi_close}, \eqref{eq:m32_max_dhi_far}, and \eqref{eq:m32_max_dhi_close}.
We conclude that
\begin{quote}
\textit{For $\mathit{f_a}$ as large as $\mathit{f_a \sim M_{\textrm{\textit{Pl}}}}$,
the CDM isocurvature constraint translates into absolute upper bounds on the gravitino mass of
$\mathit{\mathcal{O}\left(10^5\right)\,\textrm{\textit{GeV}}}$ in FHI and 
$\mathit{\mathcal{O}\left(10^9\right)\,\textrm{\textit{GeV}}}$ in DHI.}
\end{quote}

%%%%%%%%%%%%%%%%%%%%%%%%%%%%%%%%%%%%%%%%%%%%%%%%%%%%%%%%%%%%%%%%%%%%%%%%%%%%%%%%%%%%%%%%%%%%%%%%%%%%

The analytical results in Eqs.~\eqref{eq:g_max_dhi_close}
and \eqref{eq:m32_max_dhi_close} need to be compared with the fully numerical
result in Fig.~\ref{fig:bounds_dhi}.
Once again, we find excellent agreement.
Furthermore, we can use our result for $G\,\mu_{\rm CS}$ in Eq.~\eqref{eq:Gmu_dhi_close}
to determine the range of $\kappa$ values that allow to satisfy the upper bound on the
cosmic string tension.
Together with Eq.~\eqref{eq:m32_max_dhi_close}, we find the following
upper bounds on $\kappa$ and $m_{3/2}$,
\begin{align}
\label{eq:kappaCSDHI}
G\,\mu_{\rm CS} < G\,\mu_{\rm CS}^{\rm max} \qquad\Rightarrow\qquad
\kappa  \lesssim 2.1 \times 10^{-4} \,, \quad 
m_{3/2} \lesssim 2.1 \times 10^8\,\textrm{GeV} \:\bigg(\frac{g}{10^{-2}}\bigg) \,.
\end{align}
The upper bound on $\kappa$ is slightly smaller than the critical $\kappa$ value
in Eq.~\eqref{eq:kappacrit_dhi}.
This means that the cosmic string bound can only be circumvented for $\kappa$ values
deep inside the small-$\kappa$ regime.
For all other $\kappa$ values, we have to assume that no cosmic strings are
produced at the end of inflation.
Making use of Eqs.~\eqref{eq:xi_dhi_close}, \eqref{eq:v_dhi_close},
and \eqref{eq:Hinf_dhi_close}, the constraints in Eq.~\eqref{eq:kappaCSDHI} can
also be translated into
\begin{align}
v           \lesssim 2.2 \times 10^{15}\,\textrm{GeV} \,, \quad
\sqrt{\xi}  \lesssim 1.5 \times 10^{15}\,\textrm{GeV} \,, \quad
H_{\rm inf} \lesssim 4.0 \times 10^9   \,\textrm{GeV} \:\bigg(\frac{g}{10^{-2}}\bigg) \,.
\end{align}
The result is consistent with the bound on the FI mass scale $\sqrt{\xi}$
in Eq.~\eqref{eq:ximaxCS}.

%%%%%%%%%%%%%%%%%%%%%%%%%%%%%%%%%%%%%%%%%%%%%%%%%%%%%%%%%%%%%%%%%%%%%%%%%%%%%%%%%%%%%%%%%%%%%%%%%%%%

Finally, we use the expressions in Eqs.~\eqref{eq:Hinf_dhi_close} and \eqref{eq:m32_dhi_close}
to identify the region in parameter space where the PQ symmetry remains
unbroken during inflation.
In analogy to Eq.~\eqref{eq:kappa_min_fhi_close}, we find
\begin{align}
\label{eq:g_min_dhi_close}
g \gtrsim 7.1 \times 10^{-2}
\left(\frac{10^{-5}}{\kappa}\right)^{1.86}
\left(\frac{f_a}{10^8\,\textrm{GeV}}\right) \,, \quad
m_{3/2} \gtrsim 4.3 \times 10^6\,\textrm{GeV}
\:\bigg(\frac{\kappa}{10^{-5}}\bigg)^{0.07}
\left(\frac{f_a}{10^8\,\textrm{GeV}}\right) \,.
\end{align}
However, this time, we must be careful when asking for the interpretation
of these results.
For values of $g$ and $m_{3/2}$ as large as those in Eq.~\eqref{eq:g_min_dhi_close},
the critical field value $s_{\rm crit}$ begins to exceed the Planck scale
(see Eq.~\eqref{eq:g_max_scrit}).
Thus, in this part of parameter space, inflaton occurs at super-Planckian
field values where we have less control over
the SUGRA corrections to the scalar potential.
In this paper, we decided to restrict ourselves to regions in parameter
space where $s_{\rm crit} \lesssim \mathcal{O}\left(M_{\rm Pl}\right)$.
For this reason, the bounds in Eq.~\eqref{eq:g_min_dhi_close} are
irrelevant for our purposes as soon as they are in conflict
with Eq.~\eqref{eq:g_max_scrit}.

%%%%%%%%%%%%%%%%%%%%%%%%%%%%%%%%%%%%%%%%%%%%%%%%%%%%%%%%%%%%%%%%%%%%%%%%%%%%%%%%%%%%%%%%%%%%%%%%%%%%

\begin{figure}
\begin{center}

\includegraphics[width=0.49\textwidth]{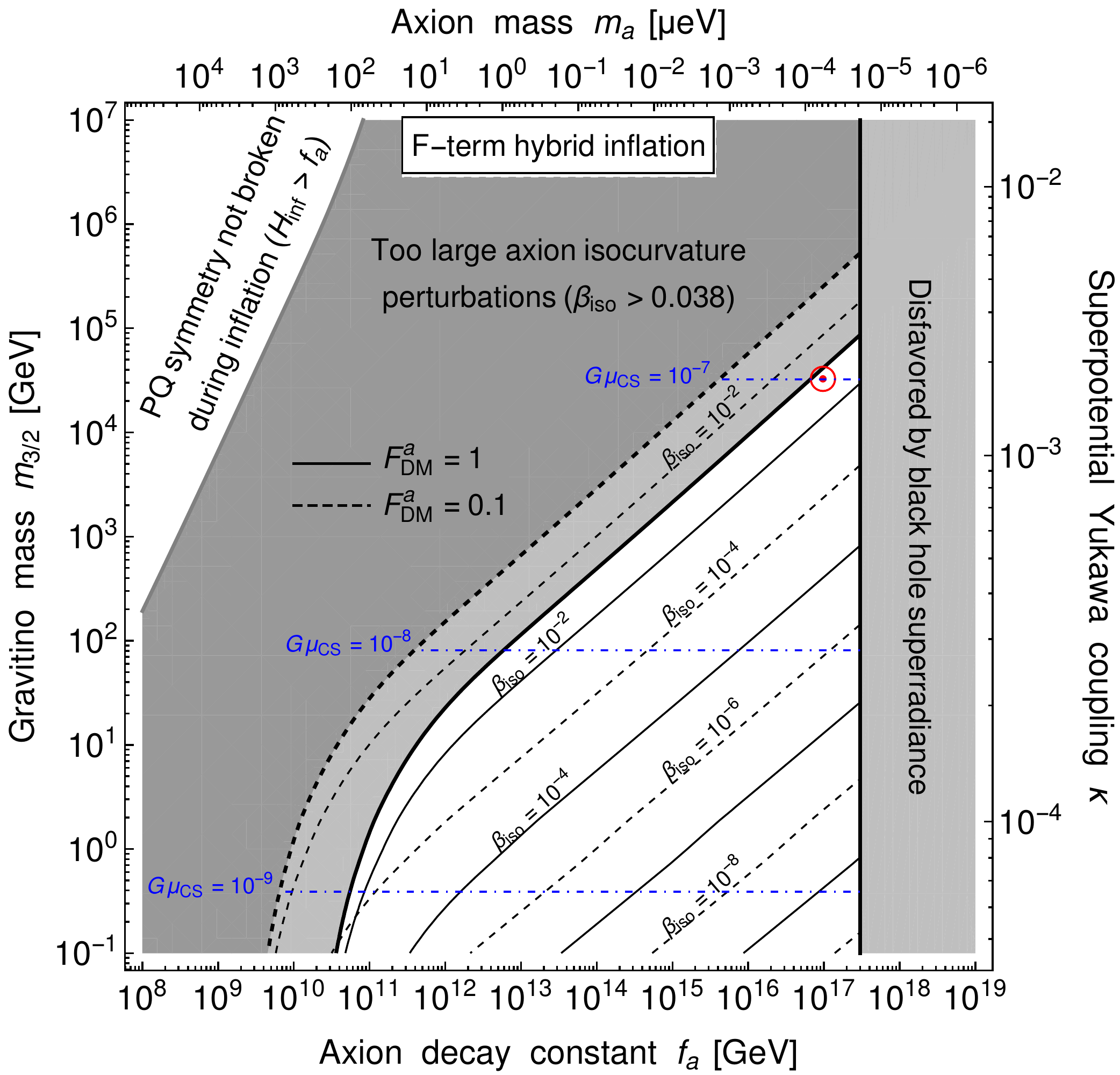}\hfill
\includegraphics[width=0.49\textwidth]{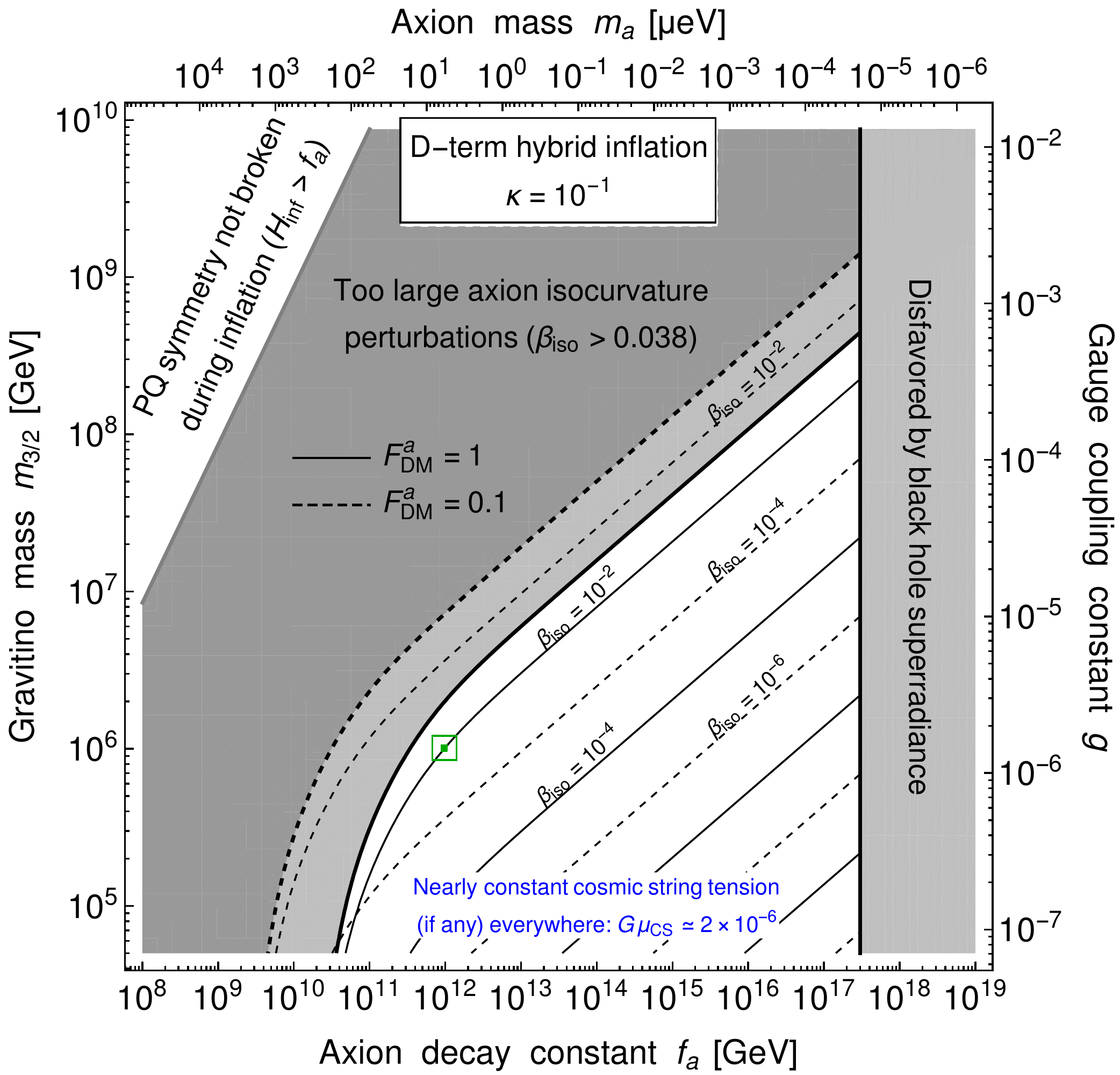}

\medskip

\includegraphics[width=0.49\textwidth]{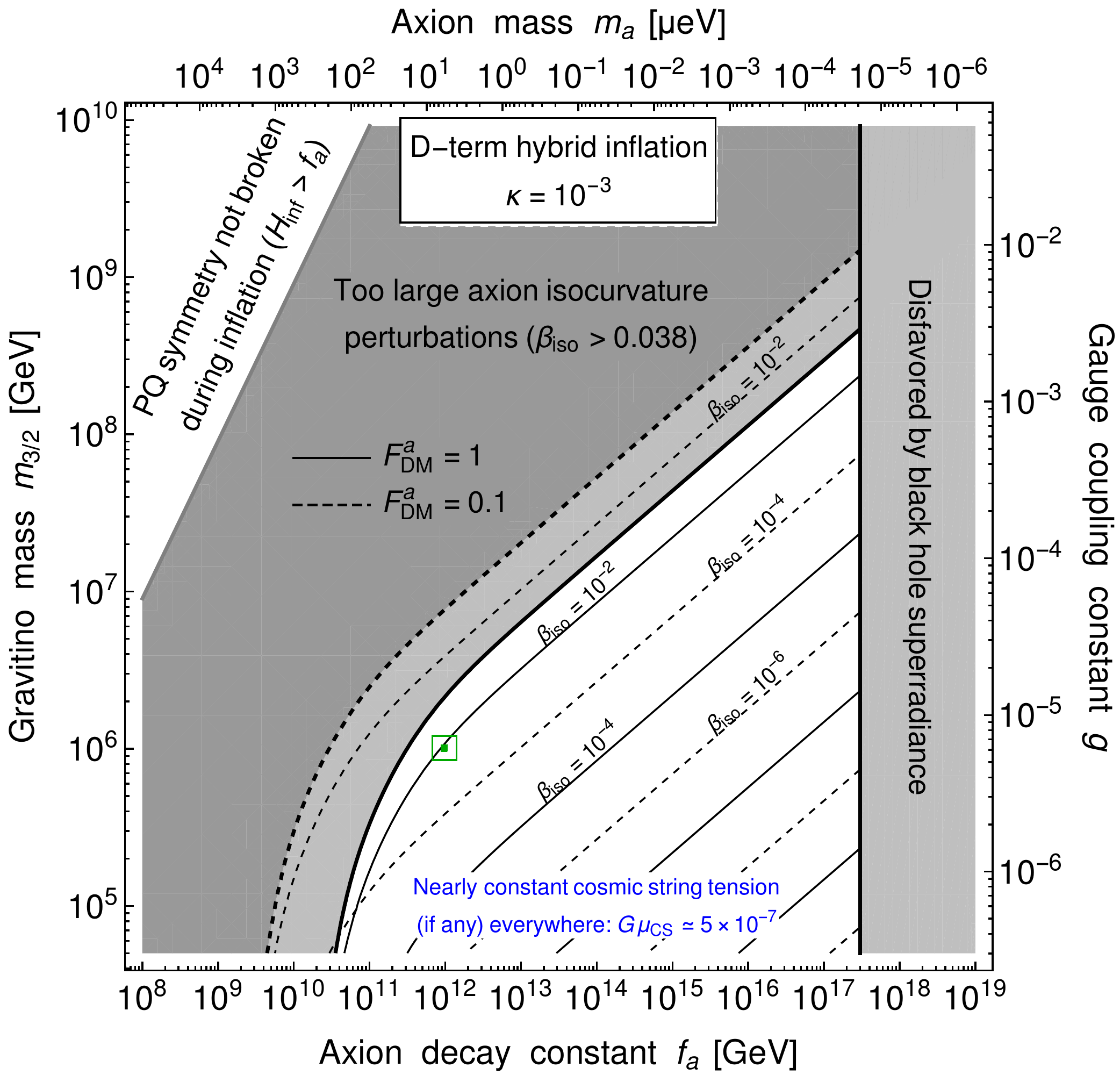}\hfill
\includegraphics[width=0.49\textwidth]{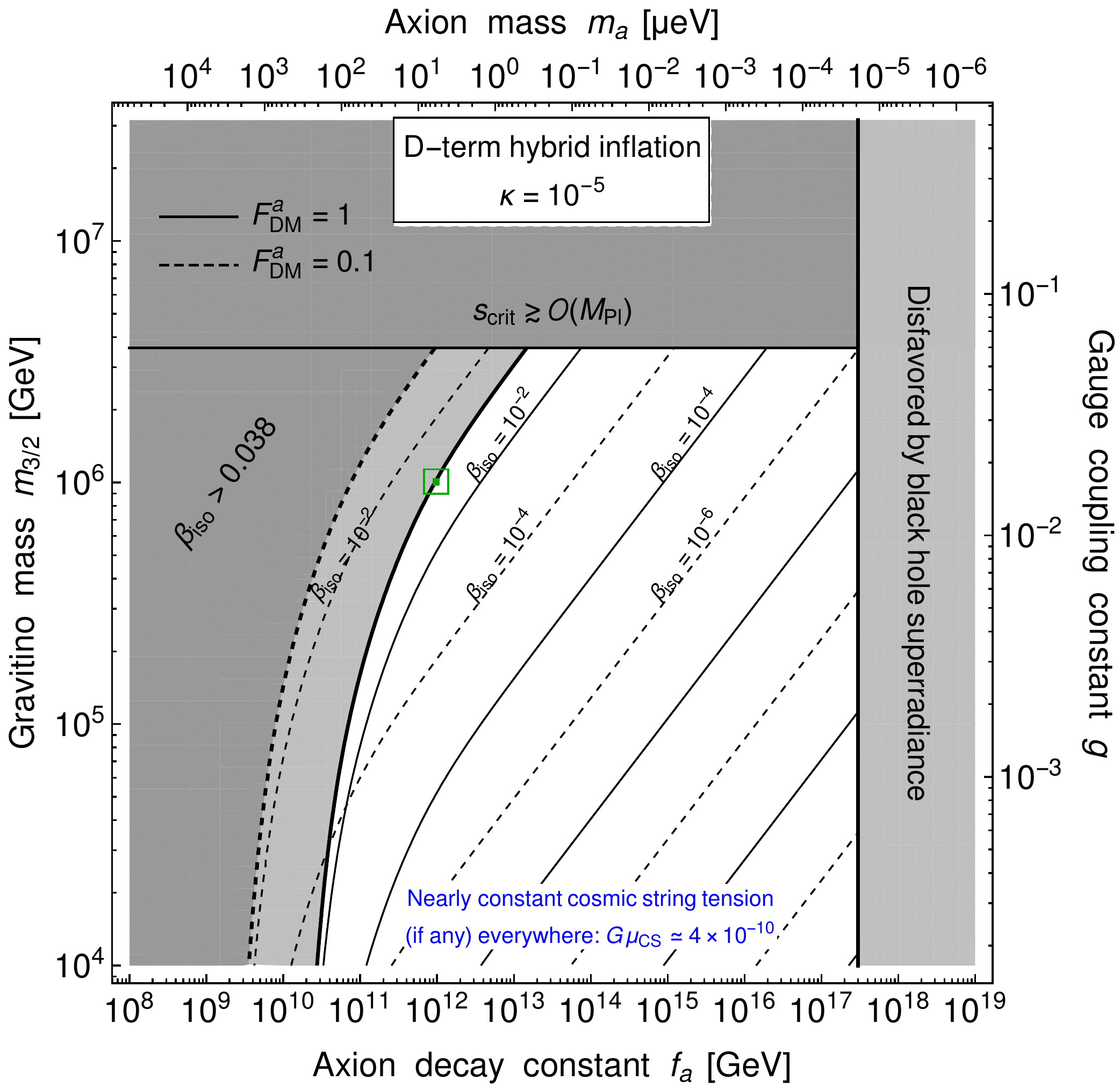}

\caption{Bounds on the parameter space of \textbf{(upper left panel)} F-term hybrid inflation
for all possible $\kappa$ values, \textbf{(upper right panel)} D-term hybrid inflation
for $\kappa = 10^{-1}$, \textbf{(lower left panel)} D-term hybrid inflation
for $\kappa = 10^{-3}$, \textbf{(lower right panel)} D-term hybrid inflation
for $\kappa = 10^{-5}$.
All plots are based on a fully numerical analysis.
The analytical expressions for the upper bounds on the gravitino
mass can be found in Eqs.~\eqref{eq:m32_max_fhi_far},
\eqref{eq:m32_max_fhi_close}, \eqref{eq:m32_max_dhi_far}, and \eqref{eq:m32_max_dhi_close}.
The lower bounds that lead to the postinflationary PQSB scenario
can be found in Eqs.~\eqref{eq:kappa_min_fhi_far}, \eqref{eq:kappa_min_fhi_close},
\eqref{eq:g_min_dhi_far}, and \eqref{eq:g_min_dhi_close}.
Analytical expressions for the cosmic string tension are contained in
Eqs.~\eqref{eq:Gmu_fhi_far}, \eqref{eq:Gmu_fhi_close},
\eqref{eq:Gmu_dhi_far}, and \eqref{eq:Gmu_dhi_close}.
The constraints on the axion decay constant from black hole superradiance,
$3 \times 10^{17} \lesssim f_a/\textrm{GeV} \lesssim 10^{19}$,
are taken from~\cite{Arvanitaki:2009fg,Arvanitaki:2010sy,Arvanitaki:2014wva}.
The red circle (in the upper left panel, at $f_a = 10^{17}\,\textrm{GeV}$ and
$m_{3/2} = 30\,\textrm{TeV}$) as well as the green square (in all other panels,
at $f_a = 10^{12}\,\textrm{GeV}$ and $m_{3/2} = 1000\,\textrm{TeV}$) denote the two 
benchmark points discussed in Sec.~\ref{sec:benchmark}.}

\label{fig:summary}

\end{center}
\end{figure}

%%%%%%%%%%%%%%%%%%%%%%%%%%%%%%%%%%%%%%%%%%%%%%%%%%%%%%%%%%%%%%%%%%%%%%%%%%%%%%%%%%%%%%%%%%%%%%%%%%%%

\section{Benchmark scenarios}
\label{sec:benchmark}

%%%%%%%%%%%%%%%%%%%%%%%%%%%%%%%%%%%%%%%%%%%%%%%%%%%%%%%%%%%%%%%%%%%%%%%%%%%%%%%%%%%%%%%%%%%%%%%%%%%%

In the two previous sections, we presented a detailed slow-roll analysis
that enabled us to assess the consequences of the CDM isocurvature
constraint in Eq.~\eqref{eq:Hinfmax} for supersymmetric hybrid inflation.
Our main results are summarized in Fig.~\ref{fig:summary}.
The four plots in this figure show the various upper and lower bounds
that we obtained throughout our analysis in dependence
of the axion decay constant $f_a$.
For both FHI and DHI, we conclude that it turns out to be quite easy to satisfy
the requirement of small axion isocurvature perturbations.
In fact, for both types of supersymmetric hybrid inflation, we find large
regions in parameter space that are consistent with all observational
constraints, including the measured values of the inflationary
CMB observables $A_s$ and $n_s$ as well as the upper bound
on the cosmic string tension $G\,\mu_{\rm CS}$.
To explore the physical implications of our results
a bit further, let us now elaborate on two characteristic benchmark scenarios
in the viable part of parameter space (see the red circle in the upper
left panel of Fig.~\ref{fig:summary} as well as the green square
in all other panels of Fig.~\ref{fig:summary}).
This discussion will help us illustrate in an exemplary fashion the
possible conclusions that one can draw from our numerical results in Fig.~\ref{fig:summary}.
For both benchmark points, we list the defining parameter values
as well as the corresponding predictions for all dependent quantities
in Tab.~\ref{tab:benchmark}.

%%%%%%%%%%%%%%%%%%%%%%%%%%%%%%%%%%%%%%%%%%%%%%%%%%%%%%%%%%%%%%%%%%%%%%%%%%%%%%%%%%%%%%%%%%%%%%%%%%%%

\paragraph*{Benchmark point I}

%%%%%%%%%%%%%%%%%%%%%%%%%%%%%%%%%%%%%%%%%%%%%%%%%%%%%%%%%%%%%%%%%%%%%%%%%%%%%%%%%%%%%%%%%%%%%%%%%%%%

First, let us consider FHI for $m_{3/2} = 30\,\textrm{TeV}$, $f_a = 10^{17}\,\textrm{GeV}$,
and $F_{\rm DM}^a = 1$.
Here, the large gravitino mass is characteristic for models of
high-scale supersymmetry that mostly rely on gravitational effects to
mediate the spontaneous breaking of supersymmetry
to the visible sector~\cite{Giudice:1998xp,Wells:2003tf,Wells:2004di}.
A minimal example for such a scenario is, \textit{e.g.}, the mediation scheme of
pure gravity mediation~\cite{Ibe:2006de,Ibe:2011aa,Ibe:2012hu}
(see also \cite{ArkaniHamed:2012gw}).
The large value of the axion decay constant is inspired
by string theory, which typically predicts $f_a$ values 
of the order of $f_a \sim 10^{16}\cdots
10^{17}\,\textrm{GeV}$~\cite{Svrcek:2006yi,Arvanitaki:2009fg,Cicoli:2012sz}.
For $f_a = 10^{17}\,\textrm{GeV}$, the axion is expected to have a tiny
mass, $m_a \simeq 5.7\times10^{-5}\,\textrm{\micro\hspace{0.5pt}eV}$
(see Eq.~\eqref{eq:ma}).
Remarkably enough, this falls into the range of masses that
might be probed by CASPEr~\cite{Budker:2013hfa}, a proposed magnetometry
experiment that aims at measuring the precession of nuclear spins induced
by their interaction with the axion DM background.
At the same time, an axion decay constant as large as 
$f_a = 10^{17}\,\textrm{GeV}$ implies that the initial misalignment
angle $\bar{\theta}_{\rm ini}$ must be fine-tuned to a relatively
small value, $\bar{\theta}_{\rm ini}/\pi \simeq 3.5\times10^{-4}$, in order to avoid
the overproduction of DM (see Eq.~\eqref{eq:thetaini}).%
\footnote{As recently pointed out in~\cite{Graham:2018jyp,Guth:2018hsa}, this conclusion
can be evaded in inflation models with an extremely small Hubble rate,
$H_{\rm inf} \lesssim \Lambda_{\rm QCD}$.
However, to realize such a small Hubble rate in the context of supersymmetric
hybrid inflation, we would have to assume tiny
coupling constants (see Eqs.~\eqref{eq:Hinf_fhi_close} and \eqref{eq:Hinf_dhi_close})
and, hence, a strongly fine-tuned initial inflaton field value.
For this reason, we shall ignore the possibility of sub-$\Lambda_{\rm QCD}$ inflation
in this paper.}
Such a small value may, \textit{e.g.}, be the outcome of
anthropic selection in a landscape of string vacua.
But irrespective of that, one should also keep in mind that tuning
$\bar{\theta}_{\rm ini}$ at the level of $1$ in $10^3$
is certainly less severe than tuning the bare vacuum angle
$\bar{\theta}$ to a value less than $10^{-10}$ by brute force.

%%%%%%%%%%%%%%%%%%%%%%%%%%%%%%%%%%%%%%%%%%%%%%%%%%%%%%%%%%%%%%%%%%%%%%%%%%%%%%%%%%%%%%%%%%%%%%%%%%%%

The two observational constraints $A_s = A_s^{\rm obs}$ and $n_s = n_s^{\rm obs}$
effectively reduce the viable parameter space of FHI to a one-dimensional hypersurface
(see Fig.~\ref{fig:bounds_fhi}).
Our choice of $m_{3/2}$ therefore fixes all other model parameters of FHI.
For $m_{3/2} = 30\,\textrm{TeV}$, consistency with the scalar CMB power spectrum
requires an inflaton Yukawa coupling $\kappa \simeq 1.7\times 10^{-3}$ and
an inflaton F-term mass scale $\mu_S \simeq 1.1 \times 10^{14}\,\textrm{GeV}$.
In view of the critical $\kappa$ value in Eq.~\eqref{eq:kappacrit_fhi},
$\kappa_0 \simeq 1.8 \times 10^{-3}$, this indicates that benchmark point I
is located just in the transition region in between the large-$\kappa$ regime and
the small-$\kappa$ regime.
As a consequence, the initial inflaton field value needs to be slightly tuned,
so as to make sure that the inflaton rolls into the correct direction
in field space (see the right panel of Fig.~\ref{fig:potential_fhi}).
However, compared to the situation deep inside the small-$\kappa$ regime,
this tuning is still comparatively mild.
The required values of $\kappa$ and $\mu_S$, moreover, imply a SSB scale during the waterfall
transition of $v \simeq 3.6\times 10^{15}\,\textrm{GeV}$.
This value lies within an order of magnitude of the GUT scale,
$\Lambda_{\rm GUT} \sim 10^{16}\,\textrm{GeV}$, which might hint
at a possible connection between FHI and grand unification.

%%%%%%%%%%%%%%%%%%%%%%%%%%%%%%%%%%%%%%%%%%%%%%%%%%%%%%%%%%%%%%%%%%%%%%%%%%%%%%%%%%%%%%%%%%%%%%%%%%%%

The required F-term mass scale $\mu_S$ also determines the Hubble rate
during inflation, $H_{\rm inf} \simeq 2.6 \times 10^9\,\textrm{GeV}$.
Given the large value of the axion decay constant, this
result complies with the CDM isocurvature constraint in Eq.~\eqref{eq:Hinfmax}.
In fact, for $H_{\rm inf} \simeq 2.6 \times 10^9\,\textrm{GeV}$,
$f_a = 10^{17}\,\textrm{GeV}$, and $F_{\rm DM}^a = 1$, we expect a primordial
isocurvature fraction $\beta_{\rm iso} \simeq 2.5 \times 10^{-2}$,
which is smaller than the current upper bound on $\beta_{\rm iso}$
by roughly $30\,\%$.
The observational sensitivity to $\beta_{\rm iso}$ is limited by cosmic
variance and predicted to be around $\beta_{\rm iso} \simeq 10^{-2}$
(see, \textit{e.g.},~\cite{Hamann:2009yf}).
An ultimate CMB experiment limited only by
cosmic variance may therefore be able
to detect the primordial axion isocurvature perturbations that contribute to the scalar
CMB power spectrum in this benchmark scenario.
These are exciting prospects that illustrate how future axion and CMB
experiments will help shed more light on the possible interplay of
supersymmetry breaking, inflation, and axion physics.

%%%%%%%%%%%%%%%%%%%%%%%%%%%%%%%%%%%%%%%%%%%%%%%%%%%%%%%%%%%%%%%%%%%%%%%%%%%%%%%%%%%%%%%%%%%%%%%%%%%%

Finally, we comment on the issue of cosmic strings in benchmark scenario I.
For $\kappa = 1.7\times10^{-3}$, we predict a cosmic string tension
$G\,\mu_{\rm CS} \simeq 9.7 \times 10^{-8}$, which just falls
short of the current upper bound in Eq.~\eqref{eq:GmuCSmax}.
Therefore, if cosmic strings should, indeed, be produced during the waterfall
transition at the end of inflation, any improvement over the current
bound in the near future should provide clues for the presence of cosmic strings.
On the other hand, we caution that a nondetection of cosmic strings
would not immediately rule out benchmark point I.
In this case, the local gauge symmetry in the waterfall sector may simply be broken
in a different sector already before the end of inflation
(see the discussion below Eq.~\eqref{eq:vCS}).
The same conclusion applies if one contrasts our prediction
$G\,\mu_{\rm CS} \simeq 9.7 \times 10^{-8}$ with less conservative bounds 
on $G\,\mu_{\rm CS}$ (see,
\textit{e.g.},~\cite{Sanidas:2012ee,Blanco-Pillado:2017rnf,Ringeval:2017eww,Mota:2014uka}). 

%%%%%%%%%%%%%%%%%%%%%%%%%%%%%%%%%%%%%%%%%%%%%%%%%%%%%%%%%%%%%%%%%%%%%%%%%%%%%%%%%%%%%%%%%%%%%%%%%%%%

\begin{table}[t]
\renewcommand{\arraystretch}{1.2}
\begin{center}
\begin{tabular}{|l|c||c|c|c|c|}
\hline
                                            &                          & Benchmark point I   & \multicolumn{3}{c|}{Benchmark point II}                       \\ \hline\hline
Inflation model                             &                          & FHI                 & \multicolumn{3}{c|}{DHI}                                      \\ \hline \hline
Gravitino mass $m_{3/2}$                    & [TeV]                    & $30$                & \multicolumn{3}{c|}{$1000$}                                   \\
Axion decay constant $f_a$                  & [GeV]                    & $10^{17}$           & \multicolumn{3}{c|}{$10^{12}$}                                \\ 
Axion DM fraction $F_{\rm DM}^a$            & [\%]                     & $100$               & \multicolumn{3}{c|}{$100$}                                    \\ \hline \hline
Axion mass $m_a$                            & [\micro\hspace{0.5pt}eV] & $5.7\times10^{-5}$  & \multicolumn{3}{c|}{$5.7$}                                    \\
Misalignment angle $\bar{\theta}_{\rm ini}$ & [$\pi$]                  & $3.5\times10^{-4}$  & \multicolumn{3}{c|}{$0.28$}                                   \\ \hline \hline
Yukawa coupling $\kappa$                    &                          & $1.7\times10^{-3}$  & $10^{-5}$           & $10^{-3}$          & $10^{-1}$          \\
Gauge coupling $g$                          &                          & $0.72$              & $1.7\times10^{-2}$  & $6.1\times10^{-6}$ & $1.4\times10^{-6}$ \\ \hline \hline
SSB scale $v$                               & [GeV]                    & $3.6\times10^{15}$  & $1.3\times10^{14}$  & $4.7\times10^{15}$ & $1.0\times10^{16}$ \\ 
F-term mass scale $\mu_S$                   & [GeV]                    & $1.1\times10^{14}$ & ------              & ------             & ------             \\ 
FI mass scale $\sqrt{\xi}$                  & [GeV]                    & ------              & $9.1\times10^{13}$  & $3.3\times10^{15}$ & $7.0\times10^{15}$ \\ 
Hubble rate $H_{\rm inf}$                   & [GeV]                    & $2.6\times10^9$     & $2.3\times10^7$     & $1.1\times10^7$    & $1.2\times10^7$    \\ \hline \hline
CS tension $G\,\mu_{\rm CS}$ (if any)       &                          & $9.7\times10^{-8}$  & $3.5\times10^{-10}$ & $4.6\times10^{-7}$ & $2.1\times10^{-6}$ \\
Isocurvature fraction $\beta_{\rm iso}$     &                          & $2.5\times10^{-2}$  & $3.5\times10^{-2}$  & $8.4\times10^{-3}$ & $9.6\times10^{-3}$ \\ \hline
\end{tabular}

\caption{Parameter values and predictions for several observables for the two benchmark
points discussed in Sec.~\ref{sec:benchmark}.
Benchmark point I is based on F-term hybrid inflation and assumes that the QCD axion
has its dynamical origin in string theory (\textit{viz.}, $f_a = 10^{17}\,\textrm{GeV}$).
Benchmark point II is, by contrast, based on D-term hybrid inflation and assumes that the QCD axion
has its dynamical origin in field theory (\textit{viz.}, $f_a = 10^{12}\,\textrm{GeV}$).
Both points are also shown in Fig.~\ref{fig:summary}
(see the red circle in the upper left panel of Fig.~\ref{fig:summary} as well as the
green square in all other panels of Fig.~\ref{fig:summary}).}

\label{tab:benchmark}

\end{center}
\end{table}

%%%%%%%%%%%%%%%%%%%%%%%%%%%%%%%%%%%%%%%%%%%%%%%%%%%%%%%%%%%%%%%%%%%%%%%%%%%%%%%%%%%%%%%%%%%%%%%%%%%%

\paragraph*{Benchmark point II}

%%%%%%%%%%%%%%%%%%%%%%%%%%%%%%%%%%%%%%%%%%%%%%%%%%%%%%%%%%%%%%%%%%%%%%%%%%%%%%%%%%%%%%%%%%%%%%%%%%%%

Next, we consider DHI for $m_{3/2} = 1000\,\textrm{TeV}$, $f_a = 10^{12}\,\textrm{GeV}$,
and $F_{\rm DM}^a = 1$.
Again, the large value of the gravitino mass is inspired by high-scale SUSY scenarios
such as pure gravity mediation.
Now, however, we choose $m_{3/2}$ towards the upper end of the expected range of values.
Such a large gravitino mass may
be instrumental in suppressing the rate of dangerous
flavor-changing neutral currents~\cite{Bhattacherjee:2012ed}.
Meanwhile, the chosen value of the axion decay constant corresponds
to the upper end of the classical axion window that allows
to generate axion DM without any fine-tuning in the initial
misalignment angle (see Eq.~\eqref{eq:Oah2anh}).
Indeed, for $f_a = 10^{12}\,\textrm{GeV}$, we require an initial
misalignment angle $\bar{\theta}_{\rm ini}/\pi \simeq 0.28$ to achieve
pure axion DM (\textit{i.e.}, $F_{\rm DM}^a = 1$),
which is a natural value.
At the same time, an axion decay constant
$f_a = 10^{12}\,\textrm{GeV}$ is a typical value that
can be easily realized in field-theoretic implementations
of the Peccei-Quinn mechanism (see, \textit{e.g.},~\cite{Harigaya:2013vja,Harigaya:2015soa}).
An important consequence of the lower axion decay constant
compared to benchmark scenario I is a correspondingly heavier axion,
$m_a \simeq 5.7\,\textrm{\micro\hspace{0.5pt}eV}$ (see Eq.~\eqref{eq:ma}).
Axion DM in this mass range will be probed by
ADMX~\cite{Stern:2016bbw} and CULTASK~\cite{Chung:2016ysi},
two microwave cavity experiments that aim at detecting the resonant conversion
of axions into photons in a strong magnetic field.

%%%%%%%%%%%%%%%%%%%%%%%%%%%%%%%%%%%%%%%%%%%%%%%%%%%%%%%%%%%%%%%%%%%%%%%%%%%%%%%%%%%%%%%%%%%%%%%%%%%%

To make use of the observational constraints $A_s = A_s^{\rm obs}$
and $n_s = n_s^{\rm obs}$, we need to fix one more model parameter.
For definiteness, we will take this parameter to be the
Yukawa coupling $\kappa$ and compare the predictions of DHI for
three different $\kappa$ values in the following, $\kappa = 10^{-5}$, $10^{-3}$, $10^{-1}$
(see the upper right and the two lower panels of Fig.~\ref{fig:summary}).
These values are chosen such that they give a characteristic impression
of the viable parameter space for small ($\kappa = 10^{-5}$),
intermediate ($\kappa = 10^{-3}$), and large $\kappa$ values ($\kappa = 10^{-1}$).
Together with our choice of $m_{3/2}$, the three benchmark values for $\kappa$
allow us to determine the gauge coupling constant $g$, the SSB scale $v$,
the FI mass scale $\sqrt{\xi}$, and the Hubble rate $H_{\rm inf}$
(see Fig.~\ref{fig:bounds_dhi} and Tab.~\ref{tab:benchmark}).
From the numerical results in Tab.~\ref{tab:benchmark}, it is evident that,
among the three $\kappa$ values under consideration, $\kappa = 10^{-1}$
is arguably the most attractive one.
Not only does it require the least tuning of the initial inflaton field value
(see Fig.~\ref{fig:potential_dhi}), but it also results in a SSB scale of exactly
$v = 1.0 \times 10^{16}\,\textrm{GeV}$.
Benchmark point II in combination with $\kappa = 10^{-1}$ therefore suggests
a possible connection between DHI and grand unification.

%%%%%%%%%%%%%%%%%%%%%%%%%%%%%%%%%%%%%%%%%%%%%%%%%%%%%%%%%%%%%%%%%%%%%%%%%%%%%%%%%%%%%%%%%%%%%%%%%%%%

The three different values of the Hubble rate in
Tab.~\ref{tab:benchmark} are all of the same order of magnitude,
$H_{\rm inf} \sim 10^7\,\textrm{GeV}$.
By construction, these values are small enough to comply with the CDM 
isocurvature constraint in Eq.~\eqref{eq:Hinfmax}.
Recall that, depending on the precise value of $\kappa$, a suppressed Hubble
rate can be either achieved by a small gauge coupling $g$ (see Eq.~\eqref{eq:Hinf_dhi_far})
or by a small Yukawa coupling $\kappa$ (see Eq.~\eqref{eq:Hinf_dhi_close}).
This is also reflected in the different $\kappa$ and $g$ values
in Tab.~\ref{tab:benchmark}.
In combination with $f_a = 10^{12}\,\textrm{GeV}$ and $F_{\rm DM}^a = 1$,
the $H_{\rm inf}$ values in Tab.~\ref{tab:benchmark} allow us to compute
the primordial isocurvature fraction $\beta_{\rm iso}$.
For $\kappa = 10^{-3}$ and $\kappa = 10^{-1}$, we find $\beta_{\rm iso} \sim 10^{-2}$,
which may or may not be within reach of an ultimate CMB experiment.
For $\kappa = 10^{-5}$, on the other hand, we obtain
$\beta_{\rm iso} \simeq 3.5 \times 10^{-2}$,
which is only roughly $8\,\%$ smaller than the current upper bound.
Here, the fact that we find different predictions for $\beta_{\rm iso}$ in dependence
of $\kappa$ is a consequence of the slightly different $\kappa$ dependence
of $H_{\rm inf}$ and $m_{3/2}$ in Eqs.~\eqref{eq:Hinf_dhi_close} and \eqref{eq:m32_dhi_close}.
From this perspective, smaller $\kappa$ values appear more favorable, as they
push $\beta_{\rm iso}$ further into the observable range.
Similarly, smaller Yukawa couplings also help suppress the cosmic string tension
(see Tab.~\ref{tab:benchmark}).
Indeed, only for $\kappa = 10^{-5}$, we find a cosmic string tension in accord with
the upper bound in Eq.~\eqref{eq:GmuCSmax}.
For $\kappa = 10^{-3}$ and $\kappa = 10^{-1}$, we have to assume again
that no cosmic strings are produced during the waterfall transition.

%%%%%%%%%%%%%%%%%%%%%%%%%%%%%%%%%%%%%%%%%%%%%%%%%%%%%%%%%%%%%%%%%%%%%%%%%%%%%%%%%%%%%%%%%%%%%%%%%%%%

\section{Conclusions}
\label{sec:conclusions}

%%%%%%%%%%%%%%%%%%%%%%%%%%%%%%%%%%%%%%%%%%%%%%%%%%%%%%%%%%%%%%%%%%%%%%%%%%%%%%%%%%%%%%%%%%%%%%%%%%%%

The PQ mechanism constitutes a well-motivated BSM scenario
that offers not only an attractive solution to the strong $CP$ problem
but also a viable particle candidate for DM\,---\,the QCD axion.
A consistent implementation of the PQ mechanism into inflationary cosmology can,
however, be challenging, depending on the details of the underlying
model of PQ symmetry breaking.
That is, if the global PQ symmetry is spontaneously broken only after inflation,
one encounters a domain wall problem, unless domain walls decay
sufficiently fast for one  reason or another.
On the other hand, if the global PQ symmetry is already broken during
inflation and not restored afterwards, quantum fluctuations of the
axion field during inflation can result in primordial CDM isocurvature
perturbations that exceed the current upper bound on the primordial isocurvature
fraction $\beta_{\rm iso}$.

%%%%%%%%%%%%%%%%%%%%%%%%%%%%%%%%%%%%%%%%%%%%%%%%%%%%%%%%%%%%%%%%%%%%%%%%%%%%%%%%%%%%%%%%%%%%%%%%%%%%

The main purpose of this paper was to demonstrate that the 
\textit{axion isocurvature perturbations problem} in the 
\textit{preinflationary PQSB scenario} can be easily solved in 
the context of supersymmetric hybrid inflation.
To this end, we studied in detail the slow-roll dynamics of both FHI and DHI.
These models represent interesting inflationary scenarios that feature a rapid
second-order phase transition at the end of inflation, which can be identified with the
spontaneous breaking of a local GUT symmetry.
For both FHI and DHI, we explicitly accounted for the effect of spontaneous
SUSY breaking in a hidden Polonyi sector, which gave us
additional control over the shape of the scalar potential.
In FHI, the leading soft contribution to the scalar potential
turns out to be a linear tadpole term, while in DHI, one obtains a quadratic mass term.
The sizes of both terms are controlled by the gravitino mass $m_{3/2}$, and the signs
of both terms can be chosen so as to partially cancel various
contributions to the scalar potential.
In the case of FHI, this means that one has to consider inflation
on the negative real axis, where the coefficient of the soft tadpole term
in Eq.~\eqref{eq:cs} turns negative.
In DHI, on the other hand, one has to assume a higher-dimensional operator
in the K\"ahler potential, $K \supset \chi/M_{\rm Pl}^2 \left|S\right|^2 \left|X\right|^2$
with a large positive coefficient, $\chi > 1/3$,
such that the soft inflaton mass becomes tachyonic.
Provided these extra assumptions, one is able to render the scalar potential
particularly flat by tuning the soft SUGRA contributions against the
radiative corrections in the effective potential.
At the same time, the inflaton F-term mass scale $\mu_S$ (in the FHI case) as well as
the FI parameter $\sqrt{\xi}$ (in the DHI case) always allow one to adjust the
total energy scale of the scalar potential and, hence, reproduce the measured
amplitude of the scalar power spectrum.
Together, these two features of supersymmetric hybrid inflation represent
a powerful mechanism to suppress the inflationary Hubble rate $H_{\rm inf}$
and, thus, solve the axion isocurvature perturbations problem.

%%%%%%%%%%%%%%%%%%%%%%%%%%%%%%%%%%%%%%%%%%%%%%%%%%%%%%%%%%%%%%%%%%%%%%%%%%%%%%%%%%%%%%%%%%%%%%%%%%%%

Both FHI and DHI can occur for small as well as for relatively large field excursions,
depending on the inflaton Yukawa coupling $\kappa$.
In our analysis, we therefore had to distinguish twice between a small-$\kappa$
regime (where $\kappa \lesssim 10^{-3}$ and $s_* \simeq s_{\rm crit}$)
and a large-$\kappa$ regime (where $\kappa \gtrsim 10^{-3}$ and
$s_* \gg s_{\rm crit}$).
In a first step, we considered FHI in the large-$\kappa$ regime.
As we were able to show, this scenario turns out to be heavily restricted
by the CDM isocurvature constraint (see Eqs.~\eqref{eq:kappa_max_fhi_far}
and \eqref{eq:m32_max_fhi_far}).
In fact, only axion decay constants of the order of the Planck
scale, $f_a \sim M_{\rm Pl}$, allow to sufficiently suppress
the isocurvature power spectrum in this case.
The reason for this is the lack of parametric
freedom in the large-$\kappa$ regime of FHI.
That is, as long as one restricts oneself to large Yukawa couplings only,
the Hubble rate automatically ends up being rather large
(see Eq.~\eqref{eq:Hinf_fhi_far}).
However, an axion decay constant as large as the Planck scale is disfavored
for various reasons.
On the theory side, string theory typically predicts
sub-Planckian values of $f_a$, while from the phenomenological perspective,
current bounds from black hole superradiance seem to exclude $f_a \sim M_{\rm Pl}$.
These issues can be avoided in the small-$\kappa$ regime of FHI, which 
offers the possibility to suppress the Hubble rate 
by means of the small Yukawa coupling $\kappa$ (see Eq.~\eqref{eq:Hinf_fhi_close}).
Consequently, the small-$\kappa$ regime of FHI complies with the CDM isocurvature
constraint for all reasonable values of $f_a$ (see
Eqs.~\eqref{eq:kappa_max_fhi_close} and \eqref{eq:m32_max_fhi_close}).
On top of that, small $\kappa$ values also suppress the tension of cosmic
strings (see Eq.~\eqref{eq:Gmu_fhi_close}), such that the production of cosmic
strings during the waterfall transition no longer poses a potential threat.
These virtues of the small-$\kappa$ regime, however, come at the cost
of a fine-tuned initial inflaton field value.
For small Yukawa couplings, one has to ensure that inflation begins on
the correct side of a local maximum in the scalar potential,
$s_{\rm crit} < s_{\rm ini} < s_{\rm max}$, where $s_{\rm crit}$ and $s_{\rm max}$
lie very close together.
Otherwise, the inflaton will roll into the wrong direction in
field space and become trapped in a wrong vacuum.
This situation is further complicated by the fact that FHI is, in reality,
a two-field model of inflation (see Eq.~\eqref{eq:cs}) that can result
in complicated trajectories in the complex inflaton plane.%
\footnote{We emphasize that the parameter bounds that we derived
in Sec.~\ref{sec:fhi} are \textit{inclusive} in the sense that they
are always applicable, irrespective of the particular inflaton trajectory in field space
(see the discussion below Eq.~\eqref{eq:cs}).}

%%%%%%%%%%%%%%%%%%%%%%%%%%%%%%%%%%%%%%%%%%%%%%%%%%%%%%%%%%%%%%%%%%%%%%%%%%%%%%%%%%%%%%%%%%%%%%%%%%%%

Because of these limitations of FHI, we turned
to DHI in Sec.~\ref{sec:dhi}.
Not only is DHI a standard single-field model of inflation, it also introduces a
larger parametric freedom through its dependence on the gauge coupling $g$.
As we were able to demonstrate, this extra freedom allows one to decrease the Hubble
rate to very small values even in the large-$\kappa$ regime (see Eqs.~\eqref{eq:Hinf_dhi_far}).
This is a characteristic advantage of DHI over FHI, which explains why
DHI in the large-$\kappa$ regime can be made consistent with the CDM
isocurvature constraint for a large range of $f_a$ values
(see Eqs.~\eqref{eq:g_max_dhi_far} and \eqref{eq:m32_max_dhi_far}).
The only remaining issue in this scenario is the possible presence of cosmic
strings with a large energy per unit length.
It may well be that the local gauge symmetry in the waterfall sector is already
broken during inflation for some reason or another (see the discussion
below Eq.~\eqref{eq:vCS}).
In this case, one does not have to worry about the production of cosmic strings.
However, if cosmic strings should, indeed, be produced at the end of inflation,
one must resort again to the small-$\kappa$ regime, so as to suppress the cosmic string tension
by means of a small Yukawa coupling (see Eq.~\eqref{eq:Gmu_dhi_close}).
Similarly to the case of FHI, this scenario allows for an efficient suppression
of the Hubble rate (see Eq.~\eqref{eq:Hinf_dhi_close}), which is why it readily satisfies
the CDM isocurvature constraint for a broad range of $f_a$ values
(see Eqs.~\eqref{eq:g_max_dhi_close} and \eqref{eq:m32_max_dhi_close}).
The only drawback in this case is the need for a fine-tuned initial inflaton field
value.
This time, however, one does not have to deal with complicated trajectories in field space.

%%%%%%%%%%%%%%%%%%%%%%%%%%%%%%%%%%%%%%%%%%%%%%%%%%%%%%%%%%%%%%%%%%%%%%%%%%%%%%%%%%%%%%%%%%%%%%%%%%%%

For both FHI and DHI, we found that at least one coupling constant needs to be
set to a value of $\mathcal{O}\left(10^{-3}\right)$ or smaller.
In the case of FHI, this coupling corresponds to the Yukawa coupling $\kappa$,
while for DHI, it typically corresponds to the gauge coupling $g$.
In both cases, we argued that such a small coupling constant is stable against quantum
corrections and, hence, technically natural.
Supersymmetric hybrid inflation is therefore able to solve the axion isocurvature
perturbations problems without any unnatural fine-tuning of model parameters.
In addition, we showed how the upper bounds on $\kappa$ and $g$ translate
into upper bounds on the gravitino mass.
For FHI, we obtained $m_{3/2} \lesssim \mathcal{O}\left(10^5\right)\,\textrm{GeV}$,
while for DHI, we obtained $m_{3/2} \lesssim \mathcal{O}\left(10^9\right)\,\textrm{GeV}$.
These observations helped us identify interesting benchmark points in parameter space
(see Sec.~\ref{sec:benchmark}), which will be probed by upcoming
axion and CMB experiments.
Possible signatures of our benchmark points include (i) an axion mass
that may be detected by axion experiments such as CASPEr, ADMX, or CULTASK,
(ii) a cosmic string tension just below the current upper bound, and
(iii) a primordial isocurvature fraction that could be measured by an
ultimate purely cosmic-variance-limited CMB experiment.

%%%%%%%%%%%%%%%%%%%%%%%%%%%%%%%%%%%%%%%%%%%%%%%%%%%%%%%%%%%%%%%%%%%%%%%%%%%%%%%%%%%%%%%%%%%%%%%%%%%%

We also emphasize that, thanks to our analysis, a future detection of axion DM with a decay
constant $f_a \sim 10^{11}\cdots10^{12}\,\textrm{GeV}$
would provide us with important clues regarding the expected scheme for the
mediation of spontaneous SUSY breaking to the visible sector.
If interpreted in the context of FHI, such a value would point towards
gravitino masses below the electroweak scale, $m_{3/2} \lesssim 1\cdots10\,\textrm{GeV}$,
which would suggest that SUSY breaking is communicated to the visible
sector via gauge mediation.
In the context of DHI, on the other hand, the detection of axion DM with
$f_a \sim 10^{11}\cdots10^{12}\,\textrm{GeV}$ would provide
with us a weaker bound on the gravitino mass,
$m_{3/2} \lesssim 10^5 \cdots 10^6\,\textrm{GeV}$.
This would, in turn, be compatible with the idea of high-scale SUSY breaking in combination
with a mediation scheme such as pure gravity mediation.
In either case, we conclude that the results of our analysis allow us
to derive highly nontrivial statements regarding the energy scale of soft
SUSY breaking from cosmological arguments.
In this sense, the CDM isocurvature constraint on $H_{\rm inf}$ in the QCD
axion scenario proves to be a remarkably powerful tool to constrain possible BSM scenarios.

%%%%%%%%%%%%%%%%%%%%%%%%%%%%%%%%%%%%%%%%%%%%%%%%%%%%%%%%%%%%%%%%%%%%%%%%%%%%%%%%%%%%%%%%%%%%%%%%%%%%

In this paper, we focused on the slow-roll dynamics of supersymmetric hybrid
inflation as well as on the implications of the CDM isocurvature constraint
on its parameter space.
At this point, it is worth stressing that our analytical results in Secs.~\ref{sec:fhi}
and \ref{sec:dhi} are, in fact, extremely general and, thus, well suited for further
investigations of supersymmetric hybrid inflation. 
Moreover, it is clear that we refrained from embedding our setup into a comprehensive
cosmological scenario that coherently describes the evolution of the Universe
from very early to very late times.
This is, \textit{e.g.}, illustrated by the fact that we merely used the gravitino
mass $m_{3/2}$ as a free input parameter.
We did not specify the dynamical origin of $m_{3/2}$, nor did we assume an explicit
scheme for the mediation of spontaneous SUSY breaking to the visible sector.
Similarly, we did not speculate about the possible composition of DM, in the case 
in which it should not consist exclusively of axions (\textit{i.e.}, for $F_{\rm DM}^a < 1$).
Any extra assumption related to these issues would prompt a more careful analysis
regarding the interplay of DM production, heavy particle decays, big bang nucleosynthesis, etc.
However, such a more complete analysis is beyond the scope of this paper and left
for future work.
We conclude our discussion by stressing once more that supersymmetric hybrid inflation
is a prime candidate for a model of inflation that offers a viable solution to the
axion isocurvature perturbations problem.

%%%%%%%%%%%%%%%%%%%%%%%%%%%%%%%%%%%%%%%%%%%%%%%%%%%%%%%%%%%%%%%%%%%%%%%%%%%%%%%%%%%%%%%%%%%%%%%%%%%%

\subsubsection*{Acknowledgements}

%%%%%%%%%%%%%%%%%%%%%%%%%%%%%%%%%%%%%%%%%%%%%%%%%%%%%%%%%%%%%%%%%%%%%%%%%%%%%%%%%%%%%%%%%%%%%%%%%%%%

The authors are grateful to the organizers of the
\textit{Kavli IPMU 10$^{\textrm{\textit{th}}}$ Anniversary Symposium}
held in October 2017 in Kashiwa, Japan, where this work was initiated.
We wish Kavli IPMU a bright future and are looking forward to its
20$^{\textrm{th}}$ anniversary.
The authors thank Marco Gorghetto, Andrew Long, and Fuminobu Takahashi
for useful discussions and valuable comments.
This project has received support from the European Union's Horizon 2020
research and innovation programme under the Marie Sk\l odowska-Curie grant
agreement No.\ 674896 (K.\,S.).
It has also been supported by Grants-in-Aid for Scientific Research from
the Ministry of Education, Culture, Sports, Science, and Technology (MEXT),
Japan, No.\ 26104001, No.\ 26104009, No.\ 16H02176, and No.\ 17H02878 (T.\,T.\,Y.)
and by the World Premier International Research Center Initiative, MEXT,
Japan (T.\,T.\,Y.).

%%%%%%%%%%%%%%%%%%%%%%%%%%%%%%%%%%%%%%%%%%%%%%%%%%%%%%%%%%%%%%%%%%%%%%%%%%%%%%%%%%%%%%%%%%%%%%%%%%%%

\bibliographystyle{JHEP}
\bibliography{arxiv_3}{}

%%%%%%%%%%%%%%%%%%%%%%%%%%%%%%%%%%%%%%%%%%%%%%%%%%%%%%%%%%%%%%%%%%%%%%%%%%%%%%%%%%%%%%%%%%%%%%%%%%%%

\end{document}